\documentclass[12pt,a4paper]{article}
\usepackage{amssymb,amsmath}
\usepackage{graphicx}
\usepackage{latexsym}
\usepackage{epic,eepic}
\usepackage[nosort]{cite}
\usepackage[hyperref,bulletsep]{collect}

\makeatletter
\let\old@makecaption=\@makecaption
\def\@makecaption{\small\old@makecaption}
\makeatother

\makeatletter \@addtoreset{equation}{section} \makeatother

\makeatletter
\let\old@startsection=\@startsection
\renewcommand{\@startsection}[6]{\old@startsection{#1}{#2}{#3}{#4}{#5}{#6\mathversion{bold}}}
\makeatother

\newcommand{\nln}{\nonumber\\}
\newcommand{\nl}{\nonumber\\&\hspace{-4\arraycolsep}&\mathord{}}

\newcommand{\earel}[1]{\mathrel{}&\hspace{-2\arraycolsep}#1\hspace{-2\arraycolsep}&\mathrel{}}
\newcommand{\eq}{\earel{=}}

\def\[{\begin{equation}}
\def\]{\end{equation}}
\def\<{\begin{eqnarray}}
\def\>{\end{eqnarray}}

\ifx\href\asklfhas\newcommand{\href}[2]{#2}\fi
\newcommand{\arxivno}[1]{\href{http://arxiv.org/abs/#1}{#1}}

\makeatletter
\def\mr@ignsp#1 {\ifx\:#1\@empty\else #1\expandafter\mr@ignsp\fi}%
\newcommand{\multiref}[1]{\begingroup%\let\protect\string%
\xdef\mr@no@sparg{\expandafter\mr@ignsp#1 \: }%
\def\mr@comma{}%
\@for\mr@refs:=\mr@no@sparg\do{\mr@comma\def\mr@comma{,}\ref{\mr@refs}}%
\endgroup}
\makeatother

\newcommand{\hypref}[2]{\ifx\href\asklfhas #2\else\href{#1}{#2}\fi}
\newcommand{\Secref}[1]{Section~\multiref{#1}}
\newcommand{\secref}[1]{Sec.~\multiref{#1}}
\newcommand{\Appref}[1]{Appendix~\multiref{#1}}
\newcommand{\appref}[1]{App.~\multiref{#1}}

\newcommand{\figref}[1]{Fig.~\multiref{#1}}
\renewcommand{\eqref}[1]{(\multiref{#1})}

\newcommand{\atopfrac}[2]{\genfrac{}{}{0pt}{}{#1}{#2}}
\newcommand{\sfrac}[2]{{\textstyle\frac{#1}{#2}}}
\newcommand{\half}{\sfrac{1}{2}}
\newcommand{\quarter}{\sfrac{1}{4}}

\newcommand{\mdl}{\mathbb{V}}
\newcommand{\alg}[1]{\mathfrak{#1}}

\newcommand{\indup}[1]{_{\mathrm{#1}}}
\newcommand{\indups}[1]{_{\mathrm{\scriptscriptstyle #1}}}
\newcommand{\superN}{\mathcal{N}}

\newcommand{\lrbrk}[1]{\left(#1\right)}
\newcommand{\bigbrk}[1]{\bigl(#1\bigr)}

\newcommand{\bigcomm}[2]{\big[#1,#2\big]}
\newcommand{\comm}[2]{[#1,#2]}

\newcommand{\acomm}[2]{\{#1,#2\}}
\newcommand{\bigacomm}[2]{\big\{#1,#2\big\}}

\newcommand{\set}[1]{\{#1\}}

\newcommand{\tr}{\mathop{\mathrm{tr}}\nolimits}

\newcommand{\beq}{\begin{equation}}
\newcommand{\eeq}{\end{equation}}
\newcommand{\beqs}{\begin{eqnarray}}
\newcommand{\eeqs}{\end{eqnarray}}
\newcommand{\vac}{|0\rangle}
\newcommand{\xxxx}{\times}
\newcommand{\xxxxd}{{\dot\times}}
\newcommand{\NNc}{N_{\mathrm{c}}}
\newcommand{\unit}{\frac{\alpha_{\mathrm{s}} \NNc}{2\pi}}
\newcommand{\as}{\alpha_{\mathrm{s}}}
\newcommand{\opident}{\mathbf{I}}
\newcommand{\opperm}{\mathbf{X}}
\newcommand{\opproj}{\mathbf{P}}
\newcommand{\opspin}{\mathbf{S}}
\newcommand{\opham}{\mathbf{H}}
\newcommand{\oplilham}{\mathbf{h}}
\newcommand{\opR}{\mathbf{R}}

\def\Spin{\mathbf{S}}
\def\Spinb{\mathbf{s}}
\def\spin{\boldsymbol{\tau}}
\def\spins{\boldsymbol{\sigma}}
\def\Scq{\Spin_l\cdot\Spin_{l+1}}

\def\eeee{{\,\rm e}\,}
\def\dddd{\partial}

\begin{document}

\renewcommand{\thefootnote}{\fnsymbol{footnote}}
\setcounter{footnote}{1}

\begin{titlepage}
\begin{flushright}
\footnotesize \texttt{\arxivno{hep-th/0412029}},\hfill \texttt{AEI
2004-088},\hfill \texttt{ITEP-TH-51/04},\hfill
\texttt{NSF-KITP-04-116},\hfill \texttt{PUTP-2139},\hfill
\texttt{UUITP-28/04} \vspace{0.3cm}
\end{flushright}

\begin{center}
{\LARGE \bfseries\mathversion{bold}
One-Loop QCD Spin Chain\\and its
Spectrum\par} \vspace{1.0cm}

{\large Niklas Beisert$^{a,b,c}$, Gabriele Ferretti$^d$,\\ Rainer Heise$^d$
and Konstantin Zarembo$^e$\footnote{Also at ITEP, Moscow, Russia}\par}
\vspace{0.8cm}

\textit{\small $^{a}$ Max-Planck-Institut f\"ur Gravitationsphysik,
Albert-Einstein-Institut,\\
Am M\"uhlenberg 1, 14476 Potsdam, Germany}
\vspace{2mm}

\textit{\small $^{b}$ Department of Physics, Princeton University,\\
Princeton, NJ 08544, USA}
\vspace{2mm}

\textit{\small $^{c}$ Kavli Institute for Theoretical Physics, University of California,\\
Santa Barbara, CA 93106, USA}
\vspace{2mm}

\textit{\small $^d$Institute for Theoretical Physics,\\
G{\"o}teborg University and Chalmers University of Technology, \\
412 96 G{\"o}teborg, Sweden}
\vspace{2mm}

\textit{\small $^e$Department of Theoretical Physics,\\
Uppsala University, 751 08 Uppsala, Sweden}
\vspace{3mm}

\texttt{nbeisert@princeton.edu}\\
\texttt{ferretti,rahe@fy.chalmers.se}\\
\texttt{konstantin.zarembo@teorfys.uu.se}
\par\vspace{0.8cm}

\end{center}

\begin{abstract}
We study the renormalization of gauge invariant operators in large~$\NNc$ QCD. We
compute the complete matrix of anomalous dimensions to leading order in the
't~Hooft coupling and study its eigenvalues. Thinking of the mixing matrix as
the Hamiltonian of a generalized spin chain we find a large integrable sector
consisting of purely gluonic operators constructed with self-dual field
strengths and an arbitrary number of derivatives. This sector contains the
true ground state of the spin chain and all the gapless excitations above it.
The ground state is essentially the anti-ferromagnetic ground state of a
XXX$_1$ spin chain and the excitations carry either a chiral spin quantum
number with relativistic dispersion relation or an anti-chiral one with
non-relativistic dispersion relation.

\end{abstract}

\end{titlepage}

\setcounter{page}{1}
\renewcommand{\thefootnote}{\arabic{footnote}}
\setcounter{footnote}{0}

%%%%%%%%%%%%%%%%%%%%%%%%%%%%%%%%%%%%%%%%%%%%%%%%%%%%%%%%%%%%%%%%%%%%%%%%%%%%%%%%
%%%%%%%%%%%%%%%%%%%%%%%%%%%%%%%%%%%%%%%%%%%%%%%%%%%%%%%%%%%%%%%%%%%%%%%%%%%%%%%%
\section{Introduction and summary}
\label{sec:1}

The computation of anomalous dimensions of composite operators is of
central importance in many areas of physics, most notably condensed matter
and particle physics. In particle physics, their phenomenological
interest stems from hadronic processes at high energy. For instance,
in the study of
deep inelastic scattering~\cite{Georgi:1951sr,Gross:1974cs}
one is led to consider the operator product expansion of two hadronic
currents.
The anomalous dimensions of twist-two operators appearing in the expansion
control the logarithmic deviations to the Bjorken scaling.
Considering operators with low twist amounts to considering only
operators made up of very few fundamental fields $F_{\mu\nu}$ and $q$
(two in most cases) and possibly many derivatives. These are the operators
that are important in collider physics.

On the other hand, the study of operators containing many fundamental fields
has received new impetus in connection with the attempts to construct a
string description of QCD at low
energies~\cite{'tHooft:1974jz,Polyakov:1987ez}.
There is probably still a long
way to go in fulfilling this dream (if indeed the idea is viable), but
for other gauge theories the answer has been found. We now have a string
theory description of the maximally supersymmetric $\superN=4$  $SU(\NNc)$ theory in terms
of type IIB strings in an $AdS_5\times S^5$
background~\cite{Maldacena:1998re, Gubser:1998bc, Witten:1998qj}.
This was rather
surprising at first since the $\superN=4$ theory is not confining and there are no
color flux tubes in the ordinary sense. The connection with string theory was
instead obtained by identifying the string spectrum with
the spectrum of anomalous
dimensions of the theory~\cite{Gubser:1998bc, Witten:1998qj}.

After many checks performed in the supergravity regime, a new important
step was taken in~\cite{Berenstein:2002jq} where it was shown that by taking
an appropriate limit of both sides of the correspondence one could match
anomalous dimensions of near BPS
operators with the corresponding string
states. On the gauge theory
side this amounts to considering long operators with large classical
dimension and R-charge. These operators are constructed by taking the
BPS operator
$\tr Z^J$, where $Z$ is one of the three chiral superfields making up the
$\superN=4$ multiplet, and adding a few ``impurities{}'', i.e. replacing some of
the $Z$ with other fields in the theory.
The analysis of near BPS states led to an agreement to all loops in
the 't Hooft coupling%
\footnote{We use the standard definition
$\as = g^2_{\mathrm{YM}}/4\pi$.}
$\as \NNc$~\cite{Santambrogio:2002sb}.

States with large quantum numbers should correspond to semiclassical
string states whether they are near BPS or not. In fact, as noticed
in~\cite{Gubser:2002tv} one can get a qualitative agreement between
the anomalous dimension of twist-two operators of large spin $S$
computed at small $\as \NNc$ and the energy of a classical rotating
string in $AdS$ space, valid for large values of $\as \NNc$. In both
cases, the anomalous dimension grows like $\log S$, one of the most
famous ``trade-marks{}'' of non-abelian gauge
theories~\cite{Georgi:1951sr,Gross:1974cs}. Of course, the
dependence of the coefficient of $\log S$ on the 't Hooft coupling
$\as \NNc$ is polynomial on the gauge theory side, as always in
perturbation theory, whereas it grows like $\sqrt{\as \NNc}$ in the
semiclassical $AdS$ analysis, a ``screening{}'' phenomenon that has
already been observed e.g. in the analysis of the Wilson
loops~\cite{Maldacena:1998im,Rey:1998ik,Erickson:2000af,Drukker:2000rr}.

The above qualitative observations can be made
quantitative by considering, once again, the $\superN=4$
theory~\cite{Frolov:2003qc,Beisert:2003xu,Frolov:2003tu,Beisert:2003ea,Engquist:2003rn,Arutyunov:2003rg,Tseytlin:2003ii,Frolov:2004bh,Serban:2004jf,Kristjansen:2004ei,Kazakov:2004qf}.
In this case the string can instead be allowed
to rotate in the compact $S^5$ dimensions. Giving the string 
two large angular momenta $J_1$ and $J_2$ on the sphere (recall
that the isometry group is $SO(6)$ and has rank three) one obtains an
expression for the energy that is analytic in $\as \NNc$ and thus can
be expanded around $\as \NNc=0$. In order to match this expansion with
the perturbative analysis one needs to compute the anomalous dimension
of operators that are far from the BPS condition. This is successfully
accomplished by noticing that the matrix of planar anomalous dimensions can
be interpreted as the Hamiltonian of an integrable spin chain --
not only at one-loop \cite{Minahan:2002ve,Beisert:2003yb},
but also at higher loops \cite{Beisert:2003tq,Beisert:2003ys}
--
and its eigenvalues can thus be obtained by applying the Bethe
ansatz~\cite{Minahan:2002ve,Beisert:2003yb,Serban:2004jf,Beisert:2004hm}.
For an introduction to this subject, c.f.~\cite{Beisert:2004ry}.
(See also \cite{Okuyama:2004bd,Roiban:2004va}
for a computation of correlation functions
{}from the Bethe ansatz.)

Following this line of thought,
Kruczenski~\cite{Kruczenski:2003gt,Kruczenski:2004kw}
showed that the agreement is already present at the level of the
effective sigma model
(see also~\cite{Hernandez:2004uw,Stefanski:2004cw,Agarwal:2004cb,Ideguchi:2004wm,Smedback:1998yn,Ryang:2004pu,Susaki:2004tg} for related work).
Namely, the effective action obtained by taking
the long wavelength limit of the one-loop spin chain Hamiltonian and
the action of the rotating string in the corresponding space-time
background are the same. This leads to the exciting possibility of
extracting information about the string dual from the perturbative
study of anomalous dimensions and was the main motivation behind the
previous work by some of us~\cite{Ferretti:2004ba} in the context of
large $\NNc$ (non-supersymmetric) QCD. In~\cite{Ferretti:2004ba} we showed
that the closed spin chains constructed with only the selfdual field strength
form an integrable subsector.
In this paper we continue this investigations in the hope that it will reveal
the nature of the excitations required in the description of the QCD string.

Before moving to describing the results of our investigation, we must stress
that the subject of integrability in the context of QCD
has a long parallel history starting with the introduction
and the study of operators on the light cone
in~\cite{Bukhvostov:1985rn}, (see also \cite{Lipatov:1994yb,Faddeev:1995zg})
and continuing with the discovery that
their renormalization is described by an integrable, non-compact spin
chain~\cite{Braun:1998id,Braun:1999te,Belitsky:1999bf}. The early papers on the
subject dealt only with ``short{}'' operators, i.e. operators
containing a small number of elementary fields but with an arbitrary
number of derivatives. We refer the reader to
the reviews~\cite{Braun:2003rp,Belitsky:2004cz}
for a survey of the literature on this subject.
Very recently, the complete matrix of anomalous dimension for
light-cone operators of arbitrary length has been given
in~\cite{Belitsky:2004sc} for theories with any number of
supersymmetries. One of the results in the present paper is the complete
one-loop matrix of anomalous dimension for large $\NNc$ QCD without the
restriction to the light cone. This was obtained by first solving the
counting problem for operators and then truncating the known $\superN=4$
answer of~\cite{Beisert:2003jj}
but it can also be thought of as a ``lifting{}'' of the light cone
results of~\cite{Belitsky:2004sc} to the full conformal group.
For earlier partial results that were useful in the initial stages
of this investigation see~\cite{Morozov:1985ef,Gracey:2002rf}. Recently,
anisotropic spin chains have also made an appearance in this
context~\cite{DiVecchia:2004jw,Belitsky:2004sc}.

The main results of the present work are the following:

We give the complete one-loop matrix of anomalous dimension for large
$\NNc$ QCD 
and study its eigenvalues. We use a particularly convenient choice of basis
for such operators that avoids the necessity of restricting oneself to
light-cone operators.
In order to use the intuition from
condensed matter we will be thinking of the matrix of anomalous
dimensions as the Hamiltonian of a spin chain.
We are particularly interested in the ``thermodynamic limit{}''
consisting of studying operators with a large number of elementary
fields.

We are able to construct the exact (anti-ferromagnetic) ground state,
corresponding to the operator with the lowest (negative) anomalous
dimension for any given length
and we show that all gapless excitations above the ground
state are described by an integrable system based on the conformal
group $SO(4,2)$.
There are in fact two equivalent ground states, corresponding to
scalar operators constructed by only using
the selfdual field strength or by only using the anti-selfdual component.
Operators constructed using both, or operators containing quarks, are
separated by a ``mass gap{}'' from the ground state, i.e. their
anomalous dimension remains higher than the ground state by a finite
amount in the thermodynamic limit.

With the above ``stringy{}'' motivation in mind we then analyze the structure
of the operators that contain ``gapless{}'' excitations. These are the QCD
equivalents of the BMN operators for $\superN=4$.

The analysis of the quantum numbers carried by the gapless excitations
reveals some surprises. Focusing on the chiral ground state for
definitiveness, we find, as already expected
{}from~\cite{Ferretti:2004ba}, excitations carrying a
(space-time) chiral index $\alpha$
corresponding to the ``spinons{}'' of the compact XXX$_1$ chain embedded
in the conformal chain ($SU(2)_L \in SO(4,2)$).
As well known from the condensed matter literature, such excitations along
the chain are characterized by a linear (``relativistic{}'') dispersion relation.

But there is another type of excitation that was not accessible by
the previous analysis which did not include derivatives. Namely, by
``spreading{}'' a covariant derivative along the ground state chain,
very much like the impurities in the BMN case, we are able to
construct new excitations along the chain that are characterized by
a quadratic (``non-relativistic{}'') dispersion relation. One would
naively expect the quantum number carried by these excitations to be
a space-time vector since a covariant derivative carries a
space-time index $\mu$.
However, by writing $\mu=(\alpha,\dot\alpha)$ and
recalling that $\alpha$ propagates in an
anti-ferromagnetic background, we are able to
show that the presence of $D_\mu$ gives rise to two independent elementary
excitations, one being the above mentioned spinon and the other
carrying a space-time anti-chiral index $\dot\alpha$.
It is well known since  the work of
Faddeev and Takhtajan \cite{Faddeev:1981ip,Faddeev:1996iy} that
scattering off background spins can modify quantum numbers of
excitations in anti-ferromagnetic spin chains. Here we found another
manifestation of this phenomenon.

The picture that emerges from the one-loop analysis has some intriguing
similarities to the twistor string theory of Witten~\cite{Witten:2003nn}
and it would be very interesting to
make the connection more explicit. What we can say is that the
thermodynamic limit of the one-loop QCD chain is described by
chiral/anti-chiral twistorial excitations with linear/quadratic
dispersion relation. Whether these are (part of) the correct degrees of
freedom required to formulate the string description of QCD,
what their space-time dynamics is and the connections (if any) with
the work of~\cite{Witten:2003nn} remain a project for the future.

The paper is organized as follows:

We begin in \Secref{sec:2} by discussing some basic preliminary facts about the
renormalization of composite operators, the simplifications occurring in the
large $\NNc$ limit and the definition of the spin chain Hamiltonian.

In \Secref{sec:3} we discuss in detail the construction
of the complete basis of
operators that form the Hilbert space of our spin chain.
Having such a complete basis makes the
restriction to the light-cone no longer necessary.

\Secref{sec:4} contains the complete one-loop chain of anomalous dimensions
expressed in terms of projectors on irreducible representations of the
conformal group.
Recall that the use of conformal symmetry to classify the operator
to one-loop is possible, even in a non-conformal field theory such as QCD, since
the beta function is of order $O((\as \NNc)^2)$ whereas the anomalous dimensions
are $O(\as \NNc)$.
The relevant mathematical formalism is summarized in
\Appref{sec:A}. The coefficients of the projection operators are obtained
by a truncation of the corresponding matrix in the $\superN=4$ theory with the
additional modifications required by the different
wave-function renormalization. We also give a table of anomalous dimensions
of all conformal primary operators up to dimension seven for pure glue and five for those involving quarks.
We conclude the section by showing briefly how the same
results could also be obtained
by ``lifting{}'' the light-cone results of~\cite{Belitsky:2004sc}.
To this purpose, we consider
for definitiveness the gluon-quark coupling. Although this
is not the way we arrived at the results, we feel it may be useful to include
it in order to establish a connection with the previous literature.

\Secref{sec:5} presents the Bethe ansatz for the chiral sector of the theory.
We show that the closed chain composed of selfdual gluon field strengths
and derivatives is integrable and we give the Bethe equations corresponding
to the conformal group $SO(4,2)$. We begin our discussion of the possible
excitations over the ground state.
As a check of our computation we present
all the Bethe roots corresponding to the conformal primaries in the integrable
sector up to dimension seven. We conclude with some remarks on the open chain,
which is also integrable for selfdual gluons in the bulk. We
present the Bethe ansatz for the case of open chains without derivatives.

\Secref{sec:6} contains a more detailed analysis of the excitations over the
ground state. First, we show that operators containing a mixture of
selfdual and anti-selfdual fields or quarks have an anomalous dimension that
remains above the previously found ground state by a finite amount (gap) in the
thermodynamic limit. This establishes that we have indeed found the true
ground state. We then move on to study the gapless excitations. As
mentioned before they are characterized by a dispersion
relation that is linear or quadratic for chiral and anti-chiral objects
respectively.

In \Appref{sec:A} we review the oscillator representation of the
conformal group and the decompositions required in the paper.

In \Appref{sec:C} we make the comparison with the length two primaries
some of which are well know
{}from the literature on deep inelastic scattering. We do this partly as
a check and partly as an illustration of how to use the Hamiltonian
obtained in \Secref{sec:4}.

In \Appref{sec:YB} we give a proof of integrability for the chiral sector
of pure YM by solving the corresponding Yang--Baxter equations.

%%%%%%%%%%%%%%%%%%%%%%%%%%%%%%%%%%%%%%%%%%%%%%%%%%%%%%%%%%%%%%%%%%%%%%%%%%%%%%%%
%%%%%%%%%%%%%%%%%%%%%%%%%%%%%%%%%%%%%%%%%%%%%%%%%%%%%%%%%%%%%%%%%%%%%%%%%%%%%%%%
\section{Preliminaries}
\label{sec:2}

In this paper we consider anomalous
dimensions of local, gauge invariant operators for
massless QCD:
\begin{equation}
     L = -\frac{1}{2}\tr F^2 + i \bar{q} \not\!\! D q,
\end{equation}
where $F_{\mu\nu} = \partial_\mu A_\nu - \partial_\nu A_\mu -
       i g_\mathrm{YM} [A_\mu, A_\nu]$ is the gauge field strength, thought of as
an $\NNc\times \NNc$ hermitian matrix, and $q$ is the quark field, transforming in
the fundamental representation of the gauge group $SU(\NNc)$ and possibly
carrying a finite number of flavors.
It will be very convenient in the following to express such field content
in the chiral basis by decomposing $F_{\mu\nu}$ into its selfdual and
anti-selfdual parts $f$ and $\bar{f}$
\begin{equation}
     F_{\mu\nu} = \sigma_{\mu\nu}^{\alpha\beta}f_{\alpha\beta} +
\bar{\sigma}_{\mu\nu}^{\dot\alpha\dot\beta}\bar{f}_{\dot\alpha\dot\beta}
\end{equation}
and the quark field in its chiral and anti-chiral parts $\psi$ and
$\bar\chi$:
\begin{equation}\label{quarkss}
     q = \begin{pmatrix}\psi_\alpha \cr \bar\chi^{\dot\alpha}\cr \end{pmatrix}
\end{equation}

We employ the `t~Hooft limit of letting the number of colors $\NNc$ go to infinity
and the usual QCD coupling $g_{\mathrm{YM}}$ to zero while holding the 't~Hooft
coupling $\as \NNc$ finite. In fact, we will deal with
one-loop perturbation theory in $\as \NNc$ which must therefore
be assumed to be small.

The latter condition would make the limit rather simple
were it not for the
fact that we apply it to the study of anomalous dimensions of
gauge invariant composite operators made out of a large number of elementary
fields where even ordinary perturbation theory becomes quite
involved.%
\footnote{Actually, with some more work it would be possible to obtain the
mixing matrix at finite~$\NNc$ but we decided not to include it in this paper as
our main motivation is to learn about the stringy description where $1/\NNc$
plays the role of the string coupling.}

Let us briefly recall how anomalous dimensions of composite
operators are defined in quantum field theory.
%although in this
%paper we will use a more powerful method, based on the dilatation
%operator, to construct them.
We use the conventions of
\cite{Peskin:1995ev} throughout the paper. In particular, let us
denote by $\mathcal{O}$ a multiplicatively renormalized operator,
i.e. an operator that can be renormalized by multiplying it with a
(divergent) renormalization constant. We set
\begin{equation}
    \mathcal{O}_\mathrm{bare} = Z_\mathcal{O} \mathcal{O}_\mathrm{ren.},
\end{equation}
where $Z_\mathcal{O}$ is constructed so that $\mathcal{O}_\mathrm{ren.}$
has finite correlation functions with the renormalized quark and gluon fields.
The anomalous dimension of $\mathcal{O}$ is defined through
\begin{equation}
      \gamma_\mathcal{O} = \mu\frac{\dddd }{\dddd\mu} \log Z_\mathcal{O}.
\end{equation}
To one-loop in $\as \NNc$ it takes the form
\begin{equation}
    \gamma_\mathcal{O} = c_\mathcal{O} \unit,
\end{equation}
where  $c_\mathcal{O}$ is
a numerical constant to be determined.

In general, operators carrying the same quantum numbers and having
the same \emph{classical} dimension will mix under renormalization and
we write:
\begin{equation}
    \mathcal{O}_\mathrm{bare}^i = Z_{\mathcal{O}j}^{\phantom{\mathcal{O}}i}
    \mathcal{O}_\mathrm{ren.}^j.
\end{equation}
The anomalous dimension also becomes a matrix and we write, in
matrix notation:
\begin{equation}
      \Gamma =  \mu\frac{\dddd }{\dddd\mu} \log Z_\mathcal{O} =
       Z_\mathcal{O}^{-1} \mu\frac{\dddd }{\dddd\mu} Z_\mathcal{O} =
       \unit H.
\end{equation}

Multiplicatively renormalizable operators correspond to eigenvectors of
$\Gamma$ and their eigenvalues correspond to the anomalous dimensions:
\begin{equation}
      \Gamma \hat{\mathcal{O}} =  \gamma_{\hat{\mathcal{O}}} \hat{\mathcal{O}}.
\end{equation}

The problem of the computation of anomalous dimensions is thus split into two
parts: First one must derive the mixing matrix $\Gamma$
and second one must find its eigenvalues.
In the past few years new techniques, inspired by the study of $\superN=4$ SYM have
been developed that bypass the computation of the divergences in the loop
diagrams and allow a direct determination of $\Gamma$. These techniques
are based on the analysis on the dilatation operator, thought of as one of
the generators of the conformal group.
The second part, involving the diagonalization of $\Gamma$, also requires
analytical work since we are mostly interested in letting the size of the
matrix go to infinity by considering ``long{}'' operators.
Fortunately, even this second problem can be handled by analytic methods
such as the Bethe ansatz for the cases of interest.

We shall refer to a generic composite operator as a
``chain{}'', where the name originates by the interpretation
of such operators as a spin chain, as we will discuss in detail later.

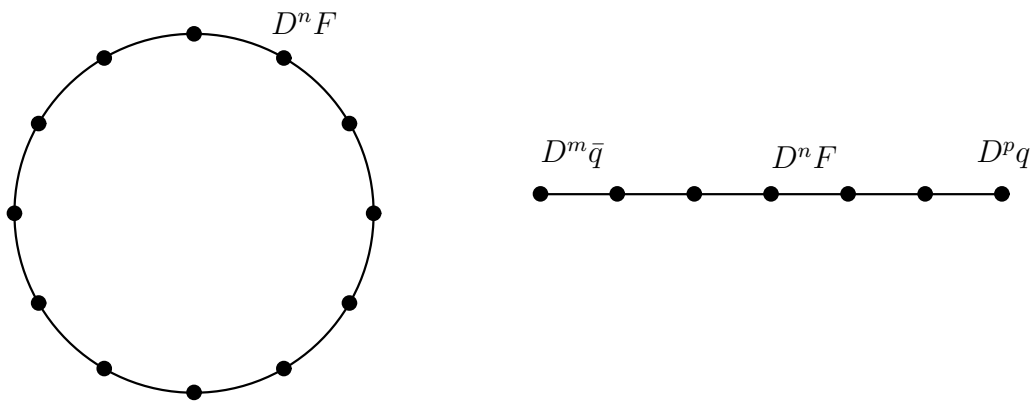
\begin{figure}[hbp]
  \begin{center}
%%%%%%%%%%%% .epic figure here
\setlength{\unitlength}{0.00066667in}
\begingroup\makeatletter\ifx\SetFigFont\undefined%
\gdef\SetFigFont#1#2#3#4#5{%
  \reset@font\fontsize{#1}{#2pt}%
  \fontfamily{#3}\fontseries{#4}\fontshape{#5}%
  \selectfont}%
\fi\endgroup%
{\renewcommand{\dashlinestretch}{30}
\begin{picture}(8200,3200)(0,100)
\thicklines
\put(1500,1500){\ellipse{2800}{2800}}
\put(1500,100){\blacken\ellipse{108}{108}}
\put(1500,2900){\blacken\ellipse{108}{108}}
\put(2900,1500){\blacken\ellipse{108}{108}}
\put(100,1500){\blacken\ellipse{108}{108}}
\put(287,2200){\blacken\ellipse{108}{108}}
\put(2712,2200){\blacken\ellipse{108}{108}}
\put(2200,2712){\blacken\ellipse{108}{108}}
\put(2200,287){\blacken\ellipse{108}{108}}
\put(2712,800){\blacken\ellipse{108}{108}}
\put(287,800){\blacken\ellipse{108}{108}}
\put(800,287){\blacken\ellipse{108}{108}}
\put(800,2712){\blacken\ellipse{108}{108}}
\put(4200,1650){\blacken\ellipse{108}{108}}
\put(4800,1650){\blacken\ellipse{108}{108}}
\put(5400,1650){\blacken\ellipse{108}{108}}
\put(6000,1650){\blacken\ellipse{108}{108}}
\put(6600,1650){\blacken\ellipse{108}{108}}
\put(7200,1650){\blacken\ellipse{108}{108}}
\put(7800,1650){\blacken\ellipse{108}{108}}
\path(4200,1650)(7800,1650)
\put(2100,2900){\makebox(0,0)[lb]{$D^n F$ }}
\put(6000,1850){\makebox(0,0)[lb]{$D^n F$ }}
\put(4200,1850){\makebox(0,0)[lb]{$D^m \bar{q}$ }}
\put(7600,1850){\makebox(0,0)[lb]{$D^p q$ }}
\end{picture}
}
%%%%%%%%%%%%
  \end{center}
  \caption{Closed and open ``spin chains{}'' corresponding to purely gluonic
operators or mesonic operators respectively.}
  \label{fig:chain}
\end{figure}

Neglecting for the moment the details of the Lorentz structure
of these operators, such chains can be grossly divided into two main groups:
(see \figref{fig:chain})

\begin{itemize}
\item Closed chains, i.e. operators schematically of the type
\begin{equation}
   \tr (D^{n_1} F) \dots (D^{n_L} F),
\end{equation}
\item Open chains, i.e. operators schematically of the type
\begin{equation}
    (D^{n_1} \bar{q}) (D^{n_2} F) \dots (D^{n_{L-1}} F) (D^{n_L} q),
\end{equation}
\end{itemize}

Notice that we will distinguish between different conformal descendants, i.e.
will not identify operators differing by a total derivative.
Therefore it does not suffice to ignore derivatives on the
first site $\bar{q}$ as is commonly done in QCD.

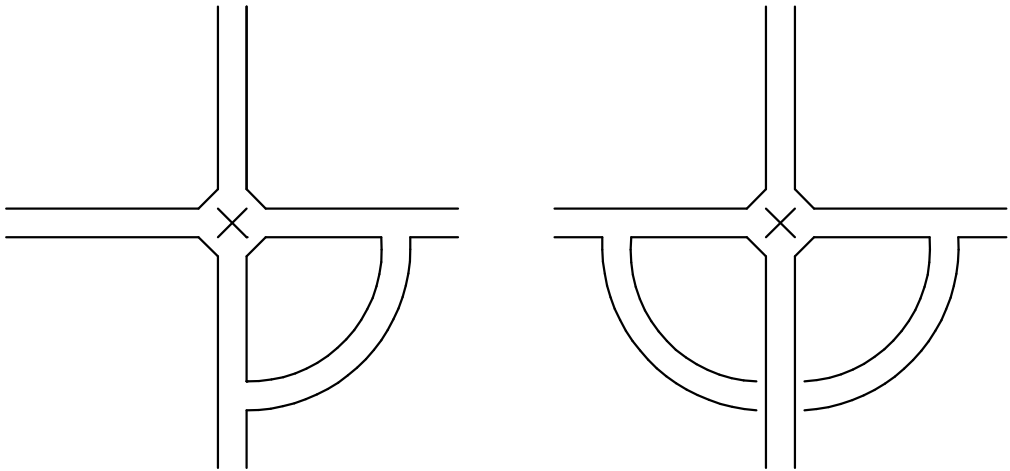
\begin{figure}[htbp]
  \begin{center}
%%%%%%%%%%%% .epic figure here
\setlength{\unitlength}{0.00066667in}
\begingroup\makeatletter\ifx\SetFigFont\undefined%
\gdef\SetFigFont#1#2#3#4#5{%
  \reset@font\fontsize{#1}{#2pt}%
  \fontfamily{#3}\fontseries{#4}\fontshape{#5}%
  \selectfont}%
\fi\endgroup%
{\renewcommand{\dashlinestretch}{30}
\begin{picture}(7866,3681)(0,-10)
\thicklines
\drawline(1908.000,708.000)(2004.418,710.421)(2100.190,721.833)
    (2194.477,742.137)(2286.457,771.156)(2375.326,808.635)
    (2460.307,854.249)(2540.658,907.597)(2615.677,968.214)
    (2684.709,1035.571)(2747.151,1109.079)(2802.457,1188.096)
    (2850.144,1271.931)(2889.795,1359.852)(2921.064,1451.092)
    (2943.677,1544.852)(2957.438,1640.314)(2962.226,1736.644)
    (2958.000,1833.000)
\drawline(1908.000,483.000)(2002.494,486.256)(2096.486,496.512)
    (2189.459,513.711)(2280.899,537.759)(2370.305,568.522)
    (2457.182,605.833)(2541.052,649.484)(2621.454,699.236)
    (2697.944,754.815)(2770.102,815.913)(2837.528,882.196)
    (2899.853,953.297)(2956.732,1028.825)(3007.852,1108.363)
    (3052.933,1191.475)(3091.724,1277.701)(3124.013,1366.566)
    (3149.622,1457.582)(3168.410,1550.247)(3180.272,1644.050)
    (3185.145,1738.474)(3183.000,1833.000)
\drawline(6258.000,483.000)(6353.071,492.170)(6447.191,508.419)
    (6539.834,531.656)(6630.481,561.751)(6718.625,598.537)
    (6803.774,641.806)(6885.452,691.318)(6963.201,746.795)
    (7036.587,807.927)(7105.199,874.372)(7168.654,945.759)
    (7226.597,1021.688)(7278.703,1101.735)(7324.682,1185.452)
    (7364.277,1272.371)(7397.265,1362.006)(7423.462,1453.855)
    (7442.722,1547.405)(7454.937,1642.133)(7460.040,1737.509)
    (7458.000,1833.000)
\drawline(6258.000,708.000)(6355.237,717.775)(6451.117,736.687)
    (6544.782,764.568)(6635.397,801.169)(6722.151,846.163)
    (6804.268,899.147)(6881.016,959.648)(6951.708,1027.126)
    (7015.712,1100.977)(7072.457,1180.542)(7121.436,1265.110)
    (7162.211,1353.925)(7194.418,1446.192)(7217.768,1541.089)
    (7232.053,1637.766)(7237.146,1735.361)(7233.001,1833.000)
\drawline(4908.000,1833.000)(4902.600,1735.051)(4906.628,1637.036)
    (4920.047,1539.860)(4942.732,1444.421)(4974.475,1351.601)
    (5014.981,1262.256)(5063.877,1177.213)(5120.711,1097.256)
    (5184.959,1023.124)(5256.026,955.503)(5333.256,895.016)
    (5415.936,842.222)(5503.303,797.609)(5594.548,761.589)
    (5688.830,734.495)(5785.278,716.576)(5883.000,707.999)
\drawline(4683.000,1833.000)(4681.728,1737.713)(4687.503,1642.594)
    (4700.296,1548.161)(4720.035,1454.933)(4746.612,1363.419)
    (4779.883,1274.120)(4819.664,1187.526)(4865.739,1104.110)
    (4917.855,1024.328)(4975.726,948.618)(5039.037,877.393)
    (5107.440,811.044)(5180.561,749.934)(5258.000,694.397)
    (5339.333,644.738)(5424.115,601.227)(5511.882,564.104)
    (5602.153,533.571)(5694.435,509.795)(5788.221,492.907)
    (5883.000,482.999)
\drawline(2058,1833)(2958,1833)
\drawline(3183,1833)(3558,1833)
\drawline(1908,1683)(1908,708)(1908,708)(1908,708)
\drawline(1908,483)(1908,33)
\drawline(1683,1683)(1683,33)
\drawline(1533,1833)(33,1833)
\drawline(1533,2058)(33,2058)
\drawline(1683,2208)(1683,3633)
\drawline(1908,2208)(1908,3633)
\drawline(5958,2208)(5958,3633)
\drawline(6183,2208)(6183,3633)
\drawline(1908,2208)(1908,3633)
\drawline(6333,2058)(7833,2058)
\drawline(7458,1833)(7833,1833)
\drawline(6333,1833)(7233,1833)
\drawline(6183,1683)(6183,33)
\drawline(5958,1683)(5958,33)
\drawline(5808,2058)(4308,2058)
\drawline(4908,1833)(5808,1833)
\drawline(4308,1833)(4683,1833)
\drawline(1908,2058)(1683,1833)
\drawline(1683,2058)(1908,1833)(1908,1833)
\drawline(5958,2058)(6183,1833)
\drawline(6183,2058)(5958,1833)
\drawline(2058,2058)(3558,2058)
\drawline(1683,2208)(1533,2058)
\drawline(1533,1833)(1683,1683)
\drawline(1908,1683)(2058,1833)
\drawline(2058,2058)(1908,2208)
\drawline(5958,2208)(5808,2058)
\drawline(5808,1833)(5958,1683)
\drawline(6183,1683)(6333,1833)
\drawline(6333,2058)(6183,2208)
\end{picture}
}
%%%%%%%%%%%%
  \end{center}
  \caption{Renormalization of a four gluon composite operator (denoted by a
cross) in double line notation. To the left is depicted a diagram connecting
two adjacent gluon legs containing a closed color loop. To the right is
shown a diagram connecting two non-adjacent gluon legs, not containing
any color loop and thus subleading in $1/\NNc$.}
  \label{ren}
\end{figure}

In the large $\NNc$ limit the following three crucial simplifications occur:
\begin{itemize}
\item If we consider two eigenvectors $\hat{\mathcal{O}}_{1}$
and $\hat{\mathcal{O}}_{2}$,
linear combinations of either type of operator
described above, with anomalous dimensions $\gamma_{\hat{\mathcal{O}}_{1}}$
and $\gamma_{\hat{\mathcal{O}}_{2}}$
respectively, then $\Gamma(\hat{\mathcal{O}}_{1}\hat{\mathcal{O}}_{2}) =
(\gamma_{\hat{\mathcal{O}}_{1}}+\gamma_{\hat{\mathcal{O}}_{2}})
(\hat{\mathcal{O}}_{1}\hat{\mathcal{O}}_{2})$. Hence, it is enough to consider
operators that cannot be split into products of gauge singlets,
commonly referred to as ``single-trace{}'' operators, although,
in the presence of quarks the name
``irreducible{}'' would be more appropriate.
\item Open and closed chains do not mix. This can be seen by noticing that with
the usual normalization, the mixing matrix between open and closed chains
becomes upper triangular. Alternatively, one can show that by scaling the
quark fields by an extra factor $\sqrt{\NNc}$ the mixing becomes block
diagonal. We shall refer to the mixing matrix for
open and closed operators as
$\Gamma_\mathrm{open}$ and $\Gamma_\mathrm{closed}$ respectively.
\item Finally, the relevant Feynman diagrams for the computation of the
anomalous dimension are only those connecting neighboring fields
(see \figref{ren}). Thus we
can identify either type of basic operator with a chain of ``spins{}'', each
spin corresponding to a basic building block $D^n f$, $D^n \psi$, etc.
The length
$L$ of the chain is conserved to one-loop and the matrix of anomalous
dimensions can be regarded as a Hamiltonian of the spin chain with only
nearest neighbor interactions.
\end{itemize}

Thus we write, for the closed chain, with the usual periodic identification
$L+1\to 1$:
\begin{equation}
     \Gamma_\mathrm{closed} = \unit \sum_{l=1}^L \opham_{l,l+1}^{\mathrm{FF}}
     \label{Hclosed}
\end{equation}
and, for the open chain ($L>2$):
\begin{equation}
     \Gamma_\mathrm{open} = \unit \left(\opham_{1,2}^{\mathrm{qF}}+
     \sum_{l=2}^{L-2} \opham^{\mathrm{FF}}_{l,l+1} +
      \opham_{L-1,L}^{\mathrm{Fq}} \right). \label{Hopen}
\end{equation}
For completeness we should also add the rather trivial Hamiltonian for an
open chain of length two
\begin{equation}
     \Gamma_{\mathrm{open},L=2}=\unit \opham^{\mathrm{qq}}_{1,2}.
\end{equation}
Obviously, only \eqref{Hclosed} and  \eqref{Hopen} admit a continuum limit
$L\to\infty$.

The matrix of anomalous dimensions is thus completely
specified if we give the expression
for the various ``link Hamiltonians{}'' connecting two neighboring sites.

%%%%%%%%%%%%%%%%%%%%%%%%%%%%%%%%%%%%%%%%%%%%%%%%%%%%%%%%%%%%%%%%%%%%%%%%%%%%%%%%
%%%%%%%%%%%%%%%%%%%%%%%%%%%%%%%%%%%%%%%%%%%%%%%%%%%%%%%%%%%%%%%%%%%%%%%%%%%%%%%%
\section{Lorentz structure}
\label{sec:3}

Before describing the link Hamiltonian, we must
carefully define the space of operators on which it acts.
This space must be constructed in such a way as to avoid double counting
or missing some allowed operator.

Seemingly different operators can be mapped into each other by
repeatedly using either of the following two identities:
\begin{equation}
       {[D_\mu, D_\nu]} = - i g_\mathrm{YM} F_{\mu\nu}\quad\mbox{and}\quad
        D_\mu F_{\nu\rho} + \mathord{}\mbox{cyclic} = 0. \label{ida}
\end{equation}
%(The second identity in \eqref{ida} is know as the Bianchi identity.)

Moreover, we must identify operators that differ by the classical
equations of motion:
\begin{equation}
      D^\mu F_{\mu\nu}^a = - g_\mathrm{YM}
      \bar{q}\gamma_\nu T^a q \quad\mbox{and}\quad
      \gamma^\mu D_\mu q = 0. \label{equazio}
\end{equation}
Equations \eqref{ida} and \eqref{equazio} will be used to maximize
the number of building blocks in a given operator as much as
possible in order to come to a unique basis. Before showing that, we
should mention the reason why the equations of motions also need to
be imposed. In \cite{Kluberg-Stern:1974rs} it is shown that if one
separates gauge invariant operators into those not vanishing by the
classical equations of motion, henceforth referred to as OS (on
shell) operators, and those vanishing by the equations of motion
(referred to as EOM operators), the full matrix of
anomalous dimensions has the form:%
\footnote{To be precise, the EOM part might
also contain BRST variations of gauge variant operators.}
\begin{equation}
    \Gamma = \begin{pmatrix}\Gamma_\mathrm{OS}^\mathrm{OS}&
                                 \Gamma_\mathrm{OS}^\mathrm{EOM}\cr
                        0 & \Gamma_\mathrm{EOM}^\mathrm{EOM} \cr
             \end{pmatrix}.
\end{equation}
Thus, although it might be necessary to add counterterms vanishing on shell
in the renormalization of ordinary OS operators, we are assured that the
former operators do not alter the anomalous dimension of the latter since
their mixing is upper triangular.
{}From now on we shall consider only $\Gamma_\mathrm{OS}^\mathrm{OS}$.

We now claim that a generic gauge singlet operator can be constructed, up
to the above identifications, out of the
following elementary building blocks:
\begin{equation}
    D^n f \in \mdl_{(n/2+1,n/2)}, \quad\mbox{and}\quad D^n \psi, D^n\chi; \in \mdl_{(n/2+1/2,n/2)}
    \label{building}
\end{equation}
together with their complex conjugates.
The fields $f$, $\psi$ and $\chi$ are the chiral components of the gluon field
strength and the quark fields discussed in the introduction.
The labels $(S_1,S_2)$
in \eqref{building}
refer to the Lorentz quantum numbers and characterize a unique
irreducible representation of the Lorentz group.
The \emph{module} $\mdl_{(S_1,S_2)}$ is a vector space on which the
representation $(S_1,S_2)$ acts.
In fact, these irreducible representations of the Lorentz group can be
combined to form infinite-dimensional irreducible
representations of the conformal group $SO(2,4)$ by putting together
all terms of the type, e.g. $D^n f$, $n=0\dots \infty$, c.f.~\Appref{sec:A}.
The use of the conformal group is what will allow us to obtain the
full matrix of anomalous dimensions to one-loop.

To be specific, $D^n f$ written in the ordinary
Lorentz basis would look like:
\begin{equation}
    D^n f= D_{(\mu_1}\dots D_{\mu_n}F^+_{\nu)\rho}
    - \mathord{}\mbox{traces} ,
\end{equation}
where $F^+_{\nu\rho}$ is the selfdual component of the field strength.
By ``$-\mbox{traces}${}'' we mean that the tensor has
been reduced to be totally traceless
and the brackets around the indices
correspond to total symmetrization.
In the chiral basis, setting
$D_\mu = \sigma_\mu^{\alpha\dot\alpha} D_{\alpha\dot\alpha}$
and
$F^+_{\mu\nu}=\sigma_{\mu\nu}^{\alpha\beta}f_{\alpha\beta}$,
one would write
\begin{equation}
    D^n f = D_{\alpha_1 \dot{\alpha}_1}\dots D_{\alpha_n \dot{\alpha}_n}
            f_{\beta\gamma} + \mathord{}\mbox{symmetrized},
\end{equation}
where ``$ + \mbox{symmetrized}$'' means that the tensor has been totally
symmetrized in the undotted and dotted indices, respectively.
One usually writes
simply $D^n f_{\alpha_1\dots\alpha_{n+2}, \dot{\alpha}_1\dots \dot{\alpha}_n}$
for the resulting irreducible component
in direct analogy with the representation labels $(n/2+1,n/2)$,
where each symmetrized spinor index contributes spin $1/2$.

To show the validity of the claim, let us look at a few examples.

Consider first an operator containing the element $D_\mu F_{\nu\rho}$.
We can decompose such an object into a totally anti-symmetric tensor,
$D_{{[}\mu} F_{\nu\rho{]}}$
vanishing by the Bianchi identity in \eqref{ida}, a vector
$\eta_{\mu\nu}D^\lambda F_{\lambda\rho} +
 \eta_{\mu\rho}D^\lambda F_{\nu\lambda}$ proportional to the gauge current
by the equations of motion,
and thus quadratic in its field content,  and a remaining tensor
belonging to $\mdl_{(3/2,1/2)} \oplus \mdl_{(1/2,3/2)}$ and
thus precisely of the type stated above. Hence:
\begin{equation}
    D_{\mu} F_{\nu\rho} = Df + D\bar{f} + O(\bar qq).
\end{equation}

Consider now an operator containing an element with two derivatives:
$D_\mu D_\nu F_{\rho\lambda}$. It is clear that the covariant derivatives
can always be symmetrized, up to terms containing an extra $F_{\mu\nu}$
due to \eqref{ida}.
Once the symmetrization has been done, the
two indices of $F_{\rho\lambda}$ cannot further
be anti-symmetrized with the ones of $D_\mu$ or $D_\nu$
because of \eqref{ida}%
\footnote{This involves splitting off further $F_{\mu\nu}$'s to bring
$D_\mu$ close to $F_{\rho\lambda}$.}
and in total three indices will have to be symmetric.
Finally, tracing any two indices will,
by the equations of motion, yield a derivative of the current and thus a
term quadratic in the fields.
Perhaps the only slightly non-trivial identity occurs when the trace is
between indices both belonging to the covariant derivative. In this case
one must use once again the Bianchi identities to recover the equations
of motion:
\begin{eqnarray}
     D^\mu D_\mu F_{\rho\lambda} \eq  -  D^\mu D_\rho F_{\lambda\mu} -
     D^\mu D_\lambda F_{\mu\rho} = - D_\rho D^\mu F_{\lambda\mu} -
     D_\lambda D^\mu F_{\mu\rho} + O(F^2)\nonumber\\
      \eq  0 + O(\bar qq) + O(F^2).
\end{eqnarray}
Again, the remaining tensors in the decomposition of the original tensor
containing two covariant derivatives belong to
the irreps $(2,1) \oplus (1,2)$.

Similarly, the element $D_\mu q$ appearing at the end of an open chain
will be equivalent to operators in $\mdl_{(1,1/2)} \oplus \mdl_{(1/2,1)}$
up to equations of motions.

One can proceed by induction and show that:
\begin{eqnarray}
      D_{\mu_1}\dots D_{\mu_n} F_{\nu\rho} \eq  D^n f + D^n \bar{f} +
      O(\bar qq) + O(F^2),\nonumber\\
       D_{\mu_1}\dots D_{\mu_n} q \eq  D^n \psi + D^n \bar\chi +
      O(\bar q qq) + O(Fq).  \label{idone}
\end{eqnarray}
Since \eqref{idone} may raise but not decrease the length of a chain, one can
proceed systematically from the chains of lowest length (two) and show that
the building blocks \eqref{building} are sufficient to construct
uniquely all the on-shell operators.

Note that we do \emph{not} identify operators that differ by a total
derivative. Although in some other applications this may be desirable, in our
case we should keep them and we will identify them as descendants
in the representation of the conformal group.

%%%%%%%%%%%%%%%%%%%%%%%%%%%%%%%%%%%%%%%%%%%%%%%%%%%%%%%%%%%%%%%%%%%%%%%%%%%%%%%%
%%%%%%%%%%%%%%%%%%%%%%%%%%%%%%%%%%%%%%%%%%%%%%%%%%%%%%%%%%%%%%%%%%%%%%%%%%%%%%%%
\section{Complete QCD spin chain}
\label{sec:4}

We are now ready to discuss the form of the link Hamiltonian.
We can think of it as a map from a generic pair of
elementary building blocks previously discussed into two others.
We can make use of Lorentz symmetry to simplify the form somewhat.
The crucial insight is
%Hi Joe... :-)
that the Hamiltonian can only map components
{}from one irreducible module to the same component of
a multiplet of the same type (or even the same multiplet).
In general this is still a complicated problem because of the large number
of Lorentz multiplets appearing in the product.
Fortunately, for one-loop anomalous dimensions,
conformal symmetry comes to the rescue by
vastly reducing the number of occurring irreps
as we shall now proceed to discuss.

Although QCD is not a conformal theory, the use of the conformal
group to classify composite operators in QCD has a long history
\cite{Makeenko:1980bh,Ohrndorf:1981qv} (see \cite{Braun:2003rp} for
a recent review). The conformal symmetry is especially useful at one
loop because the
%simplest way to understand this is to notice that the
$\beta$-function, which is responsible for the breaking of
conformal invariance, has leading term $O((\as \NNc)^2)$:
\begin{equation}
     \beta(\as \NNc) = \mu\frac{\dddd}{\dddd\mu}\as \NNc =
     -\frac{11}{3}\frac{(\as \NNc)^2}{2\pi}
\end{equation}
and thus cannot affect the one-loop anomalous dimensions which are
of order $\as \NNc$. We shall make use of the conformal symmetry to
write the full matrix of one-loop anomalous dimensions for
large~$\NNc$ QCD. (The conformal algebra and its representations are
reviewed in \Appref{sec:A}).

Each fundamental field, namely $f$, $\psi$ and $\chi$ and their conjugates,
is a primary state%
\footnote{All components of the Lorentz multiplets $D^0 f,D^0\psi,\ldots$
are conformal \emph{primaries}. The conformal \emph{highest-weight}
state is distinguished by being also the highest-weight
of the Lorentz multiplet.}
of an irreducible module
(or multiplet; a vector space transforming in
some irreducible representation) of the conformal algebra.
The remaining elements
in the module (descendants) are obtained
by acting with derivatives on the primaries.
Thus, we have the following structure for the modules:
\begin{eqnarray}
   \mdl^f \eq  \left\langle f_{\alpha_1\alpha_2},
    Df_{\alpha_1\alpha_2\alpha_3\dot{\alpha}_1},
    D^2 f_{\alpha_1\alpha_2\alpha_3\alpha_4\dot{\alpha}_1\dot{\alpha}_2}
    \dots \right\rangle, \nonumber \\
   \mdl^\psi \eq  \left\langle \psi_{\alpha_1},
    D\psi_{\alpha_1\alpha_2\dot{\alpha}_1},
    D^2 \psi_{\alpha_1\alpha_2\alpha_3\dot{\alpha}_1\dot{\alpha}_2}
    \dots \right\rangle, \nonumber \\
   \mdl^\chi \eq  \left\langle \chi_{\alpha_1},
    D\chi_{\alpha_1\alpha_2\dot{\alpha}_1},
    D^2 \chi_{\alpha_1\alpha_2\alpha_3\dot{\alpha}_1\dot{\alpha}_2}
    \dots \right\rangle,  \label{sing}
\end{eqnarray}
together with their complex conjugate modules.

The product of two of the multiplets \eqref{sing} can be decomposed into
an infinite series of irreducible modules labeled by an
extra integer $j$, the conformal spin. This is easily understood in the
oscillator representation reviewed in \Appref{sec:A}.
We need the following decompositions (and their conjugates)
\begin{eqnarray}
     &&\mdl^f \otimes \mdl^f = \sum_{j=-2}^\infty
        \mdl^{ff}_j, \qquad
     \mdl^f \otimes \mdl^{\bar f} = \sum_{j=+2}^\infty
     \mdl^{f\bar f}_j, \nonumber \\
     && \mdl^f \otimes \mdl^\psi = \sum_{j=-1}^\infty
     \mdl^{f\psi}_j, \qquad
     \mdl^f \otimes \mdl^{\bar\chi} = \sum_{j=+1}^\infty
     \mdl^{f\bar\chi}_j, \nonumber \\
     &&\mdl^{\bar f} \otimes \mdl^\psi = \sum_{j=+1}^\infty
     \mdl^{\bar{f}\psi}_j, \qquad
     \mdl^{\bar f} \otimes \mdl^{\bar\chi}
     = \sum_{j=-1}^\infty \mdl^{\bar{f}\bar\chi}_j, \nonumber\\
     &&\mdl^{\bar\psi} \otimes \mdl^\psi= \sum_{j=+1}^\infty
     \mdl^{\bar\psi\psi}_j, \qquad
     \mdl^{\chi} \otimes \mdl^\psi = \sum_{j=-1}^\infty
     \mdl^{\chi\psi}_j.
\end{eqnarray}

The explicit form of the irreducible modules appearing in the above
decompositions is discussed in \Appref{sec:A}.
The importance of these decompositions is that the
one-loop link Hamiltonian commutes with the generators of the conformal algebra
and thus assumes constant values on each of the irreducible
modules.
Moreover, each irreducible module appears only once within all
the above decompositions and there will be no mixing problem
between equivalent modules.%
\footnote{This amounts to saying that for each irreducible
module there is only a single (or $1\times 1$ matrix of) coefficients
instead of an $n\times n$ matrix in the case
of an $n$-fold occurrence.}

Introducing the projectors $\opproj_j^{ff}$ etc.~projecting on the
respective modules we can write for each glue-glue
link (not written explicitly)
\begin{equation}
     \opham^{\mathrm{FF}} = \sum_{j=-2}^\infty E^{ff}_j
     (\opproj_j^{ff} + \opproj_j^{\bar{f}\bar{f}}) +
     \sum_{j=+2}^\infty (E^{f\bar{f}, \opident}_j \opident +
     E^{f\bar{f}, \opperm}_j \opperm)(\opproj_j^{f\bar{f}}+
     \opproj_j^{\bar{f}f}), \label{gg}
\end{equation}
where the coefficients $E_j$ need to be determined and $\opident$ and
$\opperm$ are the identity operator and the exchange
operator, respectively:
\begin{eqnarray}
    \opident&:&\;\; (D^n f)(D^m \bar{f})\to(D^n f)(D^m \bar{f}), \nonumber\\
    \opperm&:&\;\; (D^n f)(D^m \bar{f})\to (D^m \bar{f})(D^n f).
\end{eqnarray}

Similarly, we write, for the quark-gluon and the gluon-quark link:
\begin{eqnarray}
     \opham^{\mathrm{qF}} \eq  \sum_{j=-1}^\infty E^{\chi f}_j
     (\opproj_j^{\chi f} +  \opproj_j^{\bar\psi \bar f}) +
       \sum_{j=+1}^\infty  E^{\chi \bar{f}}_j
     (\opproj_j^{\chi \bar f} +  \opproj_j^{\bar\psi f}),\nonumber\\
     \opham^{\mathrm{Fq}} \eq  \sum_{j=-1}^\infty  E^{\chi f}_j
     (\opproj_j^{f\psi} +  \opproj_j^{ \bar f\bar\chi}) +
       \sum_{j=+1}^\infty  E^{\chi \bar{f}}_j
     (\opproj_j^{\bar f\psi} +  \opproj_j^{f\bar\chi}) \label{gq}
\end{eqnarray}
and for the length two quark-quark Hamiltonian:
\begin{equation}
     \opham^{\mathrm{qq}} = \sum_{j=-1}^\infty
      E^{\chi \psi}_j(\opproj_j^{\chi\psi} +
     \opproj_j^{\bar\psi\bar\chi}) + \sum_{j=+1}^\infty
       E^{\bar\psi \psi}_j
     (\opproj_j^{\bar\psi\psi} + \opproj_j^{\chi\bar\chi}). \label{qq}
\end{equation}

In writing the above equations we have already made use of various symmetries
that allow us to identify the coefficients of the projectors on different
modules.
Perhaps one not so obvious symmetry is a ``chiral{}'' type of symmetry rotating
$f$ and $\bar{f}$ by
opposite phases (see~\cite{Intriligator:1998ig} for a discussion
in the context of $\superN=4$ SYM).
This symmetry is
responsible for the closure of the chiral sector. We denote
by $A$ the quantum numbers of the fields under such symmetry transformation.

The coefficients in equations \eqref{gg,gq,qq}
can be fixed in two ways, either
by truncating the $\superN=4$ results of \cite{Beisert:2003jj} or
by lifting the light cone results of
\cite{Lipatov:1994yb,Faddeev:1995zg,Braun:1998id,Braun:1999te,Belitsky:1999bf}.
Many terms can also be explicitly computed or tested
by considering chains of length two as discussed in \Appref{sec:C}.

Before giving a derivation of the results we collect below the complete set
of coefficients of the two-site Hamiltonian, yielding the full matrix
of anomalous dimensions to one-loop ($h(j) = \sum_{k=1}^j 1/k$):
\begin{eqnarray}\label{ggenergies}
E^{ff}_j \eq  2h(j+2)-\sfrac{11}{6}, \nonumber \\
E^{f\bar{f}, \opident}_j \eq  h(j-2)+h(j+2)-\sfrac{11}{6}, \nonumber \\
E^{f\bar{f}, \opperm}_j \eq
          (-1)^{j} \bigbrk{h(j-2) -4h(j-1) +6h(j) -4h(j+1) +h(j+2)}
          \nonumber \\
       \eq  \frac{6(-1)^{j+1}}{(j-1)j(j+1)(j+2)} \nonumber \\
E^{\chi f}_j \eq  h(j+2)+h(j+1)-\sfrac{5}{3} ,\nonumber \\
E^{\chi \bar{f}}_j \eq  h(j+2)+h(j-1)-\sfrac{5}{3} ,\nonumber \\
E^{\chi \psi}_j \eq  2h(j+1)-\sfrac{3}{2} ,\nonumber \\
E^{\bar\psi \psi}_j \eq   h(j+1)+h(j-1)-\sfrac{3}{2}.
\end{eqnarray}
The spectrum of operators with the lowest classical dimensions
can be found in Tables~\ref{tableone},\ref{tableoneb},\ref{tableonec}.
\renewcommand{\arraystretch}{1.1}
\begin{table}[htbp]
\centering
$\begin{array}{|ccr|c|c|l|}\hline
D&L&A&(S_1,S_2)&[p,r,q]&E^C\\%
\hline
4&2&2&(0,0)&[0,-\phantom{0}4,0]&-\frac{11}{3}^+\\%
4&2&2&(2,0)&[4,-\phantom{0}6,0]&\frac{7}{3}^+\\%
4&2& 0&(1,1)&[2,-\phantom{0}6,2]&0^+\\%
\hline
5&2& 0&(3/2,3/2)&[3,-\phantom{0}8,3]&3^+\\%
\hline
6&3&3&(0,0)&[0,-\phantom{0}6,0]&\frac{1}{2}^+ \\%
6&3&3&(1,0)&[2,-\phantom{0}7,0]&-\frac{3}{2}^- \\%
6&3&3&(3,0)&[6,-\phantom{0}9,0]&\frac{7}{2}^- \\%
6&2&2&(3,1)&[6,-10,2]&\frac{14}{3}^+ \\%
6&3&1&(1,1)&[2,-\phantom{0}8,2]&\frac{7}{6}^+ \\%
6&3&1&(0,1)&[0,-\phantom{0}7,2]&-\frac{11}{6}^- \\%
6&3&1&(2,1)&[4,-\phantom{0}9,2]&\frac{7}{6}^- \\%
6&2&0&(2,2)&[4,-10,4]&\frac{21}{5}^+\\%
\hline
7&3&3&(3/2,1/2)&[3,-\phantom{0}9,1]&2^+\mbox{ and }2^- \\%
7&3&3&(5/2,1/2)&[5,-10,1]&\frac{17}{6}^+ \\%
7&3&1&(5/2,1/2)&[5,-10,1]&\frac{17}{6}^+ \\%
7&3&1&(1/2,3/2)&[1,-\phantom{0}9,3]&\frac{23}{12}^+\mbox{ and }{-\frac{1}{12}^-}\\%
7&3&1&(3/2,3/2)&[3,-10,3]&\frac{71}{30}^+\mbox{ and } \frac{13}{6}^- \\%
7&3&1&(5/2,3/2)&[5,-11,3]&\frac{197}{60}^+\mbox{ and } \frac{41}{12}^- \\%
7&2&0&(5/2,5/2)&[5,-12,5]&\frac{26}{5}^+\\%
\hline
\end{array}$
\caption{A complete list of primary states for the purely gluonic sector of
QCD up to $D=7$. The Dynkin labels $[p,r,q]$ refer to the classical representation
of the conformal group as explained in the text. The spin labels
$(S_1,S_2)\equiv(p/2, q/2)$ are redundant but we include them for clarity.
$D$ is the classical dimension, $L$ the length of the operator and
the chirality $A$ counts
the number of $f$ minus the number of $\bar f$.
$E$ is the anomalous dimension in units of $\unit$.
$C$ denotes
charge conjugation: $C: F_{\mu\nu} \to -F_{\mu\nu}^{\scriptscriptstyle\mathrm{T}}$. In the language of
spin chains it corresponds to reversing the orientation of the chain.
For each state in the
table with $A \not = 0$ there is a corresponding state with opposite chirality,
$p$ and $q$ exchanged and with the same anomalous dimension.}
\label{tableone}
\end{table}
\begin{table}[htbp]\centering
$\begin{array}{|ccr|c|c|l|}\hline
D&L&A&(S_1,S_2)&[p,r,q]&E\\
\hline
3&2&1&(0,0)&[0,-3,0]& -\frac{3}{2}\\
3&2&1&(1,0)&[2, -4, 0]&\frac{1}{2} \\\hline
4&2&1&(3/2,1/2)&[3,-6,1]&\frac{3}{2} \\\hline
5&2&1&(2,1)&[4,-8,2]& \frac{13}{6}\\
5&3&2&(0,0)&[0,-5,0]& -\frac{4}{3}\\
5&3&2&(1,0)&[2,-6,0]& -\frac{1}{3} \mbox{ and } \frac{2}{3} \\
5&3&2&(2,0)&[4,-7,0]& \frac{5}{3}\\
5&3&0&(0,1)&[0,-6,2]& \sfrac{1}{3}\\
5&3&0&(1,1)&[2,-7,2]& \sfrac{1}{3}\\\hline
\end{array}$
\caption{A complete list of open chains with boundaries
$\chi\ldots\psi$ up to $D=5$.}
\label{tableoneb}
\end{table}

\begin{table}[htbp]\centering
$\begin{array}{|ccr|c|c|l|}\hline
D&L&A&(S_1,S_2)&[p,r,q]&E\\
\hline
3&2&0&(1/2,1/2)&[1,-4,1]& 0\\\hline
4&2&0&(1,1)&[2,-6,2]&\frac{4}{3} \\\hline%
5&2&0&(3/2,3/2)&[3,-8,3]& \frac{25}{12}\\
5&3&1&(1/2,1/2)&[1,-6,1]& -\sfrac{1}{2}\\
5&3&1&(3/2,1/2)&[3,-7,1]& 1\\\hline
\end{array}$
\caption{A complete list of open chains with boundaries
$\bar\psi\ldots\psi$ up to $D=5$.
For each state with $A \not = 0$ there is a corresponding state with opposite chirality,
$p$ and $q$ exchanged and with the same anomalous dimension.}
\label{tableonec}
\end{table}

\renewcommand{\arraystretch}{1}

%%%%%%%%%%%%%%%%%%%%%%%%%%%%%%%%%%%%%%%%%%%%%%%%%%%%%%%%%%%%%%%%%%%%%%%%%%%%%%%%%%%%
\subsection{Reduction from SYM}

All the coefficients $E^{\times\times}_j$ can be obtained from
$\superN=4$ SYM. The crucial observation is that the
set of Feynman diagrams in $\superN=4$ SYM encompasses all
the Feynman diagrams of QCD.
What is more, at the one-loop level, the additional propagating degrees of
freedom of the supersymmetric theory appear only
in a very restricted sense in the diagrams that are relevant to QCD.

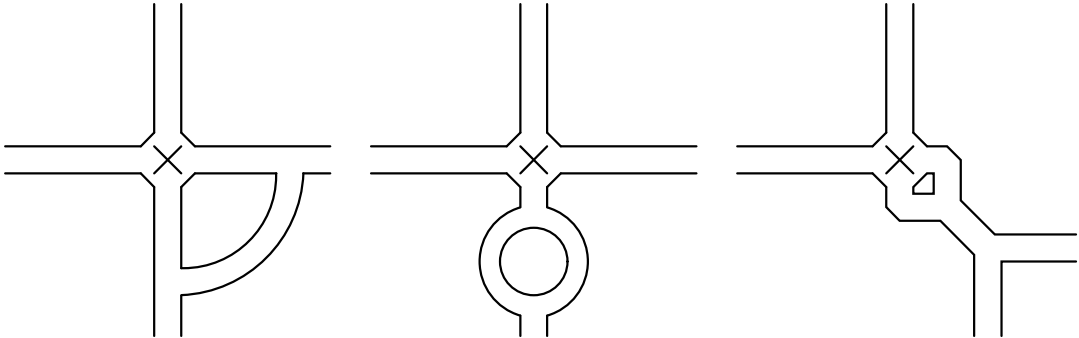
\begin{figure}[htbp]
  \begin{center}
%%%%%%%%%%%% .epic figure here
\setlength{\unitlength}{0.00047in}
\begingroup\makeatletter\ifx\SetFigFont\undefined%
\gdef\SetFigFont#1#2#3#4#5{%
  \reset@font\fontsize{#1}{#2pt}%
  \fontfamily{#3}\fontseries{#4}\fontshape{#5}%
  \selectfont}%
\fi\endgroup%
{\renewcommand{\dashlinestretch}{30}
\begin{picture}(11916,3756)(0,-10)
\thicklines
\put(5908.000,858.000){\arc{1250.000}{1.8546}{4.4286}}
\put(5858.000,858.000){\arc{1250.000}{4.9962}{7.5702}}
\put(5883,858){\ellipse{750}{750}}
\path(6183,1833)(7683,1833)
\path(6033,1683)(6183,1833)
\path(5583,2133)(4083,2133)
\path(5583,1833)(4083,1833)
\path(5583,1833)(5733,1683)
\path(6033,2283)(6033,3708)
\path(6183,2133)(7683,2133)
\path(6183,2133)(6033,2283)
\path(5733,2283)(5733,3708)
\path(5733,2283)(5583,2133)
\path(5733,2133)(6033,1833)
\path(6033,2133)(5733,1833)
\path(5733,1683)(5733,1458)
\path(5733,258)(5733,33)
\path(6033,1683)(6033,1458)
\path(6033,258)(6033,33)
\put(1924.072,1891.928){\arc{2820.320}{0.0418}{1.5290}}
\put(2008.000,1808.000){\arc{2050.610}{6.2588}{7.8784}}
\path(2133,1833)(3033,1833)
\path(3333,1833)(3633,1833)
\path(1983,1683)(1983,783)
\path(1983,483)(1983,33)
\path(1983,1683)(2133,1833)
\path(1533,2133)(33,2133)
\path(1533,1833)(33,1833)
\path(1533,1833)(1683,1683)
\path(1683,1683)(1683,33)
\path(1983,2283)(1983,3708)
\path(2133,2133)(3633,2133)
\path(2133,2133)(1983,2283)
\path(1683,2283)(1683,3708)
\path(1683,2283)(1533,2133)
\path(1683,2133)(1983,1833)
\path(1983,2133)(1683,1833)
\path(9633,2133)(8133,2133)
\path(9633,1833)(8133,1833)
\path(10083,2283)(10083,3708)
\path(9783,2283)(9783,3708)
\path(10308,1833)(10308,1608)(10083,1608)
        (10083,1683)(10233,1833)(10308,1833)
\path(10233,2133)(10458,2133)(10608,1983)
        (10608,1533)(10983,1158)(11883,1158)
\path(9783,1683)(9783,1458)(9933,1308)
        (10383,1308)(10758,933)(10758,33)
\path(9633,1833)(9783,1683)
\path(10233,2133)(10083,2283)
\path(9783,2283)(9633,2133)
\path(9783,2133)(10083,1833)
\path(10083,2133)(9783,1833)
\path(11058,33)(11058,858)(11883,858)
\end{picture}
}
%%%%%%%%%%%%
  \end{center}
  \caption{Typical 't~Hooft diagrams contributing to the one-loop scaling
dimension of a four gluon operator. There are also ``degenerate''
diagrams where
one of the vertices is connected directly to the gluon field appearing in a
covariant derivative or a commutator inside the operator.}
  \label{rain}
\end{figure}

Let us explain this in more detail,
starting with purely gluonic processes,
i.e.~interactions that couple only to gluons
within a local operator and only emit gluons.
The diagrams that contribute to the one-loop scaling dimension
are of three basic types (see \figref{rain}).

Consider first the loop-interaction connecting to
one field of the operator
(\figref{rain}, middle).
In QCD, the particles in the loop can only be gluons
(or the associated ghosts). Fundamental quarks are
suppressed in the large $\NNc$ limit.
In $\superN=4$ SYM, the gluons give precisely the same contribution,
but also the adjoint scalars and fermions do couple.
Hence we cannot simply read off the value of the gauge loop alone
{}from the final result in $\superN=4$ SYM.
We merely know that it gives the same contribution to
the scaling dimension for every gluon in the spin chain,
but we shall leave the coefficient, $E^f$, to be determined later.
Of course it can be computed (and is well-known);
at the end of this section
we will show a way to derive it from conformal and flavor symmetry.
To determine the contribution from the other two types
of interactions (\figref{rain}, left, right),
we note that scalars and fermions in the vertices of $\superN=4$ SYM
always come in pairs. This means that for purely gluonic processes they
can only contribute within internal loops of the diagram. These
two diagrams do not have such loops, hence their contribution
is exactly the same for $\superN=4$ SYM as for QCD.

We know that the complete one-loop dilatation generator
of $\superN=4$ SYM is given by
\[
\opham^{\superN=4}=
\sum_{j=0}^\infty E^{\superN=4}_j\,\opproj^{\superN=4}_{j}\qquad
\mbox{with}\qquad E^{\superN=4}_j=2h(j).
\]
The tricky part is the decomposition of the $\superN=4$ invariant projector
into projectors for the $\superN=0$ conformal group.
One finds that the module $\mdl^{\superN=4}_j$ contains
only chiral combinations of two gluons
$\mdl^{ff}_{j'},\mdl^{\bar f\bar f}_{j'}$
with definite conformal spin
$j'=j-2$. We write
\[
\opproj^{\superN=4}_j=\ldots
+\opproj^{ff}_{j-2}
+\opproj^{\bar f\bar f}_{j-2}
+\ldots\,.
\]
Consequently, the coefficient $E^{ff}_j$ equals
\[
E^{ff}_j=E^{\superN=4}_{j+2}+2E^f=2h(j+2)+2E^f
\]
where $E^f$ is the contribution from the missing loop-diagrams for scalars and fermions.
For the non-chiral combination, the determination of
$E^{f\bar f}_j$ is more involved.
The module $\mdl^{\superN=4}_j$ now contains
the non-chiral combination
$\mdl^{f\bar f}_{j'}$
with conformal spin ranging from $j'=j-2$ to $j'=j+2$.
A second complication is that the two interacting particles may or may not change
place. Finally, in $\superN=4$ SYM they can transform into particles which
are not part of QCD but carry the same quantum numbers of the conformal
group. While these contributions are clearly dropped for QCD, they
are essential for the definition of $\opproj^{\superN=4}_j$ as
a projector. We would have to treat all two-particle states
of $\superN=4$ with the same quantum numbers as $f\bar f$ to
find the proper decomposition of projectors. This is rather involved
and we just present the final result for the contributions
$E^{f\bar f,\opident}_j$ and $E^{f\bar f,\opperm}_j$
in \eqref{ggenergies}.

The situation for fundamental quarks
requires yet another insight, because
$\superN=4$ SYM contains only adjoint fields.
The point is that the gauge group structures
at the one-loop level are very restricted.
Again, they can only be of the three general forms
depicted in \figref{rain} (forgetting for the moment the
double line structure), where
a vertex represents the structure constants
$\lambda^a$ or $f^{abc}$, depending on
the type of line it attaches to (fundamental or adjoint).
The loop diagram can again not be determined from
$\superN=4$ SYM and it gives rise to the unknown coefficient $E^\psi$.
The other two structures can be distinguished by their symmetry
under interchange of the particles.
This is important, because in $\superN=4$ the second structure
can be transformed into the first one
by means of a Jacobi identity.
This means that for every contribution in the $\superN=4$ Hamiltonian
we can derive the corresponding structure.
This allows to derive also the terms for the fundamental fermions.
Note that the Yukawa-coupling to the scalars in $\superN=4$
involves two different flavors of fermions. It can therefore
be suppressed by considering only one flavor of fermions.

Finally, the unknown constants $E^f$ and $E^\psi$ can be obtained
by demanding conservation of the stress energy tensor and flavor currents.
We find $E^f=-\sfrac{11}{12}$
and $E^\psi=-\sfrac{3}{4}$ in agreement with
the $\beta$-function of large $\NNc$ QCD and results
of the next subsection.

The power of the truncation from $\superN=4$ however shows
up for integrability and the Bethe ansatz in \secref{sec:owd}.
Here we needed to do detailed computations to obtain the energy
coefficients $E^{\times\times}_j$, but the Bethe ansatz for QCD
follows from the Bethe ansatz for $\superN=4$ SYM
\cite{Beisert:2003yb} straightforwardly.

%%%%%%%%%%%%%%%%%%%%%%%%%%%%%%%%%%%%%%%%%%%%%%%%%%%%%%%%%%%%%%%%%%%%%%%%%%%%%%%%
\subsection{The lift from the light cone}

The same results \eqref{ggenergies}
can be reached by ``lifting{}'' the light cone
results~\cite{Lipatov:1994yb,Faddeev:1995zg,Braun:1998id,Braun:1999te,Belitsky:1999bf}.
The state of the art for this technique is presented
in~\cite{Belitsky:2004sc} where the full expression for the matrix
of anomalous dimensions
for the so-called ``quasi-partonic{}''
operators is given. This amounts to restricting
to the collinear subgroup $SL(2,R)$ of the full conformal group $SO(4,2)$.
Once the problem of classification and enumeration of the operators has
been solved, their answer can be lifted to the full conformal group.

Let us show how this works by lifting the expression for the
gluon-quark terms which are the ones that cannot be fixed by looking at
length-two gauge invariant operators.
One works in light-cone coordinates $+, -, \perp$ where $\perp$ denotes the
two transverse coordinates. The light cone decomposition is most
easily obtained by introducing two light cone vectors $n^\mu$ and
$\bar{n}^\mu$ satisfying the Lorentz products $n\cdot n= \bar{n}\cdot\bar{n}=0$
and $n\cdot \bar{n}=1$. One defines the $+$ and $-$ components by contracting
with $n$ and $\bar{n}$ respectively, e.g.
$F_{+-} = n^\mu\bar{n}^\nu F_{\mu\nu}$. (The notation is reviewed
in~\cite{Braun:2003rp,Belitsky:2004cz}.)

The collinear group is defined by the following three transformations,
forming a subgroup of the full conformal group:
\begin{equation}
   x_-\to \frac{x_-}{1 + 2ax_-}, \quad x_- \to x_- + c,
   \quad x_- \to \lambda x_-.
\end{equation}
It is easy to show that this subgroup is isomorphic to $SL(2,R)$. A generic
field, either spinorial or tensorial, can be split into various components
$\Phi$ carrying a dimension $d$, a spin projection $s$ (defined as
the eigenvalue of the component $\Sigma_{+-}$ of the spin operator) and a
``collinear conformal spin{}'' defined as
$\hat{\jmath} = (d+s)/2$, where we used the symbol $\hat{\jmath}$ to distinguish the
collinear conformal spin from the conformal counterpart $j$ previously used in
the full $SO(4,2)$ context. We will see now that there is a close relation
between these two quantum numbers.

A certain field component $\Phi$ can be evaluated on the light cone by
introducing a real variable $z$:
\begin{equation}
    \Phi(zn) = \sum_{k=0}^\infty \frac{z^k}{k!}D_+^k \Phi(0),
\end{equation}
where the Taylor expansion is a convenient way to keep track of the light-cone
derivatives. A generic light-cone composite operator can thus be thought
of as a polynomial in the variables $z_i$, one for each elementary field.
For instance, in~\cite{Shuryak:1981kj,Shuryak:1981pi,Belitsky:2004cz}
was considered the operator:
\begin{equation}
     S^+(z_1, z_2, z_3) = \frac{g_\mathrm{YM}}{2}\bar{q}(z_1n)(i\tilde{F}_{\perp +}(z_2n)+
      F_{\perp +}(z_2n)\gamma_5)\gamma_+ q(z_3n), \label{splus}
\end{equation}
for which the following Hamiltonian was given (changing slightly their
notation to make the comparison with our formulas clearer):
\begin{equation}
    \Gamma = \unit \left[ V_{qF}(\hat{\jmath}_{12}) + U_{Fq}(\hat{\jmath}_{23})\right].
\end{equation}
The coefficients $V_{qF}$  and $U_{Fq}$, to be related to our
$E^{\bar\psi f}$ and $E^{f \psi}$ respectively, were found to be:
\begin{eqnarray}
    V_{qF}(\hat{\jmath}) \eq {\Psi}(\hat{\jmath} + \sfrac{3}{2}) +
       {\Psi}(\hat{\jmath} - \sfrac{3}{2}) - 2 {\Psi}(1) -
       \sfrac{3}{4},
\nonumber\\
    U_{Fq}(\hat{\jmath}) \eq {\Psi}(\hat{\jmath} + \sfrac{1}{2}) +
       {\Psi}(\hat{\jmath} - \sfrac{1}{2}) - 2 {\Psi}(1) -
       \sfrac{3}{4}. \label{theirs}
\end{eqnarray}
In \eqref{theirs} we have denoted by ${\Psi}$ the logarithmic derivative of
the Gamma function $\Psi(z) = \Gamma^\prime(z)/\Gamma(z)$,
related to the harmonic sum by:
\begin{equation}
     \Psi(m) =  h(m-1) - \gamma_E,
\end{equation}
$\gamma_E$ being the Euler constant that always cancels in the final
expression.

We now must relate the collinear $\hat{\jmath}$ to our previous conformal
spin $j$. This can be done by observing that when acting on the rightmost
link (i.e. for $U_{Fq}$) the primary $ f_{\alpha\beta}\psi_\gamma$
totally symmetrized has $j=0$ and contains the
light cone component with $\hat{\jmath} = 5/2$  ($\hat{\jmath} = 3/2$ from $F$
and  $\hat{\jmath} = 1$ from $\psi$). Thus in  $U_{Fq}$ we should set
$\hat{\jmath} = j + 5/2$.
  Similarly, when acting on the leftmost link, (i.e. for
$V_{qF}$) the relation should be $\hat{\jmath} = j + 3/2$ because
now we have $j=1$.

Finally, in \eqref{splus} there is an explicit factor of
$g_\mathrm{YM}$ in front of the
operator. This means that the total anomalous dimensions
are shifted by $-11/6$ in units of $\unit$
but, since the open chain has two links one must
shift the contribution of each link \eqref{theirs} by $-11/12$.

Putting all this together yields back the previous results
for $E^{\bar\psi f}$ and $E^{f \psi}$. In general, all the
light-cone Hamiltonian can be lifted in a unique way to the full
conformal group.

%%%%%%%%%%%%%%%%%%%%%%%%%%%%%%%%%%%%%%%%%%%%%%%%%%%%%%%%%%%%%%%%%%%%%%%%%%%%%%%%
%%%%%%%%%%%%%%%%%%%%%%%%%%%%%%%%%%%%%%%%%%%%%%%%%%%%%%%%%%%%%%%%%%%%%%%%%%%%%%%%
\section{Chiral operators and Bethe ansatz}
\label{sec:5}

The conformal group provides a nice and compact bookkeeping tool for
local operators in QCD which allowed us to write down the complete
planar mixing matrix. The next obvious task is to diagonalize it.
Though the complete diagonalization is beyond our reach, it will be
possible to compute many interesting physical quantities.
In particular, we compute
the ground state of the mixing Hamiltonian
to leading order%
\footnote{In fact, one can
systematically expand in powers of $1/L$.
The coefficients of $L^0$ and $L^{-1}$ also have interesting physical
meaning, being related to the boundary energy and the central charge
respectively.}
in $L$ and analyze the spectrum
of small perturbations around it by the Bethe ansatz. Before
going into the details of the general Bethe ansatz equations, we find it
instructive to discuss a simple reduction of the full mixing
problem.

%%%%%%%%%%%%%%%%%%%%%%%%%%%%%%%%%%%%%%%%%%%%%%%%%%%%%%%%%%%%%%%%%%%%%%%%%%%%%%%%
\subsection{Chiral sector}

There are several types of operators that mix only among themselves
and thus can be treated separately. An obvious example is a set of
operators with the same quantum numbers such as spin, chirality or
classical dimension. Though the length of the chain (the number of
fields in an operator) is not a good quantum number, it is also
conserved at the one-loop level, simply because producing an extra
field requires an extra interaction vertex and is thus suppressed by
the QCD coupling. As in~\cite{Ferretti:2004ba}, we shall first
consider operators with minimal classical dimension for a given
length. Such operators are composed of the gluon field strength and
contain no derivatives:
\begin{equation}
{\cal O}=\tr F_{\mu_1\nu_1}\ldots F_{\mu_L\nu_L}\,,
\end{equation}
A yet simpler subsector consists of chiral operators which contain
only self-dual components of the field strength:
\begin{equation}\label{chirop}
{\cal O}=\tr f_{\alpha_1\beta_1}\ldots f_{\alpha_{L}\beta_{L}}\,.
\end{equation}
This relatively simple subset of operators contains the
ground state of the full mixing Hamiltonian.

Each entry $f_{\alpha_l\beta_l}$  in the operator represents one
site of the chain of length $L$. Since we are excluding derivatives,
the number of degrees of freedom
per site is now finite. The three states $f_{\alpha_l\beta_l}$ form
the spin-1 representation of $SU(2)_L$. The spin operator is
\begin{equation}\label{spinopera}
(\opspin^i\,f)_{\alpha\beta}=\frac{1}{2}\left(
\sigma_{\hphantom{i}\alpha}^{i\hphantom{\alpha}\gamma}\,
\delta_{\beta}^{\hphantom{\beta }\varepsilon } +\delta
_{\alpha}^{\hphantom{\alpha}\gamma}\, \sigma
_{\hphantom{i}\beta}^{i\hphantom{\beta }\varepsilon } \right)
f_{\gamma \varepsilon },
\end{equation}
where $\sigma_{\hphantom{i}\alpha}^{i\hphantom{\alpha}\gamma}$,
$i=1,2,3$ are the ordinary Pauli matrices. The mixing matrix
\eqref{Hclosed} acts pairwise on the adjacent sites of the chain.
A pair of spins can be in the spin-0, spin-1 or spin-2 state  and
the interaction Hamiltonian depends only on the total spin. {}From
\eqref{gg,ggenergies} we find that
\begin{equation}
\opham_{l,l+1}= \left\{  \begin{array}{ll}
-11/6,  & \mbox{if}~(\opspin_l+\opspin_{l+1})^2=0, \\
1/6,  & \mbox{if}~(\opspin_l+\opspin_{l+1})^2=2, \\
7/6,  & \mbox{if}~(\opspin_l+\opspin_{l+1})^2=6. \\
\end{array}  \right.
\end{equation}
Using $\mathbf{S}_l^2=2$, the Hamiltonian can be also written as
\cite{Ferretti:2004ba}
\begin{equation}\label{spin1}
\opham_{l,l+1} = \sfrac{7}{6}\opident+\half\Scq-\half (\Scq)^2,
\end{equation}
which is a Hamiltonian of the spin-1 quantum spin chain. Remarkably,
this Hamiltonian is
integrable~\cite{Zamolodchikov:1980ku,Kulish:1981gi,Reshetikhin:1985vd}
and its spectrum can be analyzed by
the Bethe ansatz~\cite{Bethe:1931hc,Takhtajan:1981xx,Babujian:1982ib,Babujian:1983ae,Faddeev:1981ip,Faddeev:1996iy}.

It is convenient to use the basis in which the $S^3_l$'s are diagonal:
\begin{equation}
f_+=f_{11},\qquad
f_0=\frac{1}{\sqrt{2}}\left(f_{12}+f_{21}\right),\qquad f_-=f_{22}.
\label{tripl}
\end{equation}
The operator
\begin{equation}\label{ferrgs}
{\cal O}_\Omega =\tr f_+^L
\end{equation}
is an obvious eigenstate of the Hamiltonian \eqref{spin1} with
anomalous dimension $\gamma_\Omega  =7\as \NNc L/12\pi$ and spin $S=L$. This state
is ferromagnetic (all spins aligned) and in the present case it is the
state with the highest possible energy, which is easy to understand
{}from \eqref{spin1}: anti-alignment of nearest-neighbor spins lowers
the energy, so the true ground state is anti-ferromagnetic. Quantum
anti-ferromagnets are rather complicated systems even in one
dimension, but in the present case the anti-ferromagnetic ground
state can be found with the help of the Bethe ansatz~\cite{Bethe:1931hc,Faddeev:1981ip,Faddeev:1996iy}.

It should be stressed at this point that integrability does not extend to
the full Hamiltonian \eqref{gg} due to the presence of the exchange operator
$\mathbf{X}$. Clearly, without such term,  \eqref{gg} would be the sum of two
independent integrable Hamiltonian, one for the chiral and one for the
anti-chiral sector. The presence of $\mathbf{X}$ turns the system into a spin
ladder by coupling the two sectors. However, it also spoils integrability as
can be seen explicitly by constructing the candidate higher order conserved
charges and showing that they do not commute with the Hamiltonian.
In $\superN=4$ SYM integrability is regained for the entire mixing matrix due to the
presence of extra fields. In fact, we interpret the integrability
of the chiral sector of QCD as a remnant of $\superN=4$ integrability.

%%%%%%%%%%%%%%%%%%%%%%%%%%%%%%%%%%%%%%%%%%%%%%%%%%%%%%%%%%%%%%%%%%%%%%%%%%%%%%%%
\subsection{The Bethe ansatz}

The Bethe ansatz describes all eigenstates of the spin chain in
terms of elementary excitations around the ferromagnetic
(pseudo)vacuum. The excitations close to the pseudo-vacuum
(magnons) correspond to
replacing some of the $f_+$'s in \eqref{ferrgs} by $f_0$ or $f_-$.
The magnons are thus created and annihilated by operators
\begin{equation}\label{ldo}
a^\dagger(l)
\approx\frac{1}{\sqrt{L}}\sum_{m=1}^L \eeee^{ip(l)m}\; \opspin_m^-,\qquad
a(l) \approx\frac{1}{\sqrt{L}}\sum_{m=1}^L
\eeee^{-i \bar{p}(\bar{l})m} \;\opspin_m^+.
\end{equation}
The momenta here are parameterized by rapidities:
$\eeee^{ip(l)}=(l+i)/(l-i)$, which is standard in the literature on
Bethe ansatz. Each magnon reduces the spin by 1, so the state
$a^\dagger(l_1)\ldots a^\dagger(l_M)|0\rangle$ has spin $S=L-M$. The
rapidities $l_i$ are in general complex because the magnons can form
bound states with decaying wave function. The operators \eqref{ldo}
create eigenstates of the Hamiltonian only asymptotically for very
large chains and small number of magnons, when scattering of magnons
on each other can be neglected. Nevertheless, the Hamiltonian can be
diagonalized in a purely algebraic way if the creation and
annihilation operators are appropriately deformed to take into
account the scattering. The construction of the spectrum-generating
operators, known as the algebraic Bethe ansatz, is rather involved
and we refer to the original papers
\cite{Takhtajan:1981xx,Babujian:1983ae} or to the review
\cite{Faddeev:1996iy} for the detailed derivation. Here we just
quote the results.

The rapidities of magnons should be all different and
satisfy a set of algebraic
equations:
\begin{equation}\label{bethel}
\left(\frac{l_j+i}{l_j-i}\right)^L=
\prod_{\textstyle\atopfrac{k=1}{k\neq j}}^{M}\frac{l_j-l_k+i}{l_j-l_k-i}\,,
\end{equation}
which is basically the periodicity condition for the multi-magnon
wave function. The right hand side of equation \eqref{bethel} contains
scattering phases. If scattering is neglected the equation
reduces to the quantization condition for the momenta:
\[p(l_j)\equiv\pi -2\arctan l_j=\frac{2\pi n_j}{L}.\]
Solutions of the Bethe equations $\{l_j,~j=1,\ldots ,M\}$, $0\leq
M\leq L$ parameterize all eigenstates of the spin-chain Hamiltonian.
The eigenvalues are
\begin{equation}
\gamma_{\{l_j\}}=\unit\left(
\frac{7L}{6}-\sum_{j=1}^M\frac{2}{l_j^2+1}\right).
\end{equation}
There is a useful relationship between the momentum of a Bethe root
and its contribution to the anomalous dimension:
\[
\gamma (l)=\unit\,p'(l).
\]
This relationship holds for arbitrary compounds of Bethe roots and
will be very useful in calculations.

The Bethe states which correspond to QCD operators should satisfy an
extra condition
\begin{equation}
\prod_{j=1}^M \frac{l_j+i}{l_j-i} =1,
\end{equation}
which guarantees that the state has zero total momentum and hence is
translation invariant. This condition reflects the cyclicity of
the trace in the operators \eqref{chirop}.

%%%%%%%%%%%%%%%%%%%%%%%%%%%%%%%%%%%%%%%%%%%%%%%%%%%%%%%%%%%%%%%%%%%%%%%%%%%%%%%%
\subsection{Anti-ferromagnetic ground state}

We are now in a position to construct the true, anti-ferromagnetic
vacuum. Again we omit many details that can be found in the original
papers~\cite{Takhtajan:1981xx,Babujian:1982ib,Babujian:1983ae}. The
ground state has spin zero and thus contains $L$ magnons.
All their rapidities should be different, so the construction of the
anti-ferromagnetic vacuum of the spin chain is similar to the filling of
a Fermi sea. As we mentioned earlier, magnons with the same momentum
form a bound state, so their rapidities become complex. The
left hand side of the Bethe equation \eqref{bethel} then is not a
pure phase and in the thermodynamic limit $L\rightarrow \infty $ it
either blows up or goes to zero. This has to be compensated by a
zero or a pole in the scattering amplitude on the right hand side. A
pole or a zero arises when some of the Bethe roots differ by $i$.
The bound state of $k$ magnons is thus described in the
thermodynamic limit by an array of Bethe roots with a common real
part and integer or half-integer imaginary parts. Such a compound of
Bethe roots is usually called a $k$-string. The 2-strings play the
most important role for the spin-1 chain.
They are pairs of roots at
\begin{equation}
l_{2j-1}=\lambda_j+i/2,\qquad l_{2j}=\lambda_j-i/2,
\end{equation}
where the centers of the strings $\lambda_j$ are real numbers. The ground
state of the spin-1 chain is the Fermi sea of 2-strings
~\cite{Takhtajan:1981xx,Babujian:1983ae}.

To find the distribution of the 2-strings, we first multiply the
Bethe equations for both roots and rewrite them in the logarithmic
form:
\begin{equation}\label{bethlog}
p_{2}(\lambda _j)=\frac{2\pi n_j}{L}+\frac{1}{L}\sum_{k\neq j}\delta
_{2,2}(\lambda _j-\lambda _k),
\end{equation}
where $p_2(\lambda )$ is the momentum of the 2-string:
\begin{equation}
\eeee^{ip_2(\lambda )}=\frac{\lambda +i/2}{\lambda -i/2}\,
\frac{\lambda +3i/2}{\lambda -3i/2}\,,
\end{equation}
which is equal to
\begin{equation}
p_2(\lambda )=2\pi-2\arctan 2\lambda -2\arctan\frac{2\lambda }{3}\,.
\end{equation}
The branch of the arctangent is chosen such that
$-\pi/2<\arctan\lambda<\pi/2$. The phase ambiguity in choosing the
branch is reflected in the mode numbers $n_j$ which must be
different for different strings. For the scattering phase of the two
strings we find after some calculations:
\begin{equation}\label{eq:asdfjlh}
\delta _{2,2}(\lambda)=3\pi-4\arctan\lambda -2\arctan\frac{\lambda
}{2}\,.
\end{equation}
Since in the vacuum all available one-particle states are occupied,
we can put $n_j=j$, $j=1,\ldots ,L/2$ in \eqref{bethlog}.

In the thermodynamic limit, the distribution of
 Bethe strings can be characterized by a continuous function
$\lambda=\lambda(x)$ of the variable $x=j/L$ or by the density
\[
\rho(\xi)=-\left.\frac{1}{\lambda'(x)}\right|_{\xi=\lambda(x)}
\approx\frac{L}{\lambda _{j}-\lambda _{j+1}}\,.
\]
The thermodynamic limit of the Bethe equations  for rapidities is an
integral equation for the density:
\begin{eqnarray}\label{inteqdens}
&&\frac{1}{2}\,\frac{1}{\lambda^2+1/4}
+\frac{3}{2}\,\frac{1}{\lambda^2+9/4} =\pi\rho(\lambda) \nonumber \\
&&+2\int_{-\infty}^{+\infty}d\xi\,\rho(\xi)\,
\left[\frac{1}{(\lambda-\xi)^2+1}
+\frac{1}{(\lambda-\xi)^2+4}\right].
\end{eqnarray}
This equation is derived by subtracting \eqref{bethlog} for the
$(j+1)$-th string from the equation for the $j$-th string and taking
the difference $\lambda _j-\lambda _{j+1}$ to zero. The equation can
be easily solved by the Fourier transform:
\begin{equation}
\rho(\lambda)=\int_{-\infty}^{+\infty}\frac{dq}{2\pi}\,
\,\frac{\eeee^{-iq\lambda}}{2\cosh\frac{q}{2}}
=\frac{1}{2\cosh\pi\lambda}\,. \label{rholambda}
\end{equation}
The density is normalized as
\begin{equation}
\int_{-\infty}^{+\infty}d\lambda\,\rho(\lambda)=\frac{1}{2}\,,
\end{equation}
so that the ground state contains precisely $L/2$ 2-strings or $L$
roots and is therefore a spin zero state. The ground-state energy is, to
leading order in $L$:
\begin{equation}\label{antiferrogro}
\gamma _0=\unit L  \left[\frac{7}{6}-\int_{-\infty
}^{+\infty }d\lambda \,\rho (\lambda )\left(\frac{3}{\lambda
^2+9/4}+\frac{1}{\lambda ^2+1/4}\right)\right] =-\frac{5\as \NNc L}{12\pi}\,,
\end{equation}
which is the lowest possible anomalous dimension for operators of a
given length. In the subsector without derivatives, the complete
Hamiltonian is also bounded from above by the ferromagnetic vacuum.
However, this is an artifact of the truncation -- including
derivatives it is possible to arbitrarily raise the anomalous
dimension of operators of a given length. For instance, in the case
of twist-two operators anomalous
dimensions grow like $\log S$ where $S$ is the number of
derivatives.

%%%%%%%%%%%%%%%%%%%%%%%%%%%%%%%%%%%%%%%%%%%%%%%%%%%%%%%%%%%%%%%%%%%%%%%%%%%%%%%%
\subsection{Operators with derivatives}\label{sec:owd}

We now consider adding covariant derivatives to the chiral operators
discussed above. The resulting set of operators turns
out to be the largest integrable sector in the pure Yang-Mills
theory. At the same time this sector contains all low-energy
excitations around the anti-ferromagnetic vacuum discussed in the
previous section. We will show below that the other, non-chiral
modes are separated from the vacuum by a gap. The most general
operators in the chiral sector have the form
\begin{equation}\label{oo}
{\cal O}=\tr (D^{m_1}f)\ldots (D^{m_L}f).
\end{equation}
This sector is closed under renormalization at one-loop
and integrability follows from integrability
in $\superN=4$ SYM \cite{Beisert:2003yb}
by a simple argument:
The operators \eqref{oo} also form a one-loop closed subsector
of $\superN=4$ SYM.%
\footnote{Similar ideas have been pursued by R.~Argurio
(private communication).}
Therefore not only the mixing matrix for \eqref{oo}
is inherited (up to a constant shift), but also its integrability.%
\footnote{In \Appref{sec:YB} we present an independent proof
of this statement.}
The mixing matrix can now be diagonalized by a Bethe ansatz.
As usual, the Bethe equations are completely fixed by group
theory and can be read off from the general result of
\cite{Ogievetsky:1986hu}, see also~\cite{Johannesson:1986ig}.

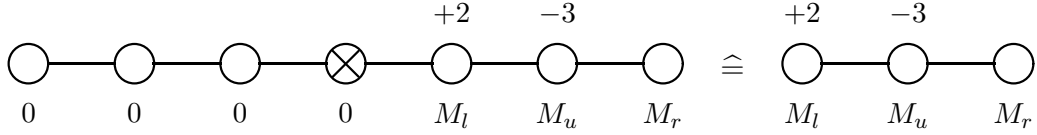
\begin{figure}\centering
\begin{minipage}{260pt}
\setlength{\unitlength}{1pt}%
\small\thicklines%
\begin{picture}(260,55)(-10,-30)
\put(  0,00){\circle{15}}%
\put(  0,10){\makebox(0,0)[b]{}}%
\put(  7,00){\line(1,0){26}}%
\put( 40,00){\circle{15}}%
\put( 40,15){\makebox(0,0)[b]{}}%
\put( 47,00){\line(1,0){26}}%
\put( 80,00){\circle{15}}%
\put( 80,15){\makebox(0,0)[b]{}}%
\put( 87,00){\line(1,0){26}}%
\put(120,00){\circle{15}}%
\put(120,15){\makebox(0,0)[b]{}}%
\put(115,-5){\line(1, 1){10}}%
\put(115, 5){\line(1,-1){10}}%
\put(127,00){\line(1,0){26}}%
\put(160,00){\circle{15}}%
\put(160,15){\makebox(0,0)[b]{$+2$}}%
\put(167,00){\line(1,0){26}}%
\put(200,00){\circle{15}}%
\put(200,15){\makebox(0,0)[b]{$-3$}}%
\put(207,00){\line(1,0){26}}%
\put(240,00){\circle{15}}%
\put(240,15){\makebox(0,0)[b]{}}%
\put(  0,-15){\makebox(0,0)[t]{$0$}}%
\put( 40,-15){\makebox(0,0)[t]{$0$}}%
\put( 80,-15){\makebox(0,0)[t]{$0$}}%
\put(120,-15){\makebox(0,0)[t]{$0$}}%
\put(160,-15){\makebox(0,0)[t]{$M_l$}}%
\put(200,-15){\makebox(0,0)[t]{$M_u$}}%
\put(240,-15){\makebox(0,0)[t]{$M_r$}}%
\end{picture}
\end{minipage}\quad$\widehat{=}$\quad
\begin{minipage}{120pt}
\setlength{\unitlength}{1pt}%
\small\thicklines%
\begin{picture}(120,55)(-10,-30)
\put(  0,00){\circle{15}}%
\put(  0,15){\makebox(0,0)[b]{$+2$}}%
\put(  7,00){\line(1,0){26}}%
\put( 40,00){\circle{15}}%
\put( 40,15){\makebox(0,0)[b]{$-3$}}%
\put( 47,00){\line(1,0){26}}%
\put( 80,00){\circle{15}}%
\put( 80,15){\makebox(0,0)[b]{}}%
\put(  0,-15){\makebox(0,0)[t]{$M_l$}}%
\put( 40,-15){\makebox(0,0)[t]{$M_u$}}%
\put( 80,-15){\makebox(0,0)[t]{$M_r$}}%
\end{picture}
\end{minipage}

\caption{Reduction of the distinguished Dynkin diagram
of $SU(4|2,2)$ to $SU(2,2)$.
Above the diagrams we have indicated the
Dynkin labels of the spin representation
and below the excitation numbers of each simple root.}

\label{fig:Dynkin}
\end{figure}

The mixing of operators \eqref{oo} is described by an $SO(4,2)$
spin chain with spins in the representation whose Dynkin
labels are $[p,r,q]=[2,-3,0]$. The left and right Dynkin indices refer to twice
the Lorentz spins $p = 2 S_1$ and $q = 2 S_2$,
whereas the central Dynkin label is given by $r=-D-S_1-S_2$.
The spectrum of the integrable spin chain
with these symmetries is characterized by three sets of Bethe roots
$u_j$, $l_j$ and $r_j$. The Bethe equations can be inferred from
\cite{Ogievetsky:1986hu}:
\begin{eqnarray}\label{bethe1}
\left(\frac{l_j+i}{l_j-i}\right)^L\eq
\prod^{M_l}_{\textstyle\atopfrac{k=1}{k\neq j}}\frac{l_j-l_k+i}{l_j-l_k-i}\,
\prod^{M_u}_{k=1}\frac{l_j-u_k-i/2}{l_j-u_k+i/2}\,,
\nonumber \\
\left(\frac{u_j-3i/2}{u_j+3i/2}\right)^L\eq
\prod^{M_u}_{\textstyle\atopfrac{k=1}{k\neq j}}\frac{u_j-u_k+i}{u_j-u_k-i}\,
\prod^{M_l}_{k=1}\frac{u_j-l_k-i/2}{u_j-l_k+i/2}\,
\prod^{M_r}_{k=1}\frac{u_j-r_k-i/2}{u_j-r_k+i/2}\,,
\nonumber \\
1\eq
\prod^{M_r}_{\textstyle\atopfrac{k=1}{k\neq j}}\frac{r_j-r_k+i}{r_j-r_k-i}\,
\prod^{M_u}_{k=1}\frac{r_j-u_k-i/2}{r_j-u_k+i/2}\,.
\end{eqnarray}
These are precisely the Bethe equations of $\superN=4$
SYM \cite{Beisert:2003yb}
when one truncates the supergroup $SU(2,2|4)$
down to the bosonic part $SU(2,2)$.
For this purpose one needs to consider the `Beast' form
in \cite{Beisert:2003yb} and remove (or simply not excite)
the fermionic node along with
the three nodes of the internal symmetry $SU(4)$
{}from the distinguished Dynkin diagram
of $SU(2,2|4)$, see \figref{fig:Dynkin}.
The remaining Cartan matrix for the construction of the Bethe ansatz
is the one of $SU(2,2)$ and the spin labels are $[2,-3,0]$,
i.e.~\eqref{bethe1}.%

If one reduces further by not exciting the
$u$ and $r$ roots, one returns to the Bethe
equations \eqref{bethe1} for the $SU(2)_L$ sector.
Equivalently, removing $l$ and $r$ roots
leads to the $SL(2)$ Bethe ansatz reviewed
in \cite{Braun:2003rp,Belitsky:2004cz}.

The derivation of the momentum constraint and of the anomalous dimension
follows exactly the same route as in the previous section.
We simply present the results, which
agree with $\superN=4$ SYM \cite{Beisert:2003yb}
up to an overall energy shift proportional to $L$.
The cyclic states are in addition subject to the
momentum constraint:
\begin{equation}
\prod_{j=1}^{M_l} \frac{l_j+i}{l_j-i}
\prod_{j=1}^{M_u} \frac{u_j-3i/2}{u_j+3i/2}=1.
\end{equation}
The anomalous dimension is
\begin{equation}\label{ADim}
\gamma=\unit\left(
\frac{7L}{6}-\sum_k\frac{2}{l_k^2+1}+\sum_k\frac{3}{u_k^2+9/4}
\right).
\end{equation}

The Bethe state with $M_l$ $l$-roots, $M_u$ $u$-roots and
$M_r$ $r$-roots is the highest weight in the $SO(4,2)$
representation  with Dynkin labels
\[
[2L+M_u-2M_l,-3L+M_l+M_r-2M_u,M_u-2M_r].
\]
In other words it has
\begin{equation}\label{spins}
D=2L+M_u,\qquad S_1=L+\half M_u-M_l,\qquad
S_2=\half M_u-M_r.
\end{equation}
The reference state with no excitations (the pseudo-vacuum)
corresponds to the operator \eqref{ferrgs} that was
identified in \Secref{sec:5} as the ferromagnetic vacuum:
\begin{equation}\label{oohw}
{\cal O}_\Omega=\tr f_{11}^L = \tr f_{+}^L.
\end{equation}
Adding a $u$-root corresponds to adding a derivative $D_{1\dot{1}}$
to the above operator. An $l$-roots flips one left spin: $1\rightarrow 2$,
and an $r$-root flips one right spin: $\dot{1}\rightarrow \dot{2}$.
Since the Lorentz spins in \eqref{spins} cannot be negative,
the numbers of roots of different types are constrained by
\begin{equation}
M_l\leq L+\half M_u,\qquad M_r\leq\half M_u\,.
\end{equation}
Keeping only $l$-roots brings us back to the Bethe equations of
the XXX$_1$ chain discussed in~\cite{Ferretti:2004ba}.

We can reproduce from the Bethe equations
all anomalous dimensions of the operators in
Table~\ref{tableone}
that belong to the chiral sector, i.e. those satisfying $A=L$.
The results are given in Table~\ref{two} for comparison.

\begin{table}\centering
$\begin{array}{|cc|c|c|c|l|}
\hline
D&L&(S_1,S_2)&{[p,r,q]} & E & \multicolumn{1}{c|}{\mbox{roots}}\\
\hline
4 & 2 & (0,0) & {[0,-4,0]}  & -11/3 &  l_{1,2} = \pm i/\sqrt{3} \\
4 & 2 & (2,0) & {[4,-6,0]}  & 7/3   & \mbox{(no roots)} \\
6 & 3 & (0,0) & {[0,-6,0]}  & 1/2   & l_1=0,\, l_{2,3} = \pm i\mbox{~(singular state)}\\
6 & 3 & (1,0) & {[2,-7,0]}  & -3/2  & l_{1,2}=\pm i/\sqrt{5}\\
6 & 3 & (3,0) & {[6,-9,0]}  & 7/2   & \mbox{(no roots)}\\
6 & 2 & (3,1) & {[6,-10,2]} & 14/3  & u_{1,2}=\pm 3/\sqrt{28}\\
7 & 3 & (3/2,1/2) & {[3,-9,1]}  & 2, 2  & u_1= \pm 1/2 \sqrt{5/7}\mbox{~(paired state)}\\
&&&&& l_1=\bar l_2=\mp\sqrt{\frac{1 + i \sqrt{399}}{30}}
\\
7 & 3 & (5/2,1/2) & {[5,-10,1]} & 17/6  & u_1=l_1=0\\
\hline
\end{array}$
\caption{Anomalous dimensions ($\gamma = \unit E$) and Bethe roots for the
lowest dimensional
chiral closed chains ($A=L$ in Table~\protect\ref{tableone}).}
\label{two}
\end{table}

Adding the derivatives does not change the ground state,
since $u$-roots give positive contribution to the energy.
In Section~\ref{sec:6}
we will show that adding $\bar{f}$ to a chiral operator raises the
energy by substantially larger amount. Thus, the ground state and
all the low-energy modes of the mixing matrix are described by an
integrable system.

\subsection{Open chains}\label{sec:opench}

Before moving on to a more detailed discussion of the spectrum of the closed
chain, we would like to make some comments on the open chain.
(For earlier work on operators of this type we refer the reader
to~\cite{Braun:1998id,Belitsky:1999bf,Derkachov:1999ze}.) 
The open chains with quarks at the ends are also integrable if the gluon
operators in the middle are chiral (see Appendix~\ref{sec:YB}).
Let us consider operators that have no derivatives:
\begin{equation}\label{kasjdf}
 {\cal O}=\bar{q}f_{\alpha _2\beta _2}\ldots
 f_{\alpha _{L-1}\beta _{L-1}} q.
\end{equation}
The mixing matrix for these operators is described by
the following open spin--1 chain:
\begin{equation}\label{wqeoifdj}
\Gamma =\unit\left\{
 \frac{4-L_r }{6}+\Spinb_1\cdot\Spin_2
 +\Spin_{L-1}\cdot \Spinb_L+ \sum_{l=2}^{L-2}
 \left[\frac{7}{6}\opident+\frac{1}{2}\Scq-\frac{1}{2} (\Scq)^2\right]
 \right\},
\end{equation}
where $L_r =0,1,2$ counts the number of anti-chiral quarks.
Here $\Spinb$ depends on the chirality of the quark \eqref{quarkss}
that is inserted at each end of the spin chain:
For chiral quarks we set $\Spinb=\spins/2$, while anti-chiral
ones have no $SU(2)\indup{L}$ spin and thus $\Spinb=0$.
Both types of boundary interaction are integrable.
In the case with two anti-chiral quarks, we can take either the spin-0 or spin-1 combination of $SU(2)_R$ without affecting the anomalous dimension.
The chain with the boundary interaction has been studied
recently in \cite{Arnaudon:2004xx} 
(see also~\cite{Chen:2004mu, Chen:2004yf} for
a discussion of open spin chains in a supersymmetric context).

The Bethe ansatz for the system described in~\eqref{wqeoifdj} is
\[
\lrbrk{\frac{l_j+i}{l_j-i}}^{2L-2-L_r }
=
\prod^{M}_{\textstyle\atopfrac{k=1}{k\neq j}}
\frac{l_j-l_k+i}{l_j-l_k-i}\,
\frac{l_j+l_k+i}{l_j+l_k-i}
\]
and its eigenvalues are given by
\begin{equation}
\gamma=\unit\left(
\frac{7L-4L_r -11}{6}-\sum_{k=1}^{M}\frac{2}{l_k^2+1}
\right).
\end{equation}
There is no momentum constraint.

We can reproduce from the Bethe equations
all anomalous dimensions of the operators in
Table~\ref{tableoneb},\ref{tableonec}
that belong to the chiral sector, i.e. those satisfying $A=L-1-L_r $.
The results are given in Table~\ref{xxx} for comparison.

We will not study the thermodynamic limit of the open
spin chain here, but we should mention that the anti-ferromagnetic phase
of open spin chains exhibit rather unusual behavior under certain
circumstances \cite{Doikou:2002wd}.

\begin{table}\centering
$\begin{array}{|ccc|c|c|c|l|}
\hline
D & L & L_r  & (S_1,S_2) & {[p,r,q]} &  E &\multicolumn{1}{c|}{\mbox{roots}}\\
\hline
5 & 3 & 0 & (0,0) & {[0,-5,0]}  & -4/3 & l_{1,2}=(1\pm i)/\sqrt[4]{12} \\
5 & 3 & 0 & (1,0) & {[2,-6,0]}  & -1/3   & l_1=0 \\
5 & 3 & 0 & (1,0) & {[2,-6,0]}  & 2/3   & l_1=1 \\
5 & 3 & 0 & (2,0) & {[4,-7,0]}  & 5/3   & \mbox{(no roots)}\\\hline
5 & 3 & 1 & (1/2,1/2) & {[1,-6,1]}  & -1/2  & l_1=1/\sqrt{3}\\
5 & 3 & 1 & (3/2,1/2) & {[3,-7,1]}  & 1   & \mbox{(no roots)}\\\hline
5 & 3 & 2 & (1,0) & {[2,-6,0]}  & 1/3  & \mbox{(no roots)}\\
5 & 3 & 2 & (1,1) & {[2,-7,2]}  & 1/3  & \mbox{(no roots)}\\
\hline
\end{array}$
\caption{Anomalous dimensions ($\gamma = \unit E$) and Bethe roots for the
lowest dimensional
open chiral chains ($A=L-1-L_r$ in Table~\protect\ref{tableoneb},\protect\ref{tableonec}).}
\label{xxx}
\end{table}

%%%%%%%%%%%%%%%%%%%%%%%%%%%%%%%%%%%%%%%%%%%%%%%%%%%%%%%%%%%%%%%%%%%%%%%%%%%%%%%%
%%%%%%%%%%%%%%%%%%%%%%%%%%%%%%%%%%%%%%%%%%%%%%%%%%%%%%%%%%%%%%%%%%%%%%%%%%%%%%%%
\section{Spectrum of excitations}
\label{sec:6}

We can now study the spectrum of excitations around the anti-ferromagnetic
ground state
of the spin chain. In part, this can be done using techniques that
are well known from the literature on one-dimensional
anti-ferromagnets. Let us first summarize our findings: There are three
types of excitations%
\footnote{In this context, an excitation refers
to a distortion of the wave-function
which carries an \emph{independent} momentum.
Physical states composed from such excitations
are subject to constraints:
A single excitation may be unphysical, see below.}
(Fig.~\ref{fig:chain2}):
\begin{itemize}
\item
Excitations with spin quantum numbers $(0,1)$ separated from the vacuum by a
finite gap: $\varepsilon \sim {\rm const}$.
These excitations correspond to changing a $f$
into a $\bar{f}$.

\item
Excitations with spin quantum numbers $(1/2,0)$ and relativistic dispersion
relation: $\varepsilon \sim p\sim 1/L$.
These modes are the low-energy states of the
anti-ferromagnetic XXX$_1$ spin chain which are usually called
spinons or spin waves.
A single spinon has fractional spin and can therefore never appear on its own
\cite{Faddeev:1981ip,Takhtajan:1981xx,Babujian:1983ae}.
Furthermore, it obeys a non-standard exchange statistics.

\item
Excitations with spin quantum numbers $(0,1/2)$ and non-relativistic
dispersion relation: $\varepsilon \sim p^2\sim 1/L^2$.
These modes have the lowest possible energy $\sim 1/L^2$
and are turned on by the insertion of a covariant derivative into the
operator.
The derivative has both a left (chiral) and a right (anti-chiral) index but,
since the left spin propagates in the anti-ferromagnetic vacuum,
the left and right excitations propagate independently.
The left excitation dissolves into the ground state
and thus the only new element is the right excitation of
classical scaling dimension $1$.
These excitations are somewhat analogous to magnons in a
ferromagnet and can form
multi-particle bound states. They are probably the most surprising and
potentially interesting result of our analysis.
\end{itemize}

\begin{figure}[htbp]
  \begin{center}
%%%%%%%%%%%% .epic figure here
\setlength{\unitlength}{0.0006in}%
\begingroup\makeatletter\ifx\SetFigFont\undefined%
\gdef\SetFigFont#1#2#3#4#5{%
  \reset@font\fontsize{#1}{#2pt}%
  \fontfamily{#3}\fontseries{#4}\fontshape{#5}%
  \selectfont}%
\fi\endgroup%
{\renewcommand{\dashlinestretch}{20}
\begin{picture}(4000,4000)(+1450,20)
\thicklines
\drawline(600,0)(600,3000)
\drawline(690,2600)(600,3000)(510,2600)
\drawline(0,0)(1200,0)
\drawline(500,20)(700,20)
\drawline(500,80)(700,80)
\drawline(500,180)(700,180)
\drawline(500,320)(700,320)
\drawline(500,500)(700,500)
%\drawline(500,360)(700,360)
%\drawline(400,400)(800,400)
\drawline(400,700)(800,700)
\drawline(500,720)(700,720)
\drawline(500,780)(700,780)
\drawline(500,880)(700,880)
\drawline(400,1000)(800,1000)
\drawline(500,1020)(700,1020)
\drawline(500,1080)(700,1080)
\drawline(500,1180)(700,1180)
\drawline(400,1300)(800,1300)
\drawline(500,1320)(700,1320)
\drawline(500,1380)(700,1380)
\drawline(500,1480)(700,1480)
\drawline(300,2000)(900,2000)
\drawline(500,2020)(700,2020)
\drawline(500,2080)(700,2080)
\drawline(400,2300)(800,2300)
\drawline(500,2320)(700,2320)
\drawline(500,2380)(700,2380)
\put(1300,-150){\makebox(0,0)[lb]{anti-ferromagnetic ground state}}
\put(1000,300){\makebox(0,0)[lb]{derivatives $\sim L^{-2}$ }}
\put(1000,900){\makebox(0,0)[lb]{spinons $\sim L^{-1}$ }}
\put(3300,400){\makebox(0,0)[lb]{$\bigg\}$ gapless excitations}}
\put(1000,1850){\makebox(0,0)[lb]{open chains and anti-chiral impurity
     $\sim L^0$ (gap) }}
\put(0,3700){\makebox(0,0)[lb]{anomalous}}
\put(0,3400){\makebox(0,0)[lb]{dimensions}}
\end{picture}
}

%%%%%%%%%%%%
  \end{center}
  \caption{Spectrum of anomalous dimensions constructed over the chiral
anti-ferromagnetic ground state. An identical tower of excitations
can be constructed over the anti-chiral anti-ferromagnetic ground
state.}
  \label{fig:chain2}
\end{figure}
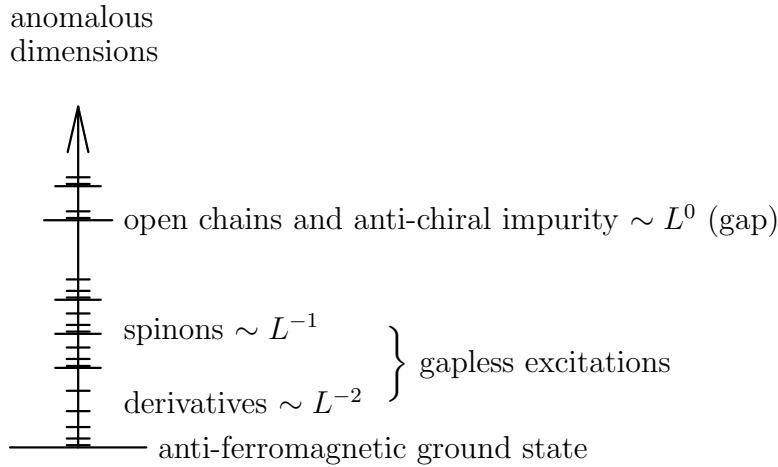

Before we move on to discuss the
nature of each excitation in the next subsections, it must be
stressed that, due to the highly non-trivial nature of the anti-ferromagnetic
vacuum and to the zero momentum constraint, not all combinations of excitations
are allowed. Only certain combinations of the elementary excitations above
correspond to physical states that can be identified with QCD operators.
This fact is already known in the case of spinons from the literature on
anti-ferromagnets and is responsible for their unusual statistic
\cite{Reshetikhin:1990jn}. We shall see more examples of this fact.

%%%%%%%%%%%%%%%%%%%%%%%%%%%%%%%%%%%%%%%%%%%%%%%%%%%%%%%%%%%%%%%%%%%%%%%%%%

\subsection{Adding an anti-chiral impurity}

We start by showing that replacing one $f$ by an $\bar{f}$ in the
chiral spin chain gives rise to a mass gap, i.e. the chains containing
such ``impurities{}'' have an energy that is larger than the vacuum by
a finite amount in the thermodynamic limit $L\to  \infty$.
For simplicity, we will work in the absence of derivatives,
but it will be obvious
{}from the discussion in Sec.~\ref{sec:adss} that the presence of derivatives
can only increase the gap.

For later use, recall that anti-ferromagnetic closed
and open spin chains (with ``free{}'' boundary conditions)
of length $L$ have a ground state energy behaving
as~\cite{Bloete:1986qm,Affleck:1986bv}:
\begin{equation}
     E_\mathrm{closed}(L) = aL+O(1/L), \;\;\;
     E_\mathrm{open}(L) = aL+b+O(1/L).
    \label{bounds}
\end{equation}
The constant $a$ depends on the constant shift in the Hamiltonian. With our
normalization it has been fixed in \eqref{antiferrogro} to be
$a = -\frac{5\as \NNc}{12\pi} <0$. The constant $b>0$ is the surface energy
(not present for the closed chain with periodic boundary conditions).
The terms of order $O(1/L)$ are also of physical interest in many
contexts, being related to the central charge of the system, but they
will not be needed in the following arguments.

In \Secref{sec:5} we have shown how to express the Hamiltonian in the chiral
sector as a spin chain. In order to study the effect of impurities, we
must now write the Hamiltonian containing both $f$ and $\bar{f}$ (but no
derivatives) as a spin chain (or better, a spin ladder).
This is the Hamiltonian found
in~\cite{Ferretti:2004ba} to describe the renormalization of operators
without derivatives. This Hamiltonian is, of course,
the restriction of our present Hamiltonian \eqref{Hclosed} to the subsector
without derivatives.

The spin system can be represented as a spin ladder.
We have already described the action of $\opspin$ on $f$
in \eqref{spinopera}. The same operator acts on $\bar{f}$ in the obvious
way.
But now we also need to describe operators switching $f$ with $\bar{f}$.
This can be done by introducing another independent spin operator
$\spin$ corresponding to a spin $1/2$ representation and acting on
each component of $f$ and $\bar{f}$ (thought of as a doublet) as:
\begin{eqnarray}
&&\spin ^3\,f=f,\qquad \spin ^+\,f=0,\qquad \spin ^-\,f=\bar{f},
\nonumber  \\*
&&\spin ^3\,\bar{f}=-\bar{f},\qquad \spin ^+\,\bar{f}=f,\qquad
\spin ^-\,\bar{f}=0.
\end{eqnarray}
The two spins $\spin_l $ and ${\bf S}_l$ can be
visualized as sitting on a spin ladder, two parallel spin chains
connected by links at each site. The mixing matrix can be written as
\begin{equation}
  \Gamma =\unit\,\sum_{l=1}^{L}\left\{
  \frac{1}{2}\left(\opident+\spin ^3_l\spin ^3_{l+1}\right)\oplilham_{l,l+1}
  +\frac{1}{8}\left(\opident-\spin_l\cdot\spin_{l+1}\right)
  \right\} \equiv\unit\, \mathrm{H}, \label{mmattix}
\end{equation}
where $\oplilham_{l,l+1}$ is the integrable spin one link
Hamiltonian \eqref{spin1}
found in the previous section:
\begin{equation}
    \oplilham_{l,l+1} =
    \left[\sfrac{7}{6}\opident+\half\Scq-\half(\Scq)^2\right].
\end{equation}
For chiral states, such as  $\mathcal{O}$ in \eqref{chirop},
$\spin^+_l\mathcal{O} = 0$, $\spin^3_l \mathcal{O} = \mathcal{O}$,
and the mixing matrix reduces to \eqref{spin1}.

Denote for convenience by $f^A$ ($A=+,0,-$) the spin one triplet
introduced in \eqref{tripl} and
write the full wave function as a linear combination
\begin{equation}
    \Psi = \sum_{n=1}^L \psi_n^{A_1\dots A_{n-1}\dot{B}A_{n+1}\dots A_L}
           \tr(f^{A_1}\dots f^{A_{n-1}}\bar{f}^{\dot{B}}
               f^{A_{n+1}}\dots f^{A_L}),
\end{equation}
in terms of some coefficients $\psi_n$.
Clearly, one could always bring the impurity
to the beginning of the trace, but it is convenient
to have it at an arbitrary position.

We want to solve the eigenvalue problem
\begin{equation}
   \mathrm{H}\, \Psi = E_\mathrm{impurity}(L)\Psi,
\end{equation}
where $E_\mathrm{impurity}(L)$ denotes the energy of the
chiral spin chain with $L$
sites, one of which is replaced by an impurity $\bar{f}$.
Decompose
\begin{equation}
    \tr(f^{A_1}\dots f^{A_{n-1}}\bar{f}^{\dot{B}}
           f^{A_{n+1}}\dots f^{A_L}) = \sum_\Omega
           C_\Omega^{A_{n+1}\dots A_L, A_1\dots A_{n-1}}
           \tr(\bar{f}^{\dot{B}}\Xi_\Omega)
\end{equation}
where $\Omega$ and $\Xi_\Omega$ are respectively the eigenvalues
and eigenvectors of the open XXX$_1$ spin chain with $L-1$
sites and $C_\Omega^{A_{n+1}\dots A_L, A_1\dots A_{n-1}}$ the
overlap coefficients.

Partially projecting the original coefficients $\psi_n$ onto the new basis:
\begin{equation}
         \chi_n^{\dot{B}} = \psi_n^{A_1\dots A_{n-1}\dot{B}A_{n+1}\dots A_L}
          C_\Omega^{A_{n+1}\dots A_L, A_1\dots A_{n-1}}
\end{equation}
and doing some index manipulations we get the secular equation
\begin{equation}
    (3+ \Omega - E_\mathrm{impurity}(L)) \chi_n^{\dot{B}} =
    \frac{3}{2}(\chi_{n-1}^{\dot{B}}+\chi_{n+1}^{\dot{B}})
\end{equation}
which has the usual plane wave solution.
Of course, one must pick a direction in the anti-chiral index,
say $ {\dot{B}} = {\dot{0}}$, and the others are obtained
by $SU(2)_R$ rotation.

Thus, setting $\chi_n^{\dot{0}} = \eeee^{i p n}$ we get the dispersion relation
\begin{equation}
       E_\mathrm{impurity}(L) = \Omega + 3(1-\cos p).
\end{equation}
The energy of the open chain on the RHS must be evaluated for $L-1$
sites because removing the impurity not only opens the chain but
also lowers the number of sites by one.
The lowest value for the energy is thus reached for $p=0$.
The important point is that we can now use the bounds (c.f.~\eqref{bounds})
together with the fact that $a<0<b$ to show that
\begin{equation}
     E_\mathrm{impurity}(L) =
     E_\mathrm{closed}(L) +O(L^0),
\end{equation}
thus establishing the presence of a mass gap for these excitations.

Let us move on to consider now the gapless excitations,
all to be found in the integrable chiral sector.

%%%%%%%%%%%%%%%%%%%%%%%%%%%%%%%%%%%%%%%%%%%%%%%%%%%%%%%%%%%%%%%%%%%%%%%%%%%%%%
\subsection{Spinons}

Here we review the spectrum of excitations around the
anti-ferromagnetic vacuum of the XXX$_1$ model \cite{Takhtajan:1981xx}.
The excitations correspond to creating holes
in the Fermi sea by removing one or more 2-strings%
\footnote{It is instructive not to think of holes as removed
  2-strings, but rather as gaps in the sequence of mode numbers
(e.g.~$n_j=j$, $n_{j+1}=j+2$)
which was defined below \eqref{eq:asdfjlh} for the ground state.
This explains why the hole excitation
carries spin $-1/2$ and not $-2$.}
{}from the vacuum and/or to adding $k$-strings with $k\neq 2$.
They obey a non-trivial exchange statistics
\cite{Reshetikhin:1990jn,Faddeev:1996iy}.

We have already obtained the density of 2-strings $\rho(\lambda)$
in the ground state in~\eqref{rholambda}.
If one of the 2-strings is removed, then instead of
\eqref{inteqdens} the density of 2-strings satisfies
\begin{eqnarray}\label{inteqdens0a}
&&\frac{1}{2}\,\frac{1}{\lambda^2+1/4}
+\frac{3}{2}\,\frac{1}{\lambda^2+9/4} =\pi\rho_{\rm hole}(\lambda)
+\frac{\pi }{L}\,\delta  (\lambda -\mu ) \nonumber \\
&&\qquad+2\int_{-\infty}^{+\infty}d\xi\,\rho_{\rm hole}(\xi)\,
\left[\frac{1}{(\lambda-\xi)^2+1}
+\frac{1}{(\lambda-\xi)^2+4}\right].
\end{eqnarray}
The density of 2-strings now has the form%
\footnote{In order to avoid
confusion, we stress that we will always be concerned with the density
of 2-strings. By the notation $\rho_{\rm hole}$ we mean the density of two
strings in the presence of a hole in the sea of two strings. Similarly for
other type of impurities.}
\begin{equation}
\rho_{\rm hole}(\lambda )=\rho (\lambda )+\frac{1}{L}\,\sigma_{\rm
hole} (\lambda -\mu )
\end{equation}
and can again be found by the Fourier transform:
\begin{eqnarray}\label{shole}
&&\sigma_{\rm hole} (\lambda )=-\int_{-\infty }^{+\infty
}\frac{dq}{2\pi }\, \,\frac{\eeee^{iq\lambda +|q|}}{4\cosh^2\frac{q}{2}} =-\delta (\lambda
)+\frac{\lambda }{2\sinh\pi \lambda } \nonumber \\
&&+\frac{1}{4\pi }\left(\Psi \left(\frac{2+i\lambda }{2}\right)+\Psi
\left(\frac{2-i\lambda }{2}\right)-\Psi \left(\frac{1+i\lambda
}{2}\right)-\Psi \left(\frac{1-i\lambda }{2}\right)\right),
\end{eqnarray}
where $\Psi (\lambda )=\Gamma' (\lambda )/\Gamma (\lambda )$. The
energy and the momentum of the hole can be readily
calculated:%
\footnote{The integrals are most easily done by using the
Fourier representation for the density and the fact that
$p'_{\rm hole}(\mu )=-2\varepsilon _{\rm hole}(\mu )$.}
\begin{eqnarray}
p_{\rm hole}(\mu )\eq -2\int_{-\infty }^{+\infty }d\lambda
\,\sigma_{\rm hole} (\lambda -\mu )\left(\pi -\arctan 2\lambda
-\arctan\frac{2\lambda }{3}\right) \nonumber \\ \eq \frac{\pi
}{2}-\arctan\sinh\pi \mu,
\nonumber \\
\varepsilon _{\rm hole}(\mu )\eq -\frac{1}{2} \int_{-\infty
}^{+\infty }d\lambda \,\sigma_{\rm hole} (\lambda -\mu )
\left(\frac{1}{\lambda ^2+1/4}+\frac{3}{\lambda ^2+9/4}\right)
 =\frac{\pi }{2\cosh\pi
\mu }\,.
\end{eqnarray}
Thus the dispersion relation,
\begin{equation}\label{dhole}
\varepsilon _{\rm hole}(p)=\frac{\pi }{2}\,\sin p,
\end{equation}
is linear for the low-energy states.

We can also compute the quantum numbers of the hole. And here we
encounter a surprise \cite{Faddeev:1981ip}. Each individual 2-string
contains two Bethe roots and thus has spin 2, but removing a
2-string from the Fermi sea distorts the distribution of the other
strings and this distortion partially screens the spin of the hole!
Naively, since we have removed a 2-string, we would expect the
correction to the density $\sigma_{\rm hole}$
to be normalized to minus one. Instead we find:
\begin{equation}
\int_{-\infty }^{+\infty }d\lambda \,\sigma_{\rm hole} (\lambda
)=-\frac{1}{4}\,,
\end{equation}
so the perturbed density contains not $L-2$, but $L-1/2$ Bethe
roots. Of course a state with a fractional number of roots or
even a fractional number of 2-strings does not
make sense and the holes in the sea of 2-strings
can only be created in sets of four.%
\footnote{We have tacitly assumed that $L$ is even.
For a spin chain with odd length $L$,
the filled Fermi sea is not physical as it contains
a half-integer number $L/2$ 2-strings.
Here, two holes would lead to a physical state.}
Yet each hole
is an independent excitation of the anti-ferromagnetic spin chain,
which has the dispersion relation \eqref{dhole} and spin 1/2
\cite{Faddeev:1981ip}. The Lorentz quantum numbers of the hole are
$(1/2,0)$.

We now want to investigate what happens if, instead of making a hole,
we add one $k$-string with $k\neq 2$. Of course, since the number of
$l$-roots is already maximized in the anti-ferromagnetic vacuum,
this by itself will not lead to a physical state but
the modification in the density of 2-strings can be investigated
independently of this issue to order $O(1/L)$.
If a $k$-string with $k\neq 2$ is added to a distribution of the
$2$-string, the equation for the density of 2-strings takes the form
(now with $\rho_{k\mathrm{-str}}(\lambda )=\rho (\lambda )+
\frac{1}{L}\,\sigma_{k\mathrm{-str}} (\lambda -\mu )$):
\begin{eqnarray}\label{inteqdens0b}
&&\frac{1}{2}\,\frac{1}{\lambda^2+1/4}
+\frac{3}{2}\,\frac{1}{\lambda^2+9/4} =\pi\rho_{k\mathrm{-str}}(\lambda)
\nonumber
\\ &&+\frac{k-2}{2L}\,\frac{1}{(\lambda -\mu
)^2+(k-2)^2/4}+\frac{k}{L}\,\frac{1}{(\lambda -\mu
)^2+k^2/4}+\frac{k+2}{2L}\,\frac{1}{(\lambda -\mu )^2+(k+2)^2/4}
\nonumber \\
&&+2\int_{-\infty}^{+\infty}d\xi\,\rho_{k\mathrm{-str}}(\xi)\,
\left[\frac{1}{(\lambda-\xi)^2+1}
+\frac{1}{(\lambda-\xi)^2+4}\right].
\end{eqnarray}
The middle line comes from the scattering of a 2-string on the
$k$-string whose rapidity $\mu$ should be determined
self-consistently and depends on the rapidities of other
excitations. The middle line can be obtained by taking the derivative of the
scattering phase for this process:
\begin{equation} \delta_{2k}(\lambda )=2\pi(2-\delta _{k1}) -2\arctan \frac{2\lambda }{k-2}
-4\arctan\frac{2\lambda }{k}-2\arctan \frac{2\lambda }{k+2}\,.
\end{equation}
For the density perturbation we find:
\begin{equation}
\sigma _{\rm 1-str}(\lambda )=-\frac{1}{2\cosh\pi\lambda}
\end{equation}
and
\begin{equation}\label{}
 \sigma _{k\rm -str}(\lambda )=-\int_{-\infty }^{+\infty }
 \frac{dq}{2\pi }\,\eeee^{iq\lambda -(n-2)|q|/2}
 =-\frac{k-2}{2\pi }\,\,\frac{1}{\lambda ^2+(k-2)^2/4}
\end{equation}
for $k>2$. The modification in the density of 2-string in the presence
of one such $k$-string is thus normalized as
\begin{equation}
\int_{-\infty }^{+\infty }d\lambda \,\sigma_{\mathrm{1-str}}(\lambda)
=-\frac{1}{2}\,,
\end{equation}
and
\begin{equation}
\int_{-\infty }^{+\infty }d\lambda \,\sigma_{k\mathrm{-str}} (\lambda
)=-1\,,
\end{equation}
so that effectively the $1$-string adds no roots ($-1/2$ of a
$2$-string plus the single Bethe root) and the $k$-string adds $k-2$
roots to a Bethe state. However, all $k$-strings do not contribute to
the energy and momentum and thus do not correspond to propagating
degrees of freedom. They are responsible for the spin and internal
degrees of freedom of the magnons.

\begin{table}\centering
$\begin{array}{|l|ccc|ccc|c|}\hline
\multicolumn{1}{|c|}{\mbox{type}}& M_l & M_u & M_2 & S_1 & S_2 & D &
\varepsilon(p) \\\hline
\mbox{hole} & -1/2 & 0 & 1/4 & 1/2 & 0 & 0 & \sim\pi |p|/2 \\
\mbox{1-string of $l$'s} & 0 & 0 & 1/2 & 0 & 0 & 0 & - \\
\mbox{$k$-string of $l$'s, $n>2$}  & k-2 & 0 & 1 & -k+2 & 0 & 0 & -\\
\hline
\mbox{1-string of $u$'s} & 1/2 & 1 & -1/4 & 0 & 1/2 & 1 & \sim p^2 /2 \\
\mbox{$k$-string of $u$'s, $n>1$} & 1 & k & -1/2 & k/2-1 & k/2 & k &
\sim p^2/2k \\\hline
\end{array}$

\caption{Chiral excitations and their quantum numbers. $M_2$
indicates the number of the $2$-strings removed from the Fermi sea.}

\label{tableroots}
\end{table}

More generally, a
physical $2N$-magnon state will contain $2N$ holes and some number of
$k$-string with $k\neq 2$.
See Table~\ref{tableroots}
for a summary of chiral excitations (including also the ones
discussed below).
How many strings of each type can be
inserted in a particular state depends on $N$ in a complicated way
and requires a careful analysis of the Bethe equations
\cite{Takhtajan:1981xx,Reshetikhin:1990jn}. The dependence on $N$
leads to a non-trivial exchange statistics of magnons so that the
Hilbert space of $2N$ magnons is not just $2N$ copies of a
single-particle space \cite{Reshetikhin:1990jn,Faddeev:1996iy}. An
obvious consistency condition on a distribution of holes and
$k$-strings (but not the only one!) is that the total number of
$2$-strings, $L/2+M_2$, should be integer and the total number of $l$-roots should
not exceed $L +M_u/2$.

%%%%%%%%%%%%%%%%%%%%%%%%%%%%%%%%%%%%%%%%%%%%%%%%%%%%%%%%%%%%%%%%%%%%%%%%%%%%%%%%
\subsection{Adding derivatives and spin separation}\label{sec:adss}

Let us finally consider the modes excited by the insertion of
a covariant derivative or, in the language of the Bethe ansatz, by
the addition of a $u$-root. Composite QCD operators associated to
physical states of this kind
contain one or more derivatives on top of the anti-ferromagnetic
vacuum of the $SU(2)_L$ spin chain. We will see that these
states are similar to
BMN operators in $\superN=4$ SYM but with important
differences which arise because of the anti-ferromagnetic nature of
the ground state.
The derivative can be thought of as an ``impurity{}'' propagating on
the background composed of $f_{\alpha\beta}$'s. Since the
anti-ferromagnetic vacuum is a Lorentz scalar, the presence of a
derivative adds a chiral and an anti-chiral index to the operator.
We write:
\begin{equation}\label{basicop}
{\cal O}^{(n)}_{\alpha\dot{\alpha}}=
\sum_{l=1}^L\eeee^{\frac{2\pi inl}{L}}\tr f_{\alpha_1\alpha_2} \ldots
D_{\alpha_{0}\dot{\alpha}}f_{\alpha_{2l-1}\alpha_{2l}} \ldots
f_{\alpha_{2L-1}\alpha_{2L}} \psi_\alpha^{\alpha_0\dots\alpha_{2L}},
\end{equation}
where we have introduced the wave-function
$\psi_\alpha^{\alpha_0\dots\alpha_{2L}}$ (linear combination of products of
$\epsilon_{\alpha_i \alpha_j}$) to denote that
all but one $\alpha_k$'s are contracted with each other and
the remaining $\alpha_k=\alpha$.
The operator \eqref{basicop} with
zero momentum ($n=0$) and
$\psi_\alpha^{\alpha_0\dots\alpha_{2L}} = \delta_\alpha^{\alpha_0}\psi_\mathrm{gs}^{\alpha_1\dots\alpha_{2L}}$, where $\psi_\mathrm{gs}^{\alpha_1\dots\alpha_{2L}}$ is the ground state wave function, is a descendant:
\[
\sum_{l=1}^L\tr f_{\alpha_1\alpha_2} \ldots
D_{\mu}f_{\alpha_{2l-1}\alpha_{2l}} \ldots
f_{\alpha_{2L-1}\alpha_{2L}}\psi_\mathrm{gs}^{\alpha_1\dots\alpha_{2L}}
=\partial_{\mu} \tr f_{\alpha_1\alpha_2} \ldots f_{\alpha_{2L-1}\alpha_{2L}}
\psi_\mathrm{gs}^{\alpha_1\dots\alpha_{2L}},
\]
and has the same anomalous dimension as the unperturbed operator.
In complete analogy with the BMN states~\cite{Berenstein:2002jq},
if the momentum of the derivative insertion
is non-zero but small, the energy will be small.
Of course, the cyclicity of the trace (zero total
momentum condition) requires taking
$\psi_\alpha^{\alpha_0\dots\alpha_{2L}}$ different from
$\delta_\alpha^{\alpha_0}\psi_\mathrm{gs}^{\alpha_1\dots\alpha_{2L}}$, i.e.
requires perturbing the anti-ferromagnetic vacuum.

Thus, we encounter an important difference between \eqref{basicop} and
BMN operators in $\superN=4$ SYM. Unlike BMN operators, \eqref{basicop}
contains not one but
two elementary excitations! One of them carries
momentum $p=2\pi n/L$ and is associated with the dotted index carried
by the derivative.
The other is associated with the extra undotted index induced by the
derivative, but propagating independently.
The reason is that the number of linearly independent
states obtained by different ways of contracting indices
$\alpha_0,\ldots,\alpha_{2L}$ is enormous for large $L$ and in the
overwhelming majority of cases the unpaired index $\alpha$ that is
left after all contractions are done is not the undotted index of
the derivative. The left spin of the derivative thus looses its
individuality and dissolves in the sea of the background spin. The
excess of the left spin is carried away by an independent
excitation. The mechanism of spin
separation just described is a direct analog of the spin-charge
separation in~\cite{Anderson:1987gf}.
In this particular case the left and right spinors
should have opposite momenta because of the trace condition, but in
a more general setup with several derivatives different excitations
are absolutely independent.

We just saw that states consisting of only one derivative must
necessarily contain a spinon and thus their anomalous dimension scales
as $1/L$.
However, an operator of the following type
with two derivatives and all undotted indices contracted
\[
{\cal O}_{\dot{\alpha}\dot{\beta}}=
\sum C_{lm}\tr f_{\alpha_1\alpha_2} \ldots
D_{\alpha_{2L+1}\dot{\alpha}}f_{\alpha_{2l-1}\alpha_{2l}} \ldots
D_{\alpha_{2L+2}\dot{\beta}}f_{\alpha_{2m-1}\alpha_{2m}} \ldots
f_{\alpha_{2L-1}\alpha_{2L}}\psi^{\alpha_1\dots\alpha_{2L+2}}
\]
might have parametrically smaller anomalous dimension $\sim 1/L^2$, since
at least the zero momentum condition can
be satisfied by making the anti-chiral
excitations propagate in opposite directions.

To investigate whether it is really possible
to have a physical state with energy scaling like $1/L^2$ over the
ground state (i.e. without spinons) we study the distortion in the
distribution of 2-strings caused by the presence of one $u$-root exactly
in the same spirit as in the previous section.

If we add a $u$-root,
the equations for the density of 2-strings is modified by the
scattering term. The scattering phase is
\begin{equation}
 \delta _{2u}(\lambda )=2\arctan\lambda-\pi ,
\end{equation}
and modifies the equation for the distribution of 2-strings as follows:
\begin{eqnarray}
&&\frac{1}{2}\,\frac{1}{\lambda^2+1/4}
+\frac{3}{2}\,\frac{1}{\lambda^2+9/4} =\pi\rho_{u}(\lambda)
\nonumber \\
&&+2\int_{-\infty}^{+\infty}d\xi\,\rho_u(\xi)\,
\left[\frac{1}{(\lambda-\xi)^2+1}
+\frac{1}{(\lambda-\xi)^2+4}\right]
-\frac{1}{L}\,\frac{1}{(\lambda-u)^2+1}\,,
\end{eqnarray}
The above equation is solved by
\begin{equation}
\rho_u(\lambda)=\rho(\lambda)+\frac{1}{L}\,\sigma_u(\lambda-u),
\end{equation}
where
\begin{equation}
\sigma_u(\lambda)=\int_{-\infty}^{+\infty}\frac{dq}{2\pi}\,
\,\frac{\eeee^{-iq\lambda}}{4\cosh^2\frac{q}{2}}
=\frac{\lambda}{2\sinh\pi\lambda}
\end{equation}
is the distortion in the sea of 2-strings
caused by the scattering off the $u$-root. The
distortion of the density is normalized as
\begin{equation}
\int_{-\infty}^{+\infty}d\lambda\,\sigma_u(\lambda)=\frac{1}{4}\,.
\end{equation}

Thus, the distortion of the Fermi sea caused by the $u$-root effectively
adds an extra 1/2 of an $l$-root to the background distribution and
the Lorentz spin of an excitation of this type turns out to be
$(0,1/2)$. The scattering off the background distribution of spins
completely screens the left spin of the derivative, so that the
derivative excitation is a right spinor, rather than a vector. Note
however that this excitation is not viable on its own due to the
fractional $l$-root. It needs to be accompanied by a fractional
excitation of type $l$ or another derivative excitation. A single
derivative excitation can only be accompanied by a hole in order for
the Fermi sea to be occupied by an integer number of $2$-strings.
 Globally, this state has spin $(1/2,1/2)$ and
dimension $L+1$ as expected when adding a derivative. However, it
carries not one, but \emph{two} independent momenta,
one for the $u$-excitation and one for the $l$-distortion.%
\footnote{In fact the two momenta are related by the momentum constraint,
which would otherwise, in the case of a single excitation, force
the momentum to be zero.}

We can now compute the energy and the momentum of the derivative
excitation:
\begin{eqnarray}
p_u(u)\eq \pi-2\arctan\frac{2u}{3}
\nonumber \\
&&-2\int_{-\infty}^{+\infty }d\lambda\,\sigma_u(\lambda-u)\,
\left(\pi - \arctan 2\lambda -\arctan\frac{2\lambda }{3} \right)
\nonumber  \\
\eq \frac{\pi}{2}-2\arctan\frac{2u}{3}+2\arctan 2u-\arctan\sinh\pi u.
\nonumber \\
\varepsilon_u(u)\eq \frac{3/2}{u^2+9/4}
-\frac{1}{2}\int_{-\infty}^{+\infty}d\lambda\,\sigma_u(\lambda-u)\,
\left(\frac{1}{\lambda^2+1/4} +\frac{3}{\lambda^2+9/4}\right)
\nonumber \\
\eq \frac{3/2}{u^2+9/4}-\frac{1/2}{u^2+1/4}+\frac{\pi}{2\cosh\pi
u}\,,
\end{eqnarray}
which gives a parametric representation of the dispersion relation
$\varepsilon=\varepsilon_u(p)$.
Thus, in the limit of small momenta,
the contribution of a right spinor with the momentum $p=2\pi
n/L$ to the anomalous dimension is given by a BMN-like formula:
\begin{equation}
\gamma(n)-\gamma _0=\frac{\pi \as \NNc n^2}{L^2}\,.
\end{equation}

Looking at the quantum number summarized in Table~\ref{tableroots} we
see that the configuration with 2~$u$-modes and a 1-string of $l$-roots
over the anti-ferromagnetic vacuum correspond to a physical state with no
holes in the 2-string vacuum. Thus there exist operators with two derivatives
and anomalous dimension of order $O(1/L^2)$ over the ground state.

For completeness let us mention that
the computation for an $k$-string of $u$-roots is similar. We obtain the
dispersion relation
\begin{eqnarray}
p_u(u)\eq 2\pi-2\arctan\frac{2u}{k+2}-2(1-\delta_{k,2})\arctan \frac{2u}{k-2}\,,
\nonumber \\
\varepsilon_u(u)\eq
\frac{(k+2)/2}{u^2+(k+2)^2/4}+\frac{(k-2)/2}{u^2+(k-2)^2/4}\,,
\end{eqnarray}
and find that also one $l$-root is effectively added to the Fermi
sea. The $k$-string describes an $k$-particle bound state of
magnons.

%%%%%%%%%%%%%%%%%%%%%%%%%%%%%%%%%%%%%%%%%%%%%%%%%%%%%%%%%%%%%%%%%%%%%%%%%%%
\subsection*{Acknowledgments}
We would like to thank F.~Anfuso, R.~Argurio, D.~Arnaudon,
A.~Belitsky, C.~Callan, A.~Doikou, V.~Fateev, A.~Gorsky,
J.A.~Gracey, M.~G\"unaydin, H.~Johannesson, N.~Kitanine,
G.~Korchemsky, J.~Minahan, S.~\"Ostlund, M.~Staudacher, M.~Strassler,
A.~Vainshtein
for discussions. K.Z. would like to thank KITP, Santa Barbara for
kind hospitality during the course of this work. The work of
N.~B.~is supported in part by the U.S.~National Science Foundation
Grants No.~PHY99-07949 and PHY02-43680. Any opinions, findings and
conclusions or recommendations expressed in this material are those
of the authors and do not necessarily reflect the views of the
National Science Foundation. The research of G.~F.~is supported by
the Swedish Research Council (Vetenskapsr{\aa}det) contract
622-2003-1124. The work of K.~Z.~was supported in part by the
Swedish Research Council under contract 621-2002-3920, by the
G{\"o}ran Gustafsson Foundation and by NSF grant PHY99-07949.

%%%%%%%%%%%%%%%%%%%%%%%%%%%%%%%%%%%%%%%%%%%%%%%%%%%%%%%%%%%%%%%%%%%%%%%%%%%%%%%%
%%%%%%%%%%%%%%%%%%%%%%%%%%%%%%%%%%%%%%%%%%%%%%%%%%%%%%%%%%%%%%%%%%%%%%%%%%%%%%%%
%%%%%%%%%%%%%%%%%%%%%%%%%%%%%%%%%%%%%%%%%%%%%%%%%%%%%%%%%%%%%%%%%%%%%%%%%%%%%%%%
\appendix

%%%%%%%%%%%%%%%%%%%%%%%%%%%%%%%%%%%%%%%%%%%%%%%%%%%%%%%%%%%%%%%%%%%%%%%%%%%%%%%%
%%%%%%%%%%%%%%%%%%%%%%%%%%%%%%%%%%%%%%%%%%%%%%%%%%%%%%%%%%%%%%%%%%%%%%%%%%%%%%%%
\section{The conformal group and QCD}
\label{sec:A}

The conformal algebra $\alg{su}(2,2)= \alg{so}(2,4)$ consists of
the Lorentz generators $\mathcal{L}^\alpha_\beta$ and
$\bar{\mathcal{L}}^{\dot\alpha}_{\dot\beta}$,
the dilatation operator
$\mathcal{D}$, the momentum $\mathcal{P}_{\alpha\dot\beta}$ and the
conformal boost $\mathcal{K}^{\alpha\dot\beta}$.
They obey the commutation relations
\begin{eqnarray}\label{sdfklj}
    &&{[\mathcal{L}^\alpha_\beta, \mathcal{L}^\delta_\gamma]} =
     \delta^\alpha_\gamma \mathcal{L}^\delta_\beta -
     \delta^\delta_\beta \mathcal{L}^\alpha_\gamma, \quad
    {[\bar{\mathcal{L}}^{\dot\alpha}_{\dot\beta},
     \mathcal{L}^\delta_\gamma]} = 0, \quad
    {[\bar{\mathcal{L}}^{\dot\alpha}_{\dot\beta},
     \bar{\mathcal{L}}^{\dot\delta}_{\dot\gamma}]} =
     \delta^{\dot\alpha}_{\dot\gamma}
     \bar{\mathcal{L}}^{\dot\delta}_{\dot\beta} -
     \delta^{\dot\delta}_{\dot\beta}
     \bar{\mathcal{L}}^{\dot\alpha}_{\dot\gamma}, \nonumber \\
    &&{[\mathcal{D}, \mathcal{L}^\alpha_\beta]} = 0,\quad
    {[\mathcal{D}, \bar{\mathcal{L}}^{\dot\alpha}_{\dot\beta}]}=0,\quad
    {[\mathcal{D}, \mathcal{P}_{\alpha\dot\beta} ]}=
     \mathcal{P}_{\alpha\dot\beta}, \quad
    {[\mathcal{D}, \mathcal{K}^{\alpha\dot\beta} ]} =
     - \mathcal{K}^{\alpha\dot\beta}, \nonumber \\
    &&{[\mathcal{L}^\alpha_\beta, \mathcal{P}_{\gamma\dot\delta}]} =
     \delta^\alpha_\gamma \mathcal{P}_{\beta\dot\delta} -
     \frac{1}{2}\delta^\alpha_\beta \mathcal{P}_{\gamma\dot\delta},\quad
    {[\mathcal{L}^\alpha_\beta, \mathcal{K}^{\gamma\dot\delta}]} =
     -\delta^\alpha_\gamma \mathcal{K}^{\beta\dot\delta} +
     \frac{1}{2}\delta^\alpha_\beta \mathcal{K}^{\gamma\dot\delta},\nonumber\\
    &&{[\bar{\mathcal{L}}^{\dot\alpha}_{\dot\beta},
     \mathcal{P}_{\gamma\dot\delta}]} = \delta^{\dot\alpha}_{\dot\delta}
     \mathcal{P}_{\gamma\dot\beta} - \frac{1}{2}
     \delta^{\dot\alpha}_{\dot\beta}\mathcal{P}_{\gamma\dot\delta}, \quad
    {[\bar{\mathcal{L}}^{\dot\alpha}_{\dot\beta},
     \mathcal{K}^{\gamma\dot\delta}]} = -\delta^{\dot\alpha}_{\dot\delta}
     \mathcal{K}^{\gamma\dot\beta} + \frac{1}{2}
     \delta^{\dot\alpha}_{\dot\beta}\mathcal{K}^{\gamma\dot\delta},\nonumber\\
    &&{[\mathcal{K}^{\alpha\dot\beta}, \mathcal{P}_{\gamma\dot\delta}]}=
     \delta^\alpha_\gamma \bar{\mathcal{L}}^{\dot\beta}_{\dot\delta} +
     \delta^{\dot\beta}_{\dot\delta} {\mathcal{L}}^\alpha_\gamma +
     \delta^\alpha_\gamma \delta^{\dot\beta}_{\dot\delta} \mathcal{D}.
\end{eqnarray}

The conformal algebra has rank three and the
Cartan subalgebra can be chosen to be
\begin{equation}
    \mathcal{J}_0 =
    \big\{\mathcal{L}^1_1,\;\; \bar{\mathcal{L}}^{\dot{1}}_{\dot{1}},
    \;\;\mathcal{D} \big\}
\end{equation}
(recall that $\mathcal{L}^1_1 = - \mathcal{L}^2_2$)
whereas raising and lowering
operators can be written respectively as
\begin{equation}
    \mathcal{J}_{+} =
    \big\{\mathcal{L}^1_2,\;\; \bar{\mathcal{L}}^{\dot{1}}_{\dot{2}},
    \;\; \mathcal{K}^{\alpha\dot\beta} \big\}\quad\hbox{and}\quad
    \mathcal{J}_{-} =
    \big\{\mathcal{L}^2_1, \;\; \bar{\mathcal{L}}^{\dot{2}}_{\dot{1}},
    \;\; \mathcal{P}_{\alpha\dot\beta} \big\}.
\end{equation}
%

%%%%%%%%%%%%%%%%%%%%%%%%%%%%%%%%%%%%%%%%%%%%%%%%%%%%%%%%%%%%%%%%%%%%%%%%%%%%%%%%
\subsection{Multiplets}

A highest weight state $|\mbox{h.w.}\rangle$
is a state that is annihilated by the operators in
$\mathcal{J}_{+}$.
By acting with the lowering generators $\mathcal{J}_{-}$
one can generate all the states of the multiplet (module)
\[
\big\{1,\mathcal{J}_-,\mathcal{J}_-^2,\ldots\big\}
|\mbox{h.w.}\rangle.
\]
We will give an example directly related to QCD below
in \appref{sec:Osc}.

A highest-weight state and
thus the corresponding multiplet is characterized by the
charge eigenvalues of the Cartan subalgebra $\mathcal{J}_0$.
A standard way of expressing these are the Dynkin labels
\begin{equation}
     {[p,r,q]}
\end{equation}
where $p$ and $q$ are \emph{twice} the $SU(2)_L\times SU(2)_R$ spins
$(S_1,S_2)$ and $r$
is a negative number related to the scaling dimension $D$ by
$r=-D-(p+q)/2$.

%%%%%%%%%%%%%%%%%%%%%%%%%%%%%%%%%%%%%%%%%%%%%%%%%%%%%%%%%%%%%%%%%%%%%%%%%%%%%%%%
\subsection{Unitarity and shortening}
\label{sec:unit}

Generic unitary representations $[p,r,q]$ of the conformal group
satisfy the unitarity bound%
\footnote{For the compact counterpart $\alg{su}(4)=\alg{so}(6)$,
all Dynkin labels would have to be non-negative integers.}
\[\label{eq:unitarity}
p+r+q\leq -2,
\]
but there are also
exceptional unitary multiplets
whose Dynkin labels are related by
\[\label{eq:unitaryfields}
p+r=-1,\quad q=0\qquad \mbox{or}\qquad
r+q=-1,\quad p=0 .
\]
In QCD, the elementary fields correspond to the second type
while local operators are of the first type.
Note that all the unitary multiplets
are infinite-dimensional as required by the
non-compact nature of the conformal group.

Generically, all combinations of the
momentum generator $\mathcal{P}$ generate new
states when acting on the highest-weight state.
Here, there are four exceptions
\<\label{eq:shortening}
\mbox{scalars} && p=0=q,\quad r=-1,\nln
\mbox{chiral fields} && p\neq 0=q,\quad r=-1-p,\nln
\mbox{anti-chiral fields} && p=0\neq q,\quad r=-1-q,\nln
\mbox{conserved currents} && p\neq 0\neq q,\quad r=-2-q-p.
\>
We know for instance that the scalars%
\footnote{Of course, there are no
elementary scalar fields in
QCD but it is just as easy to discuss the general case.}
satisfy the equations of motion (classically)
$\mathcal{P}^2|\mbox{h.w.}\rangle=0$
while the conserved currents are defined by
$\mathcal{P}^\mu|\mbox{h.w.}_\mu\rangle=0$.
The chiral and anti-chiral fields satisfy
both types of conditions.

The chiral and anti-chiral fields \eqref{eq:shortening}
are the central objects in QCD.
We will denote these multiplets by
\[
\mdl^{+k}:{[k,-1-k,0]}\qquad\mbox{and}\qquad
\mdl^{-k}:{[0,-1-k,k]},  \label{anine}
\]
where $k$ is restricted to $1$ or $2$ in the specific case of QCD.
The fields also enjoy a number of useful features
in terms of representation theory
which makes them rather easy to handle despite
the fact that they are infinite-dimensional.
For instance, they can be represented by means
of a set of harmonic oscillators as we shall
demonstrate below.

%%%%%%%%%%%%%%%%%%%%%%%%%%%%%%%%%%%%%%%%%%%%%%%%%%%%%%%%%%%%%%%%%%%%%%%%%%%%%%%%
\subsection{Oscillator representation}
\label{sec:Osc}

The conformal algebra has a nice representation
in terms of a set of bosonic oscillators
${[a^\alpha, a^{\dagger}_\beta]} = \delta^\alpha_\beta$ and
${[b^{\dot\alpha}, b^{\dagger}_{\dot\beta}]} =
   \delta^{\dot\alpha}_{\dot\beta}$
($\alpha = 1,2$, $\dot\alpha = \dot{1},\dot{2}$).
The generators take the form:
\begin{eqnarray}
 && \mathcal{L}^\alpha_\beta = a^{\dagger}_\beta a^\alpha -
   \sfrac{1}{2} \delta_\beta^\alpha a^{\dagger}_\gamma a^\gamma, \quad
 \bar{\mathcal{L}}^{\dot\alpha}_{\dot\beta} =  b^{\dagger}_{\dot\beta}
   b^{\dot\alpha} -
   \sfrac{1}{2} \delta_{\dot\beta}^{\dot\alpha}
   b^{\dagger}_{\dot\gamma} b^{\dot\gamma}\nonumber\\
  &&\mathcal{D} = 1 + \sfrac{1}{2}a^{\dagger}_\lambda a^\lambda +
        \sfrac{1}{2} b^{\dagger}_{\dot\gamma} b^{\dot\gamma}, \quad
  \mathcal{P}_{\alpha\dot\beta} =
     a^{\dagger}_\alpha b^{\dagger}_{\dot\beta}, \quad
  \mathcal{K}^{\alpha\dot\beta} = a^\alpha b^{\dot\beta}\label{SO24}
\end{eqnarray}
which can easily be seen to obey the commutation relations \eqref{sdfklj}.
(See e.g.~\cite{Bars:1982ep} for the general theory and extension
to supergroups.)
It is also possible to construct an extra $\alg{u}(1)$ generator, not part of the
conformal algebra and commuting with it:
\begin{equation}
   \mathcal{A} = \sfrac{1}{2}a^{\dagger}_\lambda a^\lambda -
        \sfrac{1}{2} b^{\dagger}_{\dot\gamma} b^{\dot\gamma}.
\end{equation}
This generator is associated to the ``chiral{}'' type symmetry transformation
discussed in the text.

If we choose a vacuum vector $\vac$ annihilated by
$a^\alpha$ and $b^{\dot\beta}$, highest weights are of type
$(a^\dagger_2)^k \vac$ or $(b^\dagger_{\dot{2}})^k\vac$. However, since the
Lorentz structure is the familiar one, we will only use
$\mathcal{K}^{\alpha\dot\beta}|\mbox{h.w.}\rangle =0$ as the relevant
condition and refer to the whole Lorentz multiplets
$a^\dagger_{\alpha_1}\dots a^\dagger_{\alpha_k}\vac$ and
$b^\dagger_{\dot{\alpha}_1}\dots b^\dagger_{\dot{\alpha}_k}\vac$
as ``primaries{}''. The modules constructed by acting with lowering
operators on the above states
provide the oscillator realization of the fundamental fields in the
Lagrangian.
Thus, the modules \eqref{anine} are spanned by:
\begin{eqnarray}
   \mdl^{+k} \eq \langle a^\dagger_{\alpha_1}\dots
    a^\dagger_{\alpha_k}\vac,
    \;\;a^\dagger_{\alpha_1}\dots a^\dagger_{\alpha_{k+1}}
    b^\dagger_{\dot{\alpha}_1} \vac,  \;\;
    a^\dagger_{\alpha_1}\dots a^\dagger_{\alpha_{k+2}}
    b^\dagger_{\dot{\alpha}_1} b^\dagger_{\dot{\alpha}_2} \vac, \dots\big\rangle,
    \nonumber \\
   \mdl^{-k} \eq  \big\langle b^\dagger_{\dot{\alpha}_1}\dots
    b^\dagger_{\dot{\alpha}_k}\vac,
    \;\;a^\dagger_{\alpha_1}b^\dagger_{\dot{\alpha}_1}\dots
    b^\dagger_{\dot{\alpha}_{k+1}} \vac,  \;\;
    a^\dagger_{\alpha_1}a^\dagger_{\alpha_2}b^\dagger_{\dot{\alpha}_1}\dots
    b^\dagger_{\dot{\alpha}_{k+2}} \vac, \dots\big\rangle. \label{osc}
\end{eqnarray}
For convenience, the modules appearing in QCD shall also be
denoted as $\mdl^\Phi$ where $\Phi = f, \bar{f}, \chi, \psi,
\bar{\chi}, \bar{\psi}$ are the physical primary fields.

Recalling that $\mathcal{P}_{\alpha\dot\beta} = -i D_{\alpha\dot\beta}$ we
write (e.g. $f_{\alpha\beta} = a^\dagger_\alpha a^\dagger_\beta\vac$):
\begin{eqnarray}
   \mdl^f \eq \mdl^{+2}= \left\langle f_{\alpha_1\alpha_2},
    Df_{\alpha_1\alpha_2\alpha_3{\dot\alpha}_1},
    D^2 f_{\alpha_1\alpha_2\alpha_3\alpha_4 {\dot\alpha}_1{\dot\alpha}_2 }
    \dots \right\rangle, \nonumber \\
   \mdl^{\bar{f}}\eq \mdl^{-2}=  \left\langle
    \bar{f}_{\dot{\alpha}_1\dot{\alpha}_2},
    D\bar{f}_{\alpha_1\dot{\alpha}_1\dot{\alpha}_2\dot{\alpha}_3},
    D^2 \bar{f}_{\alpha_1\alpha_2
       \dot{\alpha}_1\dot{\alpha}_2\dot{\alpha}_3\dot{\alpha}_4}
    \dots \right\rangle, \nonumber \\
   \mdl^\psi \eq \mdl^{+1}= \left\langle \psi_{\alpha_1},
    D\psi_{\alpha_1\alpha_2\dot{\alpha}_1},
    D^2 \psi_{\alpha_1\alpha_2\alpha_3\dot{\alpha}_1\dot{\alpha}_2}
    \dots \right\rangle, \nonumber \\
   \mdl^\chi \eq  \mdl^{+1}=\left\langle \chi_{\alpha_1},
    D\chi_{\alpha_1\alpha_2\dot{\alpha}_1},
    D^2 \chi_{\alpha_1\alpha_2\alpha_3\dot{\alpha}_1\dot{\alpha}_2}
    \dots \right\rangle, \nonumber \\
   \mdl^{\bar{\chi}} \eq  \mdl^{-1}=\left\langle \bar{\chi}_{\dot{\alpha}_1},
    D\bar{\chi}_{\alpha_1\dot{\alpha}_1\dot{\alpha}_2},
    D^2 \bar{\chi}_{\alpha_1\alpha_2
       \dot{\alpha}_1\dot{\alpha}_2\dot{\alpha}_3}
    \dots \right\rangle, \nonumber \\
   \mdl^{\bar{\psi}} \eq \mdl^{-1}= \left\langle \bar{\psi}_{\dot{\alpha}_1},
    D\bar{\psi}_{\alpha_1\dot{\alpha}_1\dot{\alpha}_2},
    D^2 \bar{\psi}_{\alpha_1\alpha_2
       \dot{\alpha}_1\dot{\alpha}_2\dot{\alpha}_3}
    \dots \right\rangle.  \label{singletons}
\end{eqnarray}
The indices of the same kind (dotted or undotted) in the above modules are
always understood as totally symmetrized.

%%%%%%%%%%%%%%%%%%%%%%%%%%%%%%%%%%%%%%%%%%%%%%%%%%%%%%%%%%%%%%%%%%%%%%%%%%%%%%%%
\subsection{Tensor products of two fields}

In computing the action of the dilatation operator on two adjacent
fields in a gauge invariant operator
we are naturally led to the problem of decomposing the
product of two of the modules \eqref{singletons}.
In the oscillator representation we would have
to introduce two commuting sets of oscillators
${[a_{(i)}^\alpha, a_{(j),\beta}^{\dagger}]} = \delta_{ij}\delta^\alpha_\beta$ and
${[b_{(i)}^{\dot\alpha}, b_{(j),{\dot\beta}}^{\dagger}]} =
   \delta_{ij}\delta^{\dot\alpha}_{\dot\beta}$ where we will only be
interested in the case $i, j = 1, 2$ but in principle one could generalize
to the product of an arbitrary number of  \eqref{singletons}.
The generators are simply the sums of the generators for each set, in
particular the conformal boost for the product of two singletons
\eqref{singletons} is
$ \mathcal{K}^{\alpha\dot\beta} = a_{(1)}^\alpha b_{(1)}^{\dot\beta} +
a_{(2)}^\alpha b_{(2)}^{\dot\beta}$. Highest weight states in the product can be
found by looking for solutions to
$\mathcal{K}^{\alpha\dot\beta}|\mathrm{\mbox{h.w.}}\rangle = 0$
amongst all linear combinations of states of the type
\begin{equation}
  a_{(1),\alpha_1}^\dagger\dots
  a_{(1),\alpha_p}^\dagger
  b_{(1),\dot{\alpha}_1}^\dagger\dots
  b_{(1),\dot{\alpha}_q}^\dagger
  a_{(2),\beta_1}^\dagger\dots
  a_{(2),\beta_m}^\dagger
  b_{(2),\dot{\beta}_1}^\dagger\dots
  b_{(2),\dot{\beta}_n}^\dagger \vac
\end{equation}
where now $\vac$ is annihilated by both sets of $a_{(i)}^\alpha$ and
$b_{(i)}^{\dot\alpha}$.

Obviously there are more possibilities to satisfy the highest weight condition
and in fact the product of two arbitrary modules  \eqref{singletons}
decomposes into an infinite sum of
irreducible components, distinguished by a quantum number $j$ named
``conformal spin{}''. After using the oscillator representation to get some
practice, one can recast the results in the more compact language of
fields or in the even more compact language of Dynkin labels.
Assuming $k\geq |k^{\prime}|$ we have
\begin{equation}
     \mdl^{\pm k} \otimes  \mdl^{k^\prime} =
     \sum_{j=\mp k^\prime}^\infty \mdl_j^{\pm k + k^\prime}.
     \label{deco}
\end{equation}
The Dynkin labels of the modules in the decomposition
\eqref{deco} are
\begin{eqnarray}
\mdl^{+s}_{j}:\begin{cases}
[s+2j,-s-2-j,0]&\mbox{for } -s/2 \leq j\leq 0,
\\
[s+j,-s-2-2j,j]&\mbox{for } j\geq 0,
\end{cases}
\nonumber\\
\mdl^{-s}_{j}:\begin{cases}
[0,-s-2-j,s+2j]&\mbox{for } -s/2 \leq j\leq 0,
\\
[j,-s-2-2j,s+j]&\mbox{for } j\geq 0.
\end{cases} \label{modules}
\end{eqnarray}
A proof of these identities is sketched in
\appref{sec:polya}.

Let us explain in practice how the decomposition works by considering
the product of two modules $\mdl^f$:
\begin{equation}
     \mdl^f \otimes \mdl^f = \sum_{j=-2}^\infty
        \mdl^{ff}_j.
\end{equation}
To understand the structure of the decomposition we
present the explicit form of the first few elements in the first few
modules. In order to do that we shall introduce the following notation:
\begin{itemize}
\item All indices of the same kind written explicitly are meant as
\emph{totally symmetrized} even if they belong to different fields.
\item \emph{Anti-symmetrization} between two indices by contraction with
$\epsilon^{\alpha\beta}$ is denoted by replacing the contacted indices
by the symbol $\xxxx$ (or $\xxxxd$).
\end{itemize}
For instance
$\epsilon^{\gamma\delta}f_{\gamma(\alpha}f_{\beta)\delta}$ is written
as $f_{\xxxx\alpha}f_{\beta\xxxx}$. As long as we consider only the products of
two modules  \eqref{singletons},
there is no room for confusion since indices can only be
anti-symmetrized between distinct fields.

Thus we have:
\begin{eqnarray}
  \mdl^{ff}_{-2} \!\eq \! \left\langle f_{\xxxx\xxxx}f_{\xxxx\xxxx}, \;\;
  Df_{\xxxx\xxxx\alpha\dot\alpha}f_{\xxxx\xxxx} + f_{\xxxx\xxxx}Df_{\xxxx\xxxx\alpha\dot\alpha},\,
  \dots \right\rangle, \nonumber \\
  \mdl^{ff}_{-1} \!\eq \! \left\langle f_{\xxxx\alpha}f_{\xxxx\beta}, \;\;
  Df_{\xxxx\alpha\gamma\dot\gamma}f_{\xxxx\beta} +
  f_{\xxxx\alpha}Df_{\xxxx\beta\gamma\dot\gamma}, \;\;
  Df_{\xxxx\xxxx\alpha\dot\alpha}f_{\xxxx\xxxx} - f_{\xxxx\xxxx}Df_{\xxxx\xxxx\alpha\dot\alpha},\,
  \dots \right\rangle, \nonumber \\
  \mdl^{ff}_{0} \!\eq \! \left\langle f_{\alpha\beta}f_{\gamma\delta}, \;\;
  Df_{\alpha\beta\eta\dot\eta}f_{\gamma\delta} +
  f_{\alpha\beta} Df_{\gamma\delta\eta\dot\eta}, \;\;
  Df_{\xxxx\alpha\beta\dot\eta}f_{\xxxx\gamma} -
  f_{\xxxx\alpha} Df_{\xxxx\beta\gamma\dot\eta},\, \dots \right\rangle, \nonumber \\
  \mdl^{ff}_{1} \!\eq \! \left\langle
  Df_{\alpha\beta\eta\dot\eta}f_{\gamma\delta} -
  f_{\alpha\beta} Df_{\gamma\delta\eta\dot\eta}, \,\dots \right\rangle
\end{eqnarray}
and so on. The first element in each module is the primary. The descendants
are obtained taking derivatives and decomposing the result into irreducible
representations. Primaries in the modules $\mdl^{ff}_{j}$ for
$j\geq 0$ have $j$ covariant derivatives. Similar decompositions hold
for all the modules. In particular:
\begin{equation}\label{eq:ffprod}
     \mdl^f \otimes \mdl^{\bar f} = \sum_{j=+2}^\infty
     \mdl^{f\bar f}_j
\end{equation}
where, for instance,
\begin{equation}
  \mdl^{f\bar{f}}_{2}=
  \langle f_{\alpha\beta}\bar{f}_{\dot\alpha\dot\beta},
  Df_{\alpha\beta\gamma\dot\alpha} \bar{f}_{\dot\beta\dot\gamma} +
  f_{\alpha\beta}D\bar{f}_{\gamma\dot\alpha\dot\beta\dot\gamma}, \dots \rangle
\end{equation}
and so on.

Similarly, we shall write
$\mdl^f \otimes \mdl^\psi = \sum_{j=-1}^\infty\mdl^{f\psi}_j$
and so on for the tensor products involving fermions.
The lower bound of the sum
can be read off from
\eqref{singletons,deco} in each case.

%%%%%%%%%%%%%%%%%%%%%%%%%%%%%%%%%%%%%%%%%%%%%%%%%%%%%%%%%%%%%%%%%%%%%%%%%%%%%%%%%%%%
\subsection{Polya counting}
\label{sec:polya}

Using Polya counting%
\footnote{See e.g.
\cite{Aharony:2003sx,Bianchi:2003wx,Beisert:2003te} for
an introduction in the context of field theory. Polya theory has also been
used in~\cite{Sundborg:1999ue} to count single-trace operators in $\superN=4$ SYM.}
we can prove the
decomposition of tensor products
\eqref{deco}.
Let us denote a state with weight $[p,r,q]$
of $SU(2,2)$ by the monomial
$a^{p} b^{q} d^{-r-p/2-q/2}$
\[\label{eq:monomial}
[p,r,q]\to
a^{p} b^{q} d^{-r-p/2-q/2}.
\]
In other words the exponents of $a,b$ are the third components of spin
and the exponent of $d$ is the dimension.
A multiplet of states is consequently written as
a polynomial, where the integer coefficients
indicate the multiplicities of states
with fixed quantum numbers.
This notation is very convenient for dealing with
the infinite dimensional modules that appear
for the conformal group. For example, consider
the field strength component $f_{11}$ and the
set of all descendants using the derivative
$D_{1\dot 1}$
\[
\set{f_{11},D_{1\dot 1} f_{11},D_{1\dot 1} D_{1\dot 1} f_{11}, \ldots}.
\]
Using the rule \eqref{eq:monomial} we can write this as the polynomial
\[
a^2d^{2}
+a^3bd^3
+a^4b^2d^4
+\ldots\,.
\]
Of course, this is just a geometric series which sums up as
\[
\frac{a^2d^{2}}{1-dab}\,.
\]
So we have found a compact way to represent $f_{11}$ along
with its $D_{1\dot 1}$ descendants.

All the states of a long unitary multiplet $[p,r,q]$
of the conformal group
are summarized in
\[\label{eq:longmulti}
[p,r,q]=\frac{a^{p}b^{q}d^{-r-p/2-q/2}(1-a^{-2p-2})(1-b^{-2q-2})}{(1-dab)(1-da/b)(1-db/a)(1-d/ab)(1-a^{-2})(1-b^{-2})}\,.
\]
In the denominator one finds the momentum and $SU(2)_L$ and $SU(2)_R$
ladder generators.
The two differences in the numerator make $SU(2)$ multiplets
finite-dimensional.

Generic unitary multiplets $[p,r,q]$ of the conformal group
satisfy the unitarity bound \eqref{eq:unitarity} but there are also
the exceptional multiplets in \eqref{eq:shortening} which need special
attention.
Let us discuss these special cases,
where we refer to the long multiplet defined in \eqref{eq:longmulti}
as $[p,r,q]\indups{L}$.
The simplest example (not directly related to QCD) is an on-shell scalar,
which has Dynkin labels $[0,-1,0]$. It turns
out that in order to remove the terms that are
reducible due to the equation of motion
$D^2\phi=\ldots$ we have to subtract a
long multiplet $[0,-3,0]\indups{L}$.%
\footnote{Not incidentally this is just the weight of $D^2\phi$.}
\[
[0,-1,0]=[0,-1,0]\indups{L}-[0,-3,0]\indups{L}.
\]
For an on-shell chiral field of spin $S_1=p/2$
with $r=-1-p$ one finds
\[
[p,r,0]=[p,r,0]\indups{L}-[p-1,r-1,1]\indups{L}+[p-2,r-1,0]\indups{L}
\]
and analogously for an anti-chiral field.
These multiplets violate the bound \eqref{eq:unitarity},
but due to their tightly restricted charges they
are indeed unitary.

A less restricted type of multiplet,
a conserved current,
sits right at the unitarity bound \eqref{eq:unitarity}.
Its state content is reduced by the
states $D^\mu J_{\mu\ldots}=0$
\[
[p,r,q]=[p,r,q]\indups{L}-[p-1,r,q-1]\indups{L}.
\]

Using these expressions and sum rules of geometric series,
it is straightforward to prove \eqref{deco}.

%%%%%%%%%%%%%%%%%%%%%%%%%%%%%%%%%%%%%%%%%%%%%%%%%%%%%%%%%%%%%%%%%%%%%%%%%%%%%%%%
%%%%%%%%%%%%%%%%%%%%%%%%%%%%%%%%%%%%%%%%%%%%%%%%%%%%%%%%%%%%%%%%%%%%%%%%%%%%%%%%
\section{Length-two primary states}
\label{sec:C}

Let us now discuss the gauge invariant primary states that can be
constructed with only two fundamental fields ($L=2$) and an
arbitrary number of  derivatives. We do this partly to present a
check of our results against the standard literature and partly as an
illustration of how to use the Hamiltonians \eqref{Hclosed} and \eqref{Hopen}.

Reducing to $L=2$ chains amounts to projecting all
modules in the decompositions
\eqref{modules} on the gauge singlet states.
In considering $L=2$ states and in order to make connection with the classical
papers on the subject is more convenient to rewrite the gluon
modules in a way that makes parity manifest.
To this end, we introduce the following projection operators:
\begin{equation}
   \opproj^{f\bar{f} \pm \bar{f}f }_{j} = \sfrac{1}{2}\left( \opident
   \pm (-)^j \opperm \right) \big(\opproj^{f\bar{f}}_j +
   \opproj^{\bar{f}f}_j \big),
\end{equation}
in terms of which the Hamiltonian \eqref{gg} reads
\begin{equation}
     \opham^{\mathrm{FF}} = \sum_{j=-2}^\infty E^{ff}_j
     \big(\opproj_j^{ff} + \opproj_j^{\bar{f}\bar{f}}\big) +
     \sum_{j=+2}^\infty \big(E^+_j   \opproj^{f\bar{f}+\bar{f}f }_{j} +
      E^-_j \opproj^{f\bar{f}-\bar{f}f }_{j}\big), \label{ggparity}
\end{equation}
where
\begin{equation}
     E^\pm_j  = h(j-2) + h(j+2) -\frac{11}{6} \mp
     \frac{6}{(j-1) j (j+1) (j+2)}\,.
\end{equation}
For sake of clarity, we give expression for the primary and the first few
descendants of these new modules:
\begin{eqnarray}
  \mdl^{f\bar{f} +\bar{f}f }_{2} \eq
  \big\langle f_{\alpha\beta}\bar{f}_{\dot\alpha\dot\beta} +
  \bar{f}_{\dot\alpha\dot\beta} f_{\alpha\beta}, \nl\quad
   Df_{\alpha\beta\gamma\dot\alpha} \bar{f}_{\dot\beta\dot\gamma} +
  f_{\alpha\beta}D\bar{f}_{\gamma\dot\alpha\dot\beta\dot\gamma} +
  \bar{f}_{\dot\alpha\dot\beta} Df_{\alpha\beta\gamma\dot\gamma} +
  D\bar{f}_{\alpha\dot\alpha\dot\beta\dot\gamma} f_{\beta\gamma},
  \nonumber \nl\quad
  Df_{\alpha\beta\gamma\xxxxd} \bar{f}_{\xxxxd\dot\alpha} -
  \bar{f}_{\xxxxd\dot\alpha}Df_{\alpha\beta\gamma\xxxxd}, \;\;
  f_{\alpha\xxxx}D\bar{f}_{\xxxx \dot\alpha\dot\beta\dot\gamma} -
  D\bar{f}_{\xxxx \dot\alpha\dot\beta\dot\gamma}f_{\alpha\xxxx},\, \dots
  \big\rangle, \nonumber\\
  \mdl^{f\bar{f} -\bar{f}f }_{2} \eq
  \big\langle f_{\alpha\beta}\bar{f}_{\dot\alpha\dot\beta} -
   \bar{f}_{\dot\alpha\dot\beta} f_{\alpha\beta}, \nonumber \nl\quad
   Df_{\alpha\beta\gamma\dot\alpha} \bar{f}_{\dot\beta\dot\gamma} +
  f_{\alpha\beta}D\bar{f}_{\gamma\dot\alpha\dot\beta\dot\gamma} -
  \bar{f}_{\dot\alpha\dot\beta} Df_{\alpha\beta\gamma\dot\gamma} -
  D\bar{f}_{\alpha\dot\alpha\dot\beta\dot\gamma} f_{\beta\gamma},
  \nonumber \nl\quad
  Df_{\alpha\beta\gamma\xxxxd} \bar{f}_{\xxxxd\dot\alpha} +
  \bar{f}_{\xxxxd\dot\alpha}Df_{\alpha\beta\gamma\xxxxd}, \;\;
  f_{\alpha\xxxx}D\bar{f}_{\xxxx \dot\alpha\dot\beta\dot\gamma} +
  D\bar{f}_{\xxxx \dot\alpha\dot\beta\dot\gamma}f_{\alpha\xxxx},\, \dots
  \big\rangle, \nonumber\\
  \mdl^{f\bar{f} +\bar{f}f }_{3} \eq
  \left\langle
  Df_{\alpha\beta\gamma\dot\alpha} \bar{f}_{\dot\beta\dot\gamma} -
  f_{\alpha\beta}D\bar{f}_{\gamma\dot\alpha\dot\beta\dot\gamma} -
  \bar{f}_{\dot\alpha\dot\beta} Df_{\alpha\beta\gamma\dot\gamma} +
  D\bar{f}_{\alpha\dot\alpha\dot\beta\dot\gamma} f_{\beta\gamma}, \, \dots
  \right\rangle, \nonumber\\
  \mdl^{f\bar{f} -\bar{f}f }_{3} \eq
  \left\langle
  Df_{\alpha\beta\gamma\dot\alpha} \bar{f}_{\dot\beta\dot\gamma} -
  f_{\alpha\beta}D\bar{f}_{\gamma\dot\alpha\dot\beta\dot\gamma} +
  \bar{f}_{\dot\alpha\dot\beta} Df_{\alpha\beta\gamma\dot\gamma} -
  D\bar{f}_{\alpha\dot\alpha\dot\beta\dot\gamma} f_{\beta\gamma}, \, \dots
  \right\rangle, \nonumber
\end{eqnarray}
where the conventions about symmetrization and anti-symmetrization are
the same as those explained in \Appref{sec:A}.

For the gluon modules the projection onto gauge singlet states
amounts to taking the trace.
All the elements in  $\mdl^{f\bar{f} - \bar{f}f }_{j}$ are traceless
(thus projected out) for $j$ even and those in  $\mdl^{ff}_{j}$,
$\mdl^{\bar{f}\bar{f}}_{j}$ and
$\mdl^{f\bar{f} + \bar{f}f }_{j}$ are
traceless for $j$ odd.
Of course, the quark-gluon modules contain
no gauge singlet at all and the quark-quark modules
contain singlets for all $j$'s.

The surviving length-two gauge invariant primaries
are summarized in Table~\ref{thisisatable}.

\begin{table}[htbp]
\begin{center}
  \begin{tabular}{|c|c|c|c|c|c|c|c|}
    \hline
    \multicolumn{1}{|c}{Name} & \multicolumn{1}{|c|}{Range} &
    \multicolumn{1}{c|}{$(S_1,S_2)$} & \multicolumn{1}{c|}{$D$} &
    \multicolumn{1}{c|}{$[p,r,q]$} &
    \multicolumn{1}{c|}{$\tau$} & \multicolumn{1}{c|}{$\chi$} &
    \multicolumn{1}{c|}{$A$}\\
    \hline
    $\mdl^{ff}_{-2}$ && $(0,0)$ & $4$ & $[0,-4,0]$ & $4$ & $0$ & $2$\\
    $\mdl^{ff}_j$ &$j=0,2,4\dots$& $(2+j/2 ,j/2)$ & $4 + j$ & $[4+j,-6-2j,j]$
          & $2$ & $4$ & $2$\\
    $\mdl^{\bar{f}\bar{f}}_{-2}$ && $(0,0)$ & $4$ & $[0,-4,0]$ & $4$ & $0$ &$ -2$\\
    $\mdl^{\bar{f}\bar{f}}_j$ &$j=0,2,4\dots$& $(j/2 ,2+j/2)$ &
        $4 + j$ & $[j,-6-2j,4+j]$ & $2$ & $-4$ & $-2$\\
    $\mdl^{f\bar{f} + \bar{f}f}_{j}$ &$j=2,4,6\dots$& $(j/2 ,j/2)$
      & $2 + j$ & $[j,-2-2j,j]$ & $2$ & $0$ & $0$\\
    $\mdl^{f\bar{f} - \bar{f}f}_j$ &$j=3,5,7\dots$& $(j/2 ,j/2)$
      & $2 + j$ & $[j,-2-2j,j]$ & $2$ & $0$ & $0$\\
    $\mdl^{\chi\psi}_{-1}$,&& $(0,0)$ & $3$ & $[0,-3,0]$ & $3$ & $0$ & $1$\\
    $\mdl^{\chi\psi}_j$ &$j=0,1,2\dots$& $(1 + j/2,j/2)$ & $3+j$ & $[2+j,-4-2j,j]$ & $2$
           & $2$ & $1$\\
    $\mdl^{\bar\psi\bar\chi}_{-1}$,&& $(0,0)$ & $3$ & $[0,-3,0]$ & $3$ & $0$ &$-1$\\
    $\mdl^{\bar\psi\bar\chi}_j$ &$j=0,1,2\dots$&
      $(j/2,1+j/2)$ & $3+j$ & $[j,-4-2j,2+j]$ & $2$ & $-2$ & $-1$\\
    $\mdl^{\bar\psi\psi}_j$ &$j=1,2,3\dots$& $(j/2,j/2)$ & $2+j$ & $[j,-2-2j,j]$
       & $2$ & $0$ & $0$\\
    $\mdl^{\chi\bar\chi}_j$&$j=1,2,3\dots$& $(j/2,j/2)$ & $2+j$ & $[j,-2-2j,j]$
     & $2$ & $0$ & $0$\\
    \hline
  \end{tabular}
\end{center}
\smallskip

\caption{Table of all gauge singlets primaries of length two.
The indices $(S_1,S_2)$ refer to the Lorentz spins,
$D$ is the classical dimension, $[p,r,q]$ are the Dynkin labels
and $A$ is the $U(1)$ charge introduced above.
The twist
is defined, as usual, by $\tau = D - (p+q)/2$.
In the presence of ``asymmetric{}'' operators, dimension and twist are
not enough to uniquely specify the Lorentz representation -- we need an extra
quantum number, e.g. the ``asymmetry{}'' factor $\chi = p-q$.
Notice the presence of exceptional cases with negative conformal spin.}
\label{thisisatable}
\end{table}

We are particularly interested in computing the one-loop anomalous
dimension of the primaries. First of all, in order
to disentangle the contribution of descendants carrying the same quantum
numbers, one always performs the computation at zero injected momentum,
for which total derivative terms simply do not contribute. Because of the
$U(1)$ symmetry relating the anomalous dimensions of the related operators,
we can restrict our attention to $A\geq 0$. Finally,
the anomalous dimensions of the primaries in the last two modules
in Table~\ref{thisisatable} are obviously the same.
We are thus led to consider the renormalization of the following operators
at zero injected momentum:%
\footnote{The exact conformal operators contain total derivatives in 
particular combinations fixed by conformal symmetry 
\cite{Makeenko:1980bh,Ohrndorf:1981qv}.}
(the range of the conformal spin $j$ is the same
as in Table~\ref{thisisatable})
\begin{eqnarray}
    \mdl^{ff}_{-2}&:& \tr\; F^{\mu\nu} F_{\mu\nu} \nonumber \\
    \mdl^{ff}_j&:& \tr\;  F_{\mu(\nu}\overleftrightarrow{D}_{\rho_1}\dots
            \overleftrightarrow{D}_{\rho_j} F_{\lambda)\sigma}
       -\hbox{traces} \nonumber \\
    \mdl^{f\bar{f} + \bar{f}f}_j&:&
         \tr\;  F^\mu_{(\rho_1}\overleftrightarrow{D}_{\rho_2}\dots
            \overleftrightarrow{D}_{\rho_{j-1}} F_{\rho_j)\mu} -\hbox{traces}\nonumber \\
    \mdl^{f\bar{f} - \bar{f}f}_j&:&
         \tr\;  \tilde{F}^\mu_{(\rho_1}\overleftrightarrow{D}_{\rho_2}\dots
            \overleftrightarrow{D}_{\rho_{j-1}} F_{\rho_j)\mu} -\hbox{traces}\nonumber \\
    \mdl^{\chi\psi}_{-1}&:& \bar{q} q\nonumber \\
    \mdl^{\chi\psi}_j&:& \bar{q} {[\gamma_\mu, \gamma_{(\nu}]}
     \overleftrightarrow{D}_{\rho_1}\dots \overleftrightarrow{D}_{\rho_{j})} q -\hbox{traces}
    \nonumber \\
    \mdl^{\bar\psi\psi}_j&:& \bar{q} \gamma_{(\rho_1}
     \overleftrightarrow{D}_{\rho_2}\dots \overleftrightarrow{D}_{\rho_{j})} q - \hbox{traces}.
\end{eqnarray}
Indices inside round brackets are totally symmetrized and by
``$- \hbox{traces}${}'' we mean that the tensor structure is
irreducible, i.e. tracing over any two indices yields zero.
This reduction will always be
understood in the following.
The operator $\tr\; F^{\mu\nu} F_{\mu\nu} $ (the Lagrangian) contains both the
selfdual and anti-selfdual part but both terms renormalize the same way
due to the $U(1)$ symmetry. Its anomalous dimension is well known:
\begin{equation}
    \tr\; F^{\mu\nu} F_{\mu\nu}:\quad \gamma = -\frac{11}{3} \unit
    \equiv 2\unit E^{ff}_{-2}.
\end{equation}
This is the first check. The factor of two comes because we must
sum the (equal)
contributions of $\opham^\mathrm{FF}_{1,2}$ and $\opham^\mathrm{FF}_{2,1}$.
Similarly, the quark mass term $ \bar{q} q$ contains both the chiral
($\chi\psi$) and anti-chiral ($\bar\chi\bar\psi$) and its anomalous
dimension is also well known:
\begin{equation}
   \bar{q} q :\quad \gamma = - \frac{3}{2}\unit \equiv\unit E^{\chi\psi}_{-1}.
\end{equation}
(This time there is no factor of two because we are considering an open chain.)

The anomalous dimensions of the primary operators of
$\mdl^{f\bar{f} + \bar{f}f}_{j}$ and
$\mdl^{\bar\psi\psi}_j$ are also well known from the classical work
of \cite{Georgi:1951sr,Gross:1974cs}. We do not write explicitly the
flavor structure but even for
flavor singlets, in the large $\NNc$ limit, the mixing between gluonic and
quark operators is suppressed.
In particular, in this limit their matrix of
anomalous dimension is upper triangular and the eigenvalues can be read off
{}from the diagonal entries. It is even possible to rescale the quark kinetic
energy in the Lagrangian by a factor $\sqrt{\NNc}$ and obtain a diagonal matrix:
\begin{eqnarray}
     &&\tr F^\mu_{(\rho_1}\overleftrightarrow{D}_{\rho_2}\dots
            \overleftrightarrow{D}_{\rho_{j-1}} F_{\rho_j)\mu}: \nonumber \\
    &&\qquad \gamma =\unit \bigg( -\frac{4}{j(j-1)} -\frac{4}{(j+1)(j+2)}
       + 4 h(j) -\frac{11}{3} \bigg) \equiv 2\unit\, E^+_j,  \nonumber \\
    &&\bar{q} \gamma_{(\rho_1}
     \overleftrightarrow{D}_{\rho_2}\dots \overleftrightarrow{D}_{\rho_{j})} q: \nonumber\\
    &&\qquad\gamma = \unit \left( h(j+1) + h(j-1) - 3/2 \right) \equiv
    \unit \,E^{\bar\psi \psi}_j.
\end{eqnarray}

In the same way, the anomalous dimension of the primary operator of
$\mdl^{f\bar{f} - \bar{f}f}_j$ can be read off from the
work of \cite{Ahmed:1976ee}:
\begin{eqnarray}
    &&\tr \tilde{F}^\mu_{(\rho_1}\overleftrightarrow{D}_{\rho_2}\dots
            \overleftrightarrow{D}_{\rho_{j-1}} F_{\rho_{j})\mu}:\nonumber \\
    &&\qquad\gamma =\unit\left( -\frac{8}{j(j+1)}
       + 4 h(j) -\frac{11}{3} \right) \equiv 2\unit\, E^-_j.
\end{eqnarray}
The work \cite{Ahmed:1976ee}
contains also the matrix of anomalous dimensions of a second
group of operators but the gluonic term appears to be a descendant, casting
some doubts on the relevance of this second set of anomalous dimensions.

We are left with the anomalous dimensions of the primaries for
$\mdl^{\chi\psi}_j$ and $\mdl^{ff}_j$. We worked
out their anomalous dimensions by diagrammatic techniques,
employing the previous trick of zero injected
momentum and the following additional ones (also well known, see
e.g.~\cite{Robertson:1990bf}):
\begin{itemize}
\item The symmetrization of the indices was accomplished by
contracting each of them with a constant vector $\Delta_{\rho_i}$ etc.
\item The tracelessness condition was enforced by dropping in the evaluation
of the Feynman diagrams all terms containing $(\Delta)^2$, $\Delta^\mu$
(and for the second case also $\Delta^\sigma$).
\item Finally, to ensure that the anti-symmetrization of any three indices
had been removed, we evaluated a particular component that is manifestly
free of such contribution, namely we set (at the end!):
$\Delta_\rho = \delta_{1 \rho}$, $\mu = 2$, (and for the
second case also $\sigma = 2$).
\end{itemize}
The resulting anomalous dimensions are:
\begin{eqnarray}
        \bar{q} {[\gamma_\mu, \gamma_{(\nu}]}
     \overleftrightarrow{D}_{\rho_1}\dots \overleftrightarrow{D}_{\rho_{j})} q
    &:&\quad \gamma =\unit\left( 2 h(j+1) -\frac{3}{2} \right)
       \equiv \unit\, E^{\chi \psi}_j, \nonumber \\
    \tr\; F_{\mu(\nu}\overleftrightarrow{D}_{\rho_1}\dots
            \overleftrightarrow{D}_{\rho_j} F_{\lambda)\sigma} &:&\quad
    \gamma = \unit \left( 4 h(j+2) - \frac{11}{3} \right)
      \equiv 2\unit\, E^{ff}_j,
\end{eqnarray} confirming the corresponding entries in the Hamiltonian. The
last two operators are ``asymmetric{}''. They
cannot be generated perturbatively in the operator product of two
currents because such product does not change chirality. They might
however be interesting for comparing with lattice
calculations~\cite{Robertson:1990bf}.

%%%%%%%%%%%%%%%%%%%%%%%%%%%%%%%%%%%%%%%%%%%%%%%%%%%%%%%%%%%%%%%%%%%%%%%%%%%%%%%%
%%%%%%%%%%%%%%%%%%%%%%%%%%%%%%%%%%%%%%%%%%%%%%%%%%%%%%%%%%%%%%%%%%%%%%%%%%%%%%%%
\section{R-matrix}
\label{sec:YB}

%%%%%%%%%%%%%%%%%%%%%%%%%%%%%%%%%%%%%%%%
\paragraph{Formalism.}

Let us give a brief introduction into the R-matrix formalism.
We define several vector spaces which we label by $k,l,\ldots$.
These are (not necessarily equal) modules $\mdl^{k,l,\ldots}$ of a symmetry algebra
and they are thus called `spins'.
In our case the symmetry algebra is the conformal algebra
and a `spin' will be a chiral field strength multiplet.
To each spin we associate a spectral parameter $u_{k,l,\ldots}$.
The scattering of two spins $k,l$ is assumed to be elastic:
Neither the spin modules $\mdl^{k,l}$ nor the spectral parameters $u_{k,l}$
are modified, there is merely a phase-shift when two
spins are interchanged.
The phase shift is described by
the R-matrix $\opR_{kl}(u_k-u_l)$,
which is a unitary bi-linear operator acting on
the two spins
\[\opR_{kl}(u_k-u_l):\mdl^k\times\mdl^l\to\mdl^k\times \mdl^l.\]
In this formalism, integrability means that the
Yang-Baxter equation is satisfied.
It ensures that the order in which
spins scatter does not matter
\[
\opR_{ik}(u_{ik})
\opR_{il}(u_{il})
\opR_{kl}(u_{kl})
=
\opR_{kl}(u_{kl})
\opR_{il}(u_{il})
\opR_{ik}(u_{ik}),
\]
where we have defined
\[
u_{kl}=u_k-u_l.
\]
The form of the R-matrix is in general very difficult, but it
can be constructed for arbitrary semi-simple algebras and
arbitrary irreducible representations.
In our case the `spins' will all be of the
type of oscillator representations. These
have special properties that allow us to compute the
R-matrix rather conveniently.

%%%%%%%%%%%%%%%%%%%%%%%%%%%%%%%%%%%%%%%%
\paragraph{R-matrix for spin oscillators.}

First of all, we introduce the compact notation $\mathcal{J}^{A}{}_{B}$
for the generators of $\alg{su}(2,2)$.
Here, an index $A$ combines a chiral and an anti-chiral index $\alpha$ and $\dot\alpha$.
We can now introduce a combined oscillator
$A^A=\set{a^\alpha,b^\dagger_{\dot\beta}}$ which behaves like a
four-component version of $a^\alpha$.
The combined generators are now written as
\[\label{eq:zweiuhr}
\mathcal{J}^{A}{}_{B}
=A^{\dagger}_B A^A-
\sfrac{1}{4}\,\delta^{A}_B A^{\dagger}_C A^C,
\qquad
\mathcal{A}=
\half A^{\dagger}_A A^A+1
\]
and satisfy the commutation relation
\[\label{eq:zweiuhrzwei}
\comm{\mathcal{J}^A{}_B}{\mathcal{J}^C{}_D}=
\delta^A_D\mathcal{J}^C{}_B-\delta^C_B\mathcal{J}^A{}_D,
\qquad
\comm{\mathcal{J}^A{}_B}{\mathcal{A}}=0.
\]
Let us now consider the R-matrix acting on the two
oscillator modules $\mdl^k$ and $\mdl^l$.
We know that the R-matrix is invariant under conformal symmetry
therefore it is useful to consider the decomposition
of the tensor product $\mdl^k\otimes\mdl^l$
into irreducible components, \eqref{deco}.
As each component $\mdl_{j}^{kl}$
appears with multiplicity one,
the R-matrix can merely assign one eigenvalue to each:
\[\label{eq:Rdeco}
\opR_{kl}(u)={\textstyle\sum}_j
\opproj^{kl}_{j} R^{kl}_j(u).
\]
We will show below that these eigenvalues
obey the recursion formula
\[\label{eq:Reigen}
R^{kl}_{j}(u_{kl})
=
\frac{u_{kl}+\sfrac{i}{4}\bigbrk{J^2_{kl,j}-J^2_{kl,j-1}}}
{u_{kl}-\sfrac{i}{4}\bigbrk{J^2_{kl,j}-J^2_{kl,j-1}}}\,
R^{kl}_{j-1}(u_{kl}).
\]
Here $J^2_{kl,j}$ denotes the eigenvalue of the
quadratic Casimir
\[
\mathcal{J}^2=
\mathcal{J}^A{}_B
\mathcal{J}^B{}_A
\]
on $\mdl^{kl}_j$.
For a highest-weight representation
with Dynkin labels $[p,r,q]$ it is given by%
\footnote{A curious observation is
that the quadratic Casimir vanishes precisely for
$[0,0,0]$ (trivial representation),
$[2,-3,0]$,
$[0,-3,2]$ (field strength components),
$[1,-4,1]$ (conserved flavor currents),
$[0,-4,0]$ (Lagrangian) and
$[1,-2,1]$.}
\[\label{eq:casimir}
J^2_{[p,r,q]}=
\half p(p+2)
+\quarter (2r+p+q)(2r+p+q+8)
+\half q(q+2).
\]
This result for the R-matrix is completely general, it holds for any two
oscillator representations, even fermionic oscillators,
for any unitary group.

%%%%%%%%%%%%%%%%%%%%%%%%%%%%%%%%%%%%%%%%
\paragraph{Closed Chiral Chain.}

Let us now consider only spins
which are chiral field strengths, i.e.~only modules of
the type $\mdl^f=\mdl_{[2,-3,0]}$.
The monodromy matrix of this system is given by
\[
\mathrm{\Omega}_k(u)=
\opR_{k,1}(u)
\opR_{k,2}(u)
\ldots
\opR_{k,L}(u)
\]
where the auxiliary spin labeled by $k$ transforms like
a field strength $\mdl^f$.
The transfer matrix is
\[
\mathrm{T}(u)=\tr_k\mathrm{\Omega}_k(u)
\]
The nearest-neighbor Hamiltonian of an
integrable system is given by the
logarithmic derivative of the R-matrix at $u=0$
\[
\opham^{ff}=i\opR^{-1}_{ff}(0)\,\frac{\partial \opR_{ff}}{\partial u}(0).
\]
The recursion formula for the eigenvalues
$E^{ff}_j$ of the two-site Hamiltonian is thus
\[
E^{ff}_{j}=
\frac{8}{J^2_{j}-J^2_{j-1}}
+E^{ff}_{j-1}.
\]
In our case the quadratic Casimir
\eqref{eq:casimir} of $\mdl_{j}^{ff}$ is given
by $J^2_j=2(j+2)(j+3)$. The recursion formula
for the energy eigenvalues is thus
\[
E^{ff}_{j}=
\frac{2}{j+2}
+E^{ff}_{j-1}=
2h(j+2)+2E^f.
\]
where $E^f$ is an energy shift that does not follow from integrability.
This agrees precisely with the result from planar QCD and thus
the sector of chiral field strengths is integrable.

%%%%%%%%%%%%%%%%%%%%%%%%%%%%%%%%%%%%%%%%
\paragraph{Open Chiral Chain.}

The integrable open chiral chain has spins
$\mdl^f$ at all sites except $1$ and $L$,
where the spin is $\mdl^{\chi}=\mdl^{\psi}=\mdl_{[1,-2,0]}$.
This system is characterized by the
monodromy matrix \cite{Arnaudon:2004xx}%
\footnote{In \cite{Arnaudon:2004xx} the system of $SU(2)$ spin $+1$ in the
bulk and $SU(2)$ spin $+1/2$ at the boundary was investigated. It
arises in the $SU(2)_L$ sector of QCD and the $SU(2,2)$ system is a
straightforward generalization.} \footnote{For the chain of length
$L=2$ we should choose
$\mathrm{\Omega}_k(u)=\opR_{k,1}(u)\opR_{k,2}(u)$ instead where the
auxiliary spin is $\mdl^{\chi}=\mdl^{\psi}$.}
\[
\mathrm{\Omega}_k(u)=
\opR_{k,1}(u\pm\sfrac{i}{2})
\opR_{k,2}(u)
\ldots
\opR_{k,L-1}(u)
\opR_{k,L}(u\pm\sfrac{i}{2})
\]
where the auxiliary spin labeled by $k$ transforms like
a field strength $\mdl^f$.
The open chain transfer matrix with $SU(2,2)$-preserving
boundary condition ($\mathbf{K}=1$) is
\[
\mathrm{T}(u)=
\tr_k
\mathrm{\Omega}_k(u)
\mathrm{\Omega}^{-1}_k(-u).
\]
The Hamiltonian in the bulk is given by the same expression as before.
At the boundaries the Hamiltonian is modified due to
the shift in the definition of $\mathrm{\Omega}_k(u)$
\[
\opham_{\chi f}=
i\opR_{\chi f}(\pm i/2)^{-1}
\frac{\partial\opR_{\chi f}}{\partial u}(\pm i/2)
\]
consequently, the energy eigenvalues obey
\[
E^{\chi f}_{j}=
\frac{4}{J^2_{j}-J^2_{j-1}+2}
+\frac{4}{J^2_{j}-J^2_{j-1}-2}
+E^{\chi f}_{j-1}.
\]
The eigenvalue of the second Casimir for $\mdl^{\chi f}_j$ is
$J^2_{\chi f,j}=2j^2+8j+\sfrac{15}{4}$; when we substitute this
we obtain
\[
E^{\chi f}_{j}=
\frac{1}{j+1}+
\frac{1}{j+2}
+E^{\chi f}_{j-1}=
h(j+2)+h(j+1)+E^\chi+E^f
\]
where $E^\chi+E^f$ is a constant we cannot determine here.

Also the chains with one or two anti-chiral fermions at the ends,
but all chiral field strengths are integrable.
We simply replace first/last R-matrix
in $\mathrm{\Omega}$ by $\opR_{k,1/L}(u\pm\sfrac{3i}{2})$.
The eigenvalue of the second Casimir for $\mdl^{\bar\chi f}_j$ is
$J^2_{\chi f,j}=2j^2+4j-\sfrac{9}{4}$ and we obtain
\[
E^{\bar\chi f}_{j}=
\frac{1}{j+2}+
\frac{1}{j-1}
+E^{\chi f}_{j-1}=
h(j+2)+h(j-1)+E^\chi+E^f.
\]
In this case an anti-chiral field is not an obstacle to
integrability. The reason why an anti-chiral field strength breaks
integrability is that it may propagate while the anti-chiral quark is fixed
at the end of the chain.
In other words, integrability breaks down for mixed field strengths
due to $E^{f\bar f,\opperm}_j\neq 0$.

%%%%%%%%%%%%%%%%%%%%%%%%%%%%%%%%%%%%%%%%
\paragraph{Proof.}

Here we will present a proof of the Yang-Baxter equation
for the R-matrix given above. It is similar to the one for
$\superN=4$ given in \cite{Beisert:2004ry}.
The three spin modules
for the YBE will be of oscillator type and one of them
will be fundamental.
First of all, consider the R-matrix of an
oscillator spin $k$ and a fundamental spin $i$.
The tensor product $\mdl^{k}\otimes\mdl\indup{fund}$
has two irreducible components which we shall denote by
$\mdl^{ki}_\pm$. The projectors are given by
\[
\opproj^{ki}_\pm=\frac{1}{2}\pm\frac{\mathcal{J}_{ki}+1-\half A_k}{2A_k-1}
\]
where $A_k$ is the value of $\mathcal{A}$ on $\mdl^k$
and $\mathcal{J}_{ki}$ is the algebra generator
acting on $\mdl^k$ whose components $\mathcal{J}_{k}{}^A{}_B$
specify the map $\mdl\indup{fund}\to\mdl\indup{fund}$.
The difference of eigenvalues of the quadratic Casimir is
\[
J^2_{ki,+}-J^2_{ki,-}=2(2A_k-1).
\]
The R-matrix \eqref{eq:Rdeco,eq:Reigen} therefore reads
\[
\opR_{ki}(u_{ki})=
\opproj^{ki}_+
+
\frac{u_{ki}-\sfrac{i}{4}(J^2_{ki,+}-J^2_{ki,-})}
     {u_{ki}+\sfrac{i}{4}(J^2_{ki,+}-J^2_{ki,-})}\,
\opproj^{ki}_-
=
\frac{u_{ki}+i-\sfrac{i}{2} A_k+i \mathcal{J}_{ki}}{u_{ki}+\sfrac{i}{2}(2A_k-1)}\,.
\]

Now consider the YBE for two oscillator spins $k,l$
and a fundamental spin $i$
and expand the R-matrices
$\opR_{ik}(u_{ik}),\opR_{il}(u_{il})$ involving
a fundamental spin
\<
0\earel{\stackrel{!}{=}}
\opR_{ik}(u_{ik})\opR_{il}(u_{il})\opR_{kl}(u_{kl})
-\opR_{kl}(u_{kl})\opR_{il}(u_{il})\opR_{ik}(u_{ik})
\nln\eq
\frac{i(u_{ik}+u_{il}+2i-\sfrac{i}{2} A_k-\sfrac{i}{2} A_l)}
{2\bigbrk{u_{ik}+\sfrac{i}{2}(2A_k-1)}\bigbrk{u_{il}+\sfrac{i}{2}(2A_l-1)}}\,
\bigcomm{\mathcal{J}_{ki}+\mathcal{J}_{li}}{\opR_{kl}(u_{kl})}
\nl
+\frac{i(u_{kl}+\sfrac{i}{2} A_k-\sfrac{i}{2} A_l)}
{2\bigbrk{u_{ik}+\sfrac{i}{2}(2A_k-1)}\bigbrk{u_{il}+\sfrac{i}{2}(2A_l-1)}}\,
\bigcomm{\mathcal{J}_{ki}-\mathcal{J}_{li}}{\opR_{kl}(u_{kl})}
\nl
-\frac{1}{2\bigbrk{u_{ik}+\sfrac{i}{2}(2A_k-1)}\bigbrk{u_{il}+\sfrac{i}{2}(2A_l-1)}}\,
\bigacomm{\comm{\mathcal{J}_{ki}}{\mathcal{J}_{li}}}{\opR_{kl}(u_{kl})}
\nl
-\frac{1}{2\bigbrk{u_{ik}+\sfrac{i}{2}(2A_k-1)}\bigbrk{u_{il}+\sfrac{i}{2}(2A_l-1)}}\,
\bigcomm{\acomm{\mathcal{J}_{ki}}{\mathcal{J}_{li}}}{\opR_{kl}(u_{kl})}.
\>
The first term vanishes because
\[
\mathcal{J}_{ki}+\mathcal{J}_{li}
=\mathcal{J}_{kl,i}
\]
is the rotation generator acting on $\mdl^k\otimes\mdl^l$
which clearly commutes with the covariant R-matrix.
Let us now introduce the symbol $\mathbf{C}_{kl,i}$
for the combination of generators in the second term
\[
\mathcal{J}_{ki}-\mathcal{J}_{li}
=\mathbf{C}_{kl,i}.
\]
Then we can express the commutator in the third term
in terms of $\mathbf{C}_{kl,i}$
\[
\comm{\mathcal{J}_{ki}}{\mathcal{J}_{li}}
=\quarter \comm{\mathcal{J}_{kl}^2}{\mathbf{C}_{kl,i}}
\]
and the quadratic Casimir
$\mathcal{J}_{kl}^2$ on $\mdl^k\times\mdl^l$.
Finally, we can write the anti-commutator in the fourth term
using squared generators%
\footnote{Note that we have to distinguish between
$(\mathcal{J}_i)^2=\mathcal{J}_i\mathcal{J}_i$
and the quadratic Casimir
$\mathcal{J}^2=\tr_i \mathcal{J}_i\mathcal{J}_i$.}
\[\label{eq:ziemlichanti}
\acomm{\mathcal{J}_{ki}}{\mathcal{J}_{li}}
=(\mathcal{J}_{kl,i})^2
-(\mathcal{J}_{k,i})^2
-(\mathcal{J}_{l,i})^2.
\]
Using the rules \eqref{eq:zweiuhr,eq:zweiuhrzwei}
for oscillator representations, we can compute
the contracted product of two generators
\[\label{eq:osciprod}
\mathcal{J}^A{}_B
\mathcal{J}^B{}_C
=
(\mathcal{A}-2)\mathcal{J}^A{}_C
+\sfrac{3}{4}(\mathcal{A}^2-1)\delta^A_C.
\]
It is a special property of oscillator representations that
$\mathcal{J}^A{}_B\mathcal{J}^B{}_C$
can be written as a linear combination of
$\mathcal{J}^A{}_C$ and $\delta^A_C$.%
\footnote{For generic representations only a four-fold
product of generators of the above sort
can be written as something simpler.}
We can now simplify the anti-commutator \eqref{eq:ziemlichanti}
\<\acomm{\mathcal{J}_{ki}}{\mathcal{J}_{li}}
\eq
(\mathcal{J}_{kl,i})^2
-(A_k-2)\mathcal{J}_{k,i}
-(A_l-2)\mathcal{J}_{l,i}
-\sfrac{3}{4}(A_k^2+A_l^2-2)
\\\nonumber\eq
(\mathcal{J}_{kl,i})^2
-\half(A_k+A_l-4)\mathcal{J}_{kl,i}
-\sfrac{3}{4}(A_k^2+A_l^2-2)
-\half(A_k-A_l)\mathbf{C}_{kl,i}.
\>
Again all terms except the one proportional to $\mathbf{C}_{kl,i}$
drop out in commutators with $\mathbf{R}_{kl}$.

We can now put everything together and find
\<\label{eq:Rfinal}
0\earel{\stackrel{!}{=}}
\opR_{ik}(u_{ik})\opR_{il}(u_{il})\opR_{kc}(u_{kl})
-\opR_{kl}(u_{kl})\opR_{il}(u_{il})\opR_{ik}(u_{ik})
\nln\eq
i\,\frac{u_{kl}\bigcomm{\mathbf{C}_{kl,i}}{\opR_{kl}(u_{kl})}
+\sfrac{i}{4}\bigcomm{\mathcal{J}_{kl}^2}{\acomm{\mathbf{C}_{kl,i}}{\opR_{kl}(u_{kl})}}}
{2\bigbrk{u_{ki}+\sfrac{i}{2}(2A_k-1)}\bigbrk{u_{li}+\sfrac{i}{2}(2A_l-1)}}\,
\>
To proceed further, let us investigate the form of
the operator $\mathbf{C}_{kl,i}$.
It is a map $\mdl^{k}\otimes \mdl^{l}\otimes \mdl^i\to\mdl^{k}\otimes \mdl^{l}\otimes \mdl^i$.
By construction it is invariant under simultaneous rotation of
all modules by the same amount. When we combine the (oscillator)
modules into irreducible components, we can write
\[\mathbf{C}_{kl,i}:
\mdl^{kl}_j\otimes \mdl\indup{fund}\to
{\textstyle\sum}_{j'}
\mdl^{kl}_{j'}\otimes \mdl\indup{fund}.
\]
By comparing the highest weight vectors of
$\mdl^{kl}_{j}$ and $\mdl^{kl}_{j'}$ we see
that $\mathbf{C}_{kl,i}$ can only be invariant
when $j'\in \set{j-1,j,j+1}$.
Furthermore, $\mathbf{C}_{kl,i}$ has negative
parity under interchange of $k$ and $l$.
As the parity is alternating with $j$, the cases
$j'=j\pm 1$ are allowed while $j'=j$ is not.
In conclusion we can write
\[
\mathbf{C}_{kl,i}:
\mdl^{kl}_j\otimes \mdl\indup{fund}\to
\mdl^{kl}_{j-1}\otimes \mdl\indup{fund}
\,\,\oplus\,\,\mdl^{kl}_{j+1}\otimes \mdl\indup{fund}
\]
This is another generic feature of tensor products
of oscillator representations.
The recursion relation \eqref{eq:Reigen}
\[\label{eq:Reigen2}
R^{kl}_{j}(u_{kl})
=
\frac{u_{kl}+\sfrac{i}{4}\bigbrk{J^2_{kl,j}-J^2_{kl,j-1}}}
{u_{kl}-\sfrac{i}{4}\bigbrk{J^2_{kl,j}-J^2_{kl,j-1}}}\,
R^{kl}_{j-1}(u_{kl})
\]
follows from projecting \eqref{eq:Rfinal}
to the various irreducible components.
Similarly, for a different ordering of the
three spins we obtain the same relation
{}from the YBE
\<
0\earel{\stackrel{!}{=}}
\opR_{ki}(u_{ki})\opR_{kl}(u_{kl})\opR_{il}(u_{il})
-\opR_{il}(u_{il})\opR_{kl}(u_{kl})\opR_{ki}(u_{ki})
\nln\eq
\frac{i\bigbrk{u_{il}-u_{ki}-\sfrac{i}{2} A_l+\sfrac{i}{2} A_k}}
{2\bigbrk{u_{ki}+\sfrac{i}{2}(2A_k-1)}\bigbrk{u_{il}+\sfrac{i}{2}(2A_l-1)}}
\,\bigcomm{\mathcal{J}_{ki}+\mathcal{J}_{li}}{\opR_{kl}(u_{kl})}
\nl
+\frac{i\bigbrk{u_{kl}+2i-\sfrac{i}{2} A_k-\sfrac{i}{2} A_l}}
{2\bigbrk{u_{ki}+\sfrac{i}{2}(2A_k-1)}\bigbrk{u_{il}+\sfrac{i}{2}(2A_l-1)}}
\,\bigcomm{\mathcal{J}_{ki}-\mathcal{J}_{li}}{\opR_{kl}(u_{kl})}
\nl
+\frac{1}
{2\bigbrk{u_{ki}+\sfrac{i}{2}(2A_k-1)}\bigbrk{u_{il}+\sfrac{i}{2}(2A_l-1)}}
\bigacomm{\mathcal{J}_{ki}-\mathcal{J}_{li}}{\bigcomm{\mathcal{J}_{ki}+\mathcal{J}_{li}}{\opR_{kl}(u_{kl})}}
\nl
-\frac{1}
{2\bigbrk{u_{ki}+\sfrac{i}{2}(2A_k-1)}\bigbrk{u_{il}+\sfrac{i}{2}(2A_l-1)}}
\bigcomm{\mathcal{J}_{ki}\mathcal{J}_{ki}-\mathcal{J}_{li}\mathcal{J}_{li}}{\opR_{kl}(u_{kl})}
\nl
-\frac{1}
{2\bigbrk{u_{ki}+\sfrac{i}{2}(2A_k-1)}\bigbrk{u_{il}+\sfrac{i}{2}(2A_l-1)}}
\bigacomm{\comm{\mathcal{J}_{ki}}{\mathcal{J}_{li}}}{\opR_{kl}(u_{kl})}
\nln\eq
i\,\frac{u_{kl}\bigcomm{\mathbf{C}_{kl,i}}{\opR_{kl}(u_{kl})}
+\frac{i}{4}\bigacomm{\comm{\mathcal{J}_{kl}^2}{\mathbf{C}_{kl,i}}}{\opR_{kl}(u_{kl})}}
{2\bigbrk{u_{ki}+\sfrac{i}{2}(2A_k-1)}\bigbrk{u_{il}+\sfrac{i}{2}(2A_l-1)}}\,.
\>

\bibliography{qcd1loop}
\bibliographystyle{nb}

\end{document}